%% file: 02_main_birthday.tex
\documentclass[11pt]{article}
\usepackage[utf8]{inputenc}
\usepackage{amsmath,setspace,geometry}
\usepackage{amsthm}
\usepackage{amsfonts}
\usepackage{amssymb}
\usepackage[shortlabels]{enumitem}
\usepackage{rotating}
\usepackage{pdflscape}
\usepackage{graphicx}
\usepackage{bbm}
\usepackage{comment}
\usepackage[dvipsnames]{xcolor}
\usepackage{hyperref}
\hypersetup{colorlinks=true, linkcolor= BrickRed, citecolor = BrickRed, filecolor = BrickRed, urlcolor = BrickRed, hypertexnames = true}
\usepackage[]{natbib} 
\bibpunct[:]{(}{)}{,}{a}{}{,}
\usepackage[english]{babel}
\usepackage{float}
\usepackage{caption}
\usepackage{subcaption}
\usepackage{etoolbox}
\AtBeginEnvironment{center}{\setlength{\topsep}{0pt}\setlength{\partopsep}{0pt}}
\usepackage{booktabs}
\usepackage{pdfpages}
\usepackage{threeparttable}
\usepackage{lscape}
\usepackage{bm}
\usepackage{tikz}
\usetikzlibrary{positioning, shapes, arrows}
\usepackage{setspace}
\begin{document}

\title{Happy Birthday? Age Labels, Search Criteria,\\ and Matching from Dating to Marriage}
\author{Suguru Otani\thanks{\href{mailto:suguru.otani@e.u-tokyo.ac.jp}{suguru.otani@e.u-tokyo.ac.jp}, Market Design Center, Department of Economics, The University of Tokyo\\
 We thank Pierre-Andr{\'e} Chiappori, Fuhito Kojima, Kentaro Nakajima, Hirofumi Okoshi, Takashi Shimizu, and Kazuharu Yanagimoto for helpful comments. We thank Shuto Fukuda for providing the data and sharing technical and institutional knowledge of the IBJ platform. This work was supported by JST ERATO Grant Number JPMJER2301, Japan. }}
\date{
First version: July 24, 2026
}
\maketitle

\begin{abstract}
Age is a match trait and a prominent label on search platforms. Using confidential records from a large Japanese marriage platform, I study how a birthday age update affects consideration, applications, relationship progression, and engagement. Only displayed age updates at birthdays. Receivers enter some acceptable-age ranges as they exit others, leaving only modest overall eligibility changes. Application counts often increase. Applications shift from younger to older suitors. Proposal counts fall at nine of ten female receiver ages \(31\text{--}40\), with losses concentrated at entry into early dating. \textcolor{black}{In stage-specific accounting for receivers in their thirties, the no-birthday proposal count is \(12.9\) percent above the benchmark for female receivers and \(6.7\) percent below it for male receivers; the younger-man share in the female-receiver channel is \(4.8\) percentage points higher. Age labels and filters are not neutral windows onto preferences: they govern who marries whom and how many marriages form.}

$\quad$ \\
\textbf{Keywords}: Marriage market, age labels, search filters, consideration sets, online dating \\
\textbf{JEL codes}: D47, D83, J12
\end{abstract}


\newpage

\section{Introduction}

Digital matching platforms reduce search frictions by helping users screen large markets. Yet the same tools can automate exclusion: filters may remove potential counterparts before they are seen. Age is a canonical example in dating and marriage. It is universally observed, central to partner choice, and closely tied to fertility and family formation \citep{choo2006marries,siow1998differntial,low2024human,doepke2023economics}. Online platforms make age a visible label, a filterable attribute, and a sorting criterion before other partner information is learned \citep{rosenfeld2019disintermediating,hwang2026rise}.

This margin is quantitatively important: \citet{hwang2026rise} report that only 19 percent of U.S. online daters use no filters and that age is the most common filter. Yet we know little about its consequences for marriage markets. Does a one-year change in displayed age reduce access to potential partners, reallocate attention across partner types, or affect continuation from first contact to engagement? Existing datasets rarely observe both premarital consideration sets and the full relationship funnel: registry and census data begin at marriage, while typical platform data end near first contact or cannot link verified attributes to marriage outcomes. They therefore provide limited evidence on how age labels and search constraints shape the path from search to marriage.

This paper addresses these questions using confidential records from IBJ, a large consultant-mediated marriage platform in Japan. In 2024, engagements across IBJ's services accounted for about 3.4\% of Japanese marriages. IBJ verifies age, income, education, and marital status; records premarital preferences over children and household time allocation; and tracks applications, meetings, non-exclusive dating, exclusive dating, and accepted proposals. I call the member receiving an application the \emph{receiver} and the member sending it the \emph{proposer}. Age is visible before a click, screenable through ranges, and updated automatically at birthdays. IBJ also records stated age preferences and saved search conditions, allowing stated acceptable ages to be compared with saved age bounds. Its high-compliance environment records confirmed engagements and exits unusually well. Although matched IBJ couples are older and more educated than national benchmarks, the assortative-matching structure of realized matches---especially for age and education---is broadly comparable to representative data once marginal distributions are netted out with likelihood ratios. \textcolor{black}{This comparison concerns realized sorting; search behavior may remain platform-specific.}

The paper makes two main contributions. First, it identifies how the annual displayed-age update affects early-stage attention and follows the resulting pairs through engagement. Second, it measures the opportunity sets implied by stated filters and quantifies their mechanical role in the level and composition of matching outcomes. \textcolor{black}{Combining the two, the paper reports stage-specific accounting contrasts for retaining the pre-birthday label, covering both accepted-proposal counts and pair composition.}

These contributions rely on observing both sides of a process that is usually latent. A stable aggregate application count can conceal offsetting entry and exit at filter boundaries, as well as re-sorting within the remaining pool; final matches alone cannot reveal which margin generated the change. Linking stated bounds, applications, bilateral continuation decisions, and engagements distinguishes filter-implied eligibility, realized applications, and subsequent continuation.

Conceptually, the analysis separates three margins that are often conflated in platform data. The \emph{eligibility margin} asks whether stated filters admit a profile; the \emph{application margin} combines exposure within the eligible set with the decision to initiate; and the \emph{continuation margin} begins after a pair has formed. The birthday can move all three, but \textcolor{black}{among observed objects} only stated-filter eligibility changes mechanically\textcolor{black}{; exposure under age-ordered browsing or narrower unobserved filters may also shift mechanically, but it is not observed}. This distinction disciplines what can be learned from each stage of the funnel.

The empirical strategy exploits the automatic one-year increase in displayed age at the birthday. No other verified attribute is mechanically updated then, and these attributes are assumed to evolve smoothly in the narrow window. I construct a birthday-relative panel using three calendar segments per month and compare the same member before and after the update. Member fixed effects absorb permanent heterogeneity, calendar-year effects absorb common shocks, within-month-position effects absorb the monthly activity cycle, and flexible tenure-segment fixed effects absorb the sharp new-member activity decline. Potential partners observe age in years and birth month but not the exact birthday, limiting precise timing around the cutoff.

The first result concerns age eligibility at the top of the funnel. A birthday moves a receiver out of ranges whose upper bound is crossed and into ranges whose lower bound is newly satisfied. These margins often offset, but their balance flips with age: entry dominates through the mid-thirties for male receivers, expanding the number of potential proposers by up to \(10.7\%\) of its pre-birthday level, and through age 30 for female receivers, with gains of up to \(6.5\%\); exit dominates at older ages, shrinking the number of potential proposers by up to \(8.1\%\) for men and \(8.3\%\) for women. The estimated application change is also small and usually positive, so a broad contraction in stated age eligibility does not explain the downstream patterns.

The eligibility decomposition explains the offset. Applications fall from proposers whose stated upper bound is crossed and rise from those whose lower bound is newly satisfied. For female receivers, however, most of the positive application jump comes from proposers whose stated ranges contain the receiver on both sides of the birthday. The annual update therefore combines a mechanical boundary change with an additional response inside the broad stated-eligible pool.

The second result is compositional. Among proposers whose stated acceptable range includes the receiver on both sides of the birthday, applications shift toward older suitors and away from younger suitors, with no systematic re-sorting by proposer income or height. This pattern is consistent with age-proximity preferences \citep{choo2006marries}, although the design does not separate a direct response to age proximity from changes in age-based exposure. In the application-tracked funnel sample, proposal counts decline at nine of the ten female receiver ages from 31 to 40; five declines exceed \(10\%\), and the largest is about \(22\%\), at age 35. For male receivers, proposal declines are confined to ages 31, 36, and 40. The funnel decomposition locates the transition from applications to non-exclusive dating as the recurring negative component for women, while the application-rate component is small at many central ages.

Two complementary decompositions sharpen this mechanism. For female receivers aged 31--40, re-sorting toward older suitors, whose applications convert at lower rates, accounts for \(22\%\) of the decline in the transition rate from applications to non-exclusive dating. Reply data locate the broader decline at meeting formation: held meetings per application fall by about \(17\%\), while continuation conditional on a held meeting is essentially unchanged.

The third result identifies a general measurement problem in two-sided matching markets: treating all counterparts on the other side of the market as a user's opportunity set can substantially overstate effective matching opportunities. \textcolor{black}{On IBJ, the all-active pool exceeds the stated-eligible pool by \(32.5\%\) for male receivers and \(33.9\%\) for female receivers. Applying fixed rates to this all-active benchmark raises stage counts mechanically by roughly \(30\%\); the directly measured pool wedge is more than an order of magnitude larger than the aggregate mechanical birthday change.}

\textcolor{black}{Within the filtered market, the no-birthday accounting differs sharply across receiver channels. For receivers in their thirties, the proposal-stage count is \(12.9\%\) above the benchmark in the female-receiver channel and \(6.7\%\) below it in the male-receiver channel; the younger-man share in the former rises from \(17.1\%\) to \(21.9\%\). Across male receivers aged 25--45, the younger-woman share rises from \(77.5\%\) to \(84.1\%\). These application-origin channels do not sum to totals by sex because every accepted proposal pairs a man and a woman. Age labels and filters therefore reshape observed opportunities, proposal volume by initiation channel, and accepted-pair composition.}

\subsection{Related Literature}

The paper speaks to three literatures: dating and marriage markets; multi-stage matching; and search, consideration sets, and platform design.

First, the paper is closely related to work on the value of age in dating and marriage markets. In the dating stage, \citet{low2024pricing} provide experimental evidence that age causally reduces women's appeal by asking real online daters to rate hypothetical profiles with randomly assigned ages. Other online-dating studies estimate preferences and search behavior from clicks, messages, or field experiments that vary profile attributes such as income, height, and education \citep{hitsch2010matching,hitsch2010makes,ong2015income,ong2015height,ong2016education}. \citet{lee2016effect} is a rare exception linking online-dating behavior to eventual marriage outcomes. In the marriage stage, age is usually studied through age assortativeness and spousal age gaps. Building on the single-dimensional age-assortativeness framework of \citet{choo2006marries}, structural matching models provide tools for estimating sorting on multiple traits, including age, anthropometric characteristics, personality, and household-type margins \citep{chiappori2010hedonic,chiappori2012fatter,dupuy2014personality,ciscato2020like,chiappori2024analyzing}, and related dynamic work studies gender and the age gap directly \citep{shephard2019marriage}. Relative to this literature, this paper combines verified age, premarital income, education, and physical-attribute data with stated age preferences, saved search conditions, and stage-by-stage relationship outcomes. This makes it possible to study age not only as a final matching trait but also as a verified, searchable signal that affects stated eligibility, applications, and subsequent continuation.

The paper also contributes to work on multi-stage two-sided matching and funnel propagation. In labor-market settings, several studies analyze how information, wages, profile characteristics, ranking scores, platform thresholds, and interface changes affect particular stages of the search process, especially views, clicks, saves, and applications \citep{belot2022wage,belot2019providing,fradkin2025competition,sockin2023s}. Other studies follow transitions from applications to interviews, offers, hires, and match outcomes \citep{stanton2025benefits,roussille2023bidding,kambayashi2025decomposing,zuchuat2023duration}. Comparable marriage-market evidence is limited because data rarely link premarital search to engagement. This paper follows applications through early dating, exclusive dating, and accepted proposals. The birthday response is not a simple decline in top-of-funnel attention: entry and exit from stated age ranges largely offset, applications re-sort across proposer age types, and downstream proposal losses for female receivers concentrate in the transition into early dating.

Finally, the paper contributes to research on search, consideration sets, and platform interface design. Matching platforms can shape outcomes not only through preferences, information, or recommendation algorithms \citep{rios2023improving,chen2023reducing,manshadi2025redesigning,sekiya2026designing,yoo2023search}, but also by restricting which options users consider \citep{einav2016peer,halaburda2018lunch}. This paper studies age labels visible before a click and age-range filters that can be saved and applied during search. Unlike birthdate and school-entry-age designs, in which a rule changes the treated individual's schooling or exposure (for example, \citealp{menezesfilho2024effect}), this cutoff changes how the opposite side classifies and screens the member. Locally, the birthday changes stated eligibility and re-sorts applications within it; quantitatively, the accounting exercises compare this annual update with the broader wedge implied by stated age filters. The contribution is therefore to measure how platform-created opportunity sets implied by filters shape both the level and composition of matching outcomes.

\section{Data and Institutional Background}

This section establishes the institutional facts behind the birthday design: age is verified but displayed in years, search begins in an interface where age is immediately visible, and the platform records the full funnel from applications to accepted engagements. I describe the data, platform coverage, matching stages, and descriptive evidence for age, income, and height as the central margins along which birthdays can reshape inbound attention; Appendix~\ref{app:data_details} reports additional platform, coverage, and raw-data details.

\subsection{Data}
The analysis uses confidential individual-level records from IBJ, a major marriage matching platform in Japan, with search-and-matching actions spanning June 2020 to December 2024. In 2024, IBJ reported 16,398 engagements across its services, equivalent to about 3.4\% of Japanese marriages.\footnote{This total includes engagements formed outside the platform by IBJ users and through other group services; the confidential analytical extract records 7,587 within-platform matches in 2024 under the project definition. See \citet{otani2025nonparametric} for platform-level trends.} Although this share is smaller than the coverage of national administrative marriage records, the IBJ data have four advantages over commonly used alternatives, including dating apps \citep{hitsch2010matching,ong2015income,bapna2016one,huber2017political,egebark2021brains,rios2023improving,buyukeren2026impact,oyer2014everything,rudder2014dataclysm}, speed-dating studies \citep{fisman2006gender,fisman2008racial,belot2013dating,bhargava2014contrast}, and census or registry data \citep{chiappori2017partner}.

First, IBJ verifies core attributes---age, education, income, and marital status---against official documents, limiting profile inflation common on dating apps \citep{toma2008separating}. \textcolor{black}{Verified premarital income also avoids the endogeneity of post-marital income to labor supply \citep{chiappori2026assortative}.} Second, the platform records time-stamped meeting requests and relationship transitions and identifies confirmed engagements in its official within-platform engagement records. I date an engagement by the first partner's formal engagement withdrawal. Third, users are selected into high-intent searches for long-term partners: participation requires screening and substantial fees, which discourage casual use. Finally, IBJ is a closed, consultant-mediated platform. Confirmed engagements require bilateral consent, exits carry status codes, and contractual rules discourage off-platform circumvention and low-compliance behavior, improving the reliability of recorded relationship status.\footnote{For example, exiting without filing the platform's formal engagement-withdrawal application and then marrying a partner met through the platform is treated as a contract violation and is subject to sanctions. The same contractual framework imposes strong penalties for last-minute cancellation of scheduled meetings and intimate contact before the formal engagement-exit process. Appendix~\ref{app:marriage_agency_platform} describes the broader environment.}

\subsection{Platform Coverage}\label{sec:platform_coverage}

IBJ covers a sizable and policy-relevant segment of the Japanese marriage market, but its users are selected. Relative to national benchmarks, matched IBJ couples are older and more highly educated, consistent with the platform's position as a high-verification, fee-based service for later and high-intent entrants. The external-validity concern is therefore compositional: IBJ is not meant to reproduce the population distribution of marriages. The relevant question is whether the matching structure within this selected market is unusual. After netting out marginal distributions using likelihood ratios, the assortative-matching structure---especially for age and education---is broadly comparable to that in representative data. I therefore interpret IBJ patterns as reflecting both selection into a high-intent platform and genuine matching behavior within that market. Appendix~\ref{app:platform_volume_coverage} reports platform volume and the full comparison with Vital Statistics, the National Fertility Survey, and the Employment Status Survey; \citet{inoue2026marital} provides the detailed coverage analysis.

\subsection{IBJ User Behavior}\label{sec:ibj_user_behavior}

Table~\ref{tb:dating_process} summarizes the matching funnel and fixes the stage definitions used throughout the paper. The process begins at the Application stage: a proposer browses candidate profiles, may click to inspect a profile, and can initiate contact by sending an omiai (formal meeting) request. Both men and women can act as proposers. If the receiver accepts the request, the pair moves to the Omiai Meeting stage, which consists of an in-person or virtual meeting followed by a bilateral decision about whether to continue. \textcolor{black}{If both sides submit positive post-meeting replies, the platform records entry into the Pre-relationship stage. Entry therefore requires a completed synchronous omiai meeting and bilateral agreement to begin trial dating; it is not a message-only interaction and does not require attendance at the first trial date, which occurs subsequently.} During this phase, users can pursue multiple trial relationships in parallel while repeatedly deciding whether to continue with each match. Counselors monitor both stages, and the platform records each party's replies and continuation decisions. These actions consequently represent high-commitment choices, unlike the low-cost messages and clicks typically observed on dating apps.

If both participants agree to move beyond trial dating, they enter the Serious relationship stage, which marks the beginning of exclusivity. This stage starts with a committed date and continues through repeated bilateral continuation decisions. The final stage is Proposal, defined as an accepted marriage proposal. This staged structure is central to the empirical design: it shows whether the birthday shock operates only by changing who is contacted at the top of the funnel or continues to affect progression after pairs have interacted.

\begin{table}[!htbp]
\caption{Dating Process by Phase, Agent, and Action}
\label{tb:dating_process}
\begin{center}
    \resizebox{\textwidth}{!}{%
    \begin{tabular}{@{}llll@{}}
    \toprule
    Stage & Agent & Action & Note \\
    \midrule
    1. Application & Proposer & Browse candidate profiles & Set filter conditions \\
     & Proposer & Click on candidate profiles & Inspect profile details \\
     & Proposer & Send omiai request & Initiates contact \\
     & Receiver & Accept or reject omiai request & \\
    \midrule
    Omiai Meeting & Both & Conduct omiai meeting & Face-to-face or virtual meeting \\
     & Both & Accept or reject trial dating & Decision to continue or exit \\
    \midrule
    2. Pre-relationship & Both & \textcolor{black}{Agree to begin trial dating} & \textcolor{black}{Bilateral positive replies; non-exclusive}\\
     & Both & Accept or reject second trial date & Decision to continue or exit \\
     & Both & Proceed with additional trial dates & Repeated decision process \\
     & Both & Accept or reject committed dating & Transition to exclusive dating \\
    \midrule
    3. Serious relationship & Both & Begin first committed date & Exclusive\\
     & Both & Accept or reject second date & Decision to continue \\
     & Both & Proceed with additional dates & Repeated decision process \\
    \midrule
    4. Proposal & Both & Accept marriage proposal & \\
    \bottomrule
\end{tabular}
}
\end{center}
\footnotesize

\end{table}

Throughout the analysis, outcomes count applications received and the resulting pairs reaching each stage: (1) \emph{Application} (omiai requests received), (2) \emph{Pre-relationship} \textcolor{black}{(bilateral agreement to begin non-exclusive trial dating)}, (3) \emph{Serious relationship} (exclusive committed dating), and (4) \emph{Proposal} (accepted engagements). Each downstream outcome is assigned to the birthday-relative segment of its initiating application, so it measures how many applications begun in that segment subsequently reach the indicated stage.\footnote{From the pre-relationship stage onward, a transition records bilateral agreement and should not be attributed to either side without using the parties' separate replies. Appendix~\ref{app:who_ended} uses those replies to decompose the application-to-pre-relationship margin.}

\subsection{User Interface and Search Conditions}\label{sec:user_interface}

Matching begins when users browse candidate lists and decide which profiles to click at the Application stage. The user interface is therefore the first object to examine: it determines which profiles enter the proposer's view, which attributes are visible before a click, and how a discrete age label can affect the top of the funnel.

\begin{figure}[!t]
  \captionsetup{skip=3pt}
  \centering
  \subfloat[Pre-click search list]{\includegraphics[width = 0.44\textwidth]{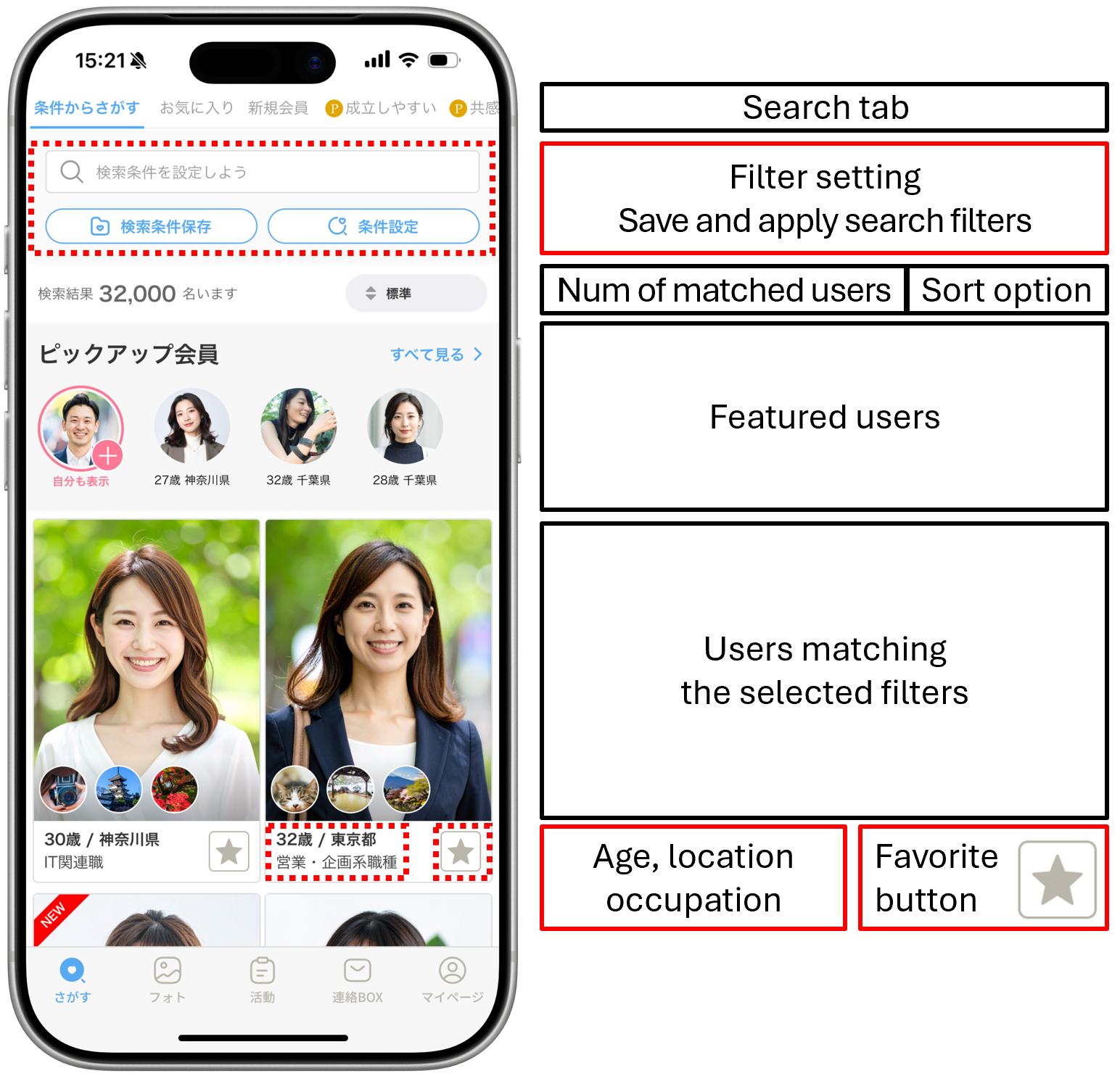}}
  \hfill
  \subfloat[Profile header after click]{\includegraphics[width = 0.44\textwidth]{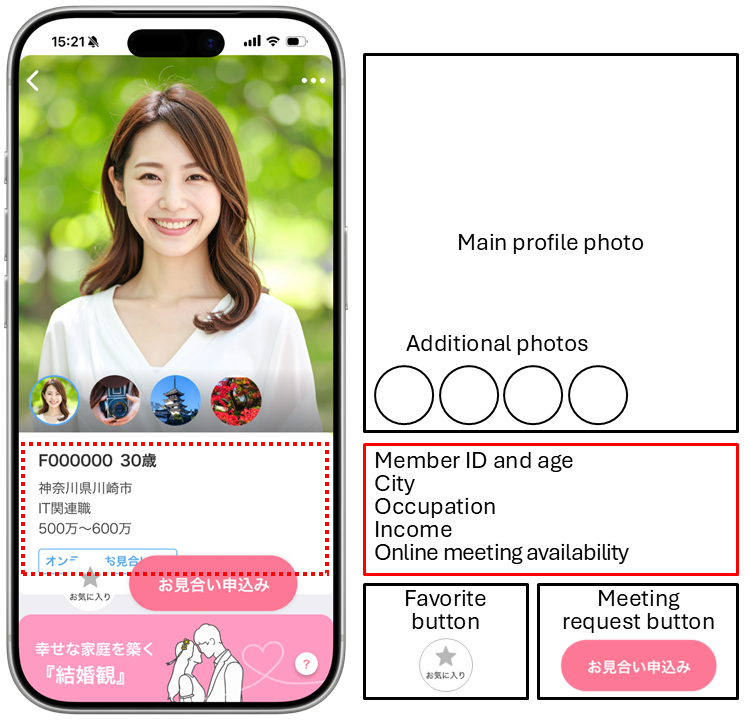}}\\[0.2in]
  \subfloat[Birth year/month and verified attributes]{\includegraphics[width = 0.44\textwidth]{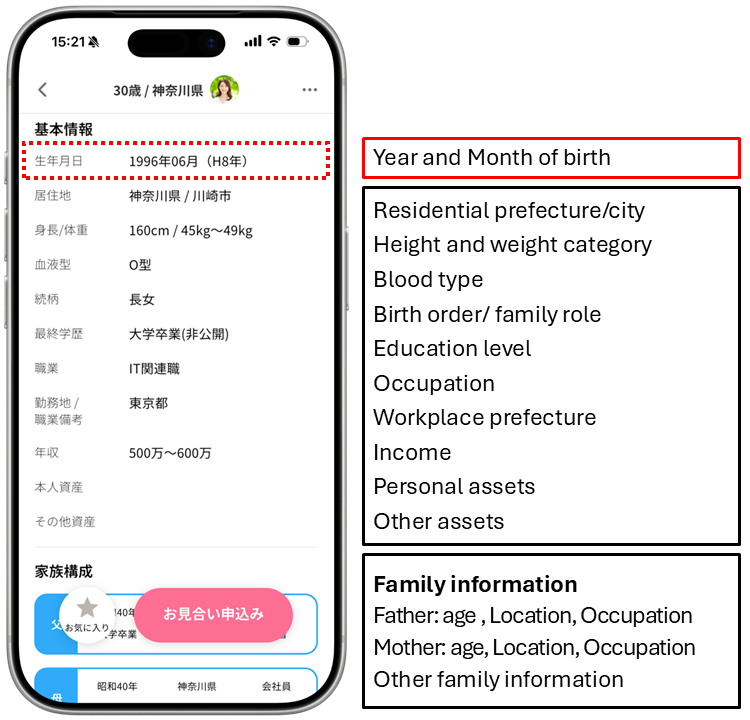}}
  
  \caption{User Interface: Pre-click and Post-click Information}
  \label{fg:ibj_view_figure} 
  \par\raggedright\footnotesize
  Note: The figure uses official IBJ interface screenshots provided directly by IBJ to the author.
\end{figure}

Figure~\ref{fg:ibj_view_figure} shows three views of this interface. Users reach potential partners through several tabs: new members, algorithmic recommendations, favorites, and detailed condition search. The analysis focuses on condition search, where users specify, save, apply, and sort profile criteria; favorites contain profiles users have already screened and bookmarked.

In condition search (Panel (a) of Figure~\ref{fg:ibj_view_figure}), a user first chooses criteria for partner attributes---such as age range, location, income, and education---and then selects a sorting rule: newest first, age order, or the platform's standard recommended order. Users can save multiple sets of search conditions and apply one of the saved sets when browsing; these saved search conditions are recorded in the data, as shown in Section~\ref{subsec:actual_filters}. The system displays profiles satisfying the selected criteria in a scrollable list. Before clicking, the proposer observes only a limited set of information for each candidate: profile photo, age in years, residential location, occupation, and whether the candidate has been marked as a favorite. Clicking opens the detailed profile. The post-click header (Panel (b)) displays the main and additional profile photos, member identifier, age, city, occupation, income, online-meeting availability, and the meeting-request button. Scrolling through the profile, the user first encounters a lengthy self-introduction and then the additional verified attributes shown in Panel (c), including birth year and month, residence, height and weight category, education, occupation, workplace prefecture, income, assets, and family information.

This interface is useful for identification. First, displayed age updates automatically at the birthday, whereas other verified attributes---income, education, occupation, and residence---are difficult to change flexibly and are effectively stable in a narrow birthday window.\footnote{Updating income requires a withholding-tax certificate, typically issued annually. Users cannot revise verified fields directly: changes require review by a consultant and the platform. The data do not record profile-update timestamps, so I cannot verify every revision's timing.} The discontinuity in the proposer's information set is therefore concentrated in the receiver's age label. Second, proposers observe age in years before clicking and in the post-click header, and birth year and month after scrolling, but not the exact birthday date, limiting strategic timing of applications just before the cutoff. Finally, although query-level filters are not observed for every search, the data record each user's stated minimum and maximum acceptable partner ages, discussed in Section~\ref{sec:summary_statistics}. These bounds provide a broad, single-record proxy for each user's consideration set and make it possible to analyze how a birthday moves a receiver inside or outside potential partners' age criteria. Section~\ref{sec:potential_proposers} examines this potential-proposer channel in detail.

\subsection{Summary Statistics}\label{sec:summary_statistics}

Table~\ref{tb:summary_statistics_female_matched} reports members with at least one observed application in the relevant role, not the event-time sample; Section~\ref{sec:empirical_strategy} and Appendix~\ref{app:event_time_support} define and document the latter. The table highlights the first fact needed for the mechanism: age screens are gendered and are recorded in the data. Panel (a) shows that male receivers average 39.40 years of age, an income-category upper-edge value of 7.54 million Japanese yen (JPY), and 171.44 cm in height, compared with 35.71 years, 4.91 million JPY, and 158.81 cm for female receivers. \textcolor{black}{Average stated lower and upper acceptable-age bounds, each computed among members reporting that bound, are 29.47 and 38.92 for male receivers and 32.40 and 43.27 for female receivers.} These receiver-side bounds are descriptive; the birthday mechanism instead runs through the opposite side of the market, because proposers set their own acceptable age ranges and a one-year birthday increment can move a receiver across many proposers' age screens.

The second fact is funnel compression. Male receivers obtain substantially fewer applications than female receivers (70.05 versus 122.21 per person), but downstream gaps are smaller: male and female receivers average 2.06 and 1.96 pre-relationships, 0.20 and 0.17 serious relationships, and 0.13 and 0.11 proposals, respectively. Panel (b) presents the same variables from the proposer side; the receiver and proposer panels are two role-based summaries of an overlapping platform population.\footnote{The sender-side and receiver-side action aggregates need not coincide: in-sample proposers and receivers are overlapping but not identical sets, and a counterparty on either side of an action can fall outside the sampled population or study window, so total applications sent and total applications received differ. The panels are role-based descriptive summaries, not a closed sent-equals-received accounting.} Male proposers average 38.88 years of age and an income-category upper-edge value of 7.34 million JPY, compared with 35.38 years and 4.99 million JPY for female proposers. \textcolor{black}{Average stated lower and upper bounds, computed analogously, are 29.04 and 38.72 for male proposers and 32.10 and 43.01 for female proposers.} Male proposers send 169.85 applications on average, compared with 63.16 for female proposers; these contacts translate into 2.72 versus 1.85 pre-relationships, 0.23 versus 0.18 serious relationships, and 0.15 versus 0.11 proposals.\footnote{Additional categorical variables are also exploited in \cite{inoue2026marital} in a multidimensional matching framework. \textcolor{black}{Axiomatic assortativeness measures are examined by \cite{imamura2025note}; a forthcoming revision applies them to the IBJ platform.}}

\begin{table}[!htbp]
\caption{Summary Statistics of Individual-level Variables by Gender}
\label{tb:summary_statistics_female_matched}
\begin{center}
    \subfloat[Receiver]{\resizebox{\textwidth}{!}{\input{figuretable/birthday_project/summary_statistics_male_female_receiver}}}\\[0.2in]
  \subfloat[Proposer]{\resizebox{\textwidth}{!}{\input{figuretable/birthday_project/summary_statistics_male_female_proposer}}}
\end{center}
\footnotesize
Note: The sample comprises members with an observed application in the relevant role. (N) is the number of members, SD is the standard deviation, and Min and Max are the sample minimum and maximum. Age is measured at each member's first observed application in the relevant role; action counts aggregate over the sample period. The upper and lower age-preference variables report the member's stated maximum and minimum acceptable partner ages. Eligibility calculations treat a missing upper bound as no upper age restriction and a missing lower bound analogously. Income is the upper limit of the member's income category, measured in units of 10{,}000 JPY (man-en). For example, 700 denotes income of at least 6.0 million and less than 7.0 million JPY.

\end{table}

\begin{figure}[!htbp]
  \begin{center}
  \includegraphics[width = 1.0\textwidth]{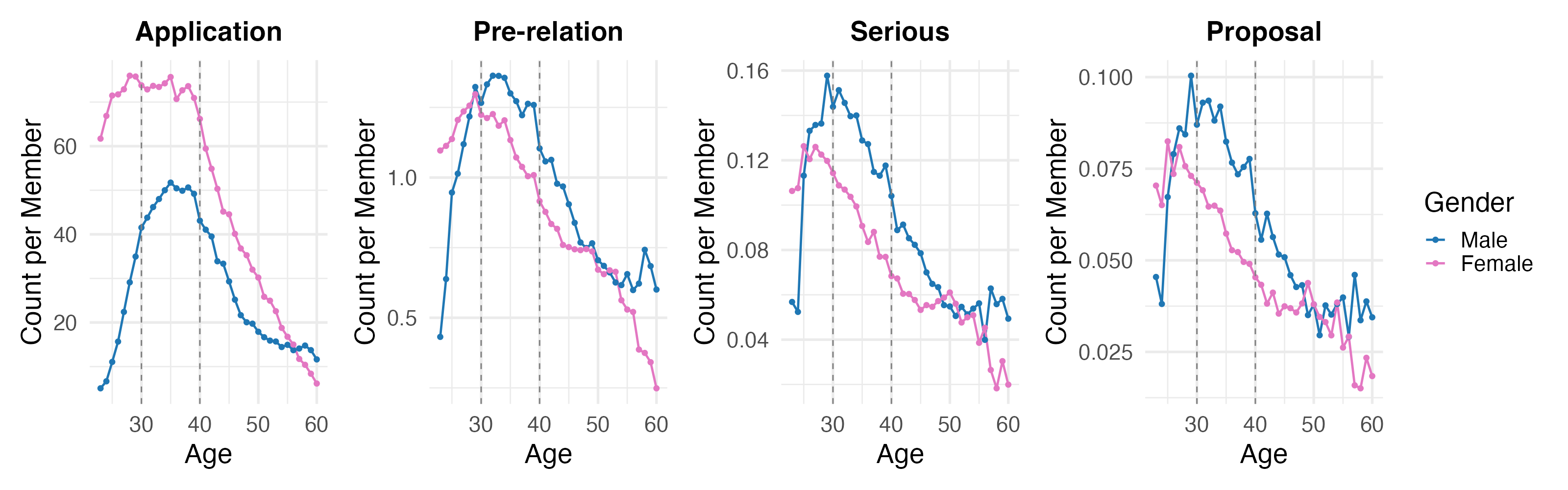}
  \includegraphics[width = 1.0\textwidth]{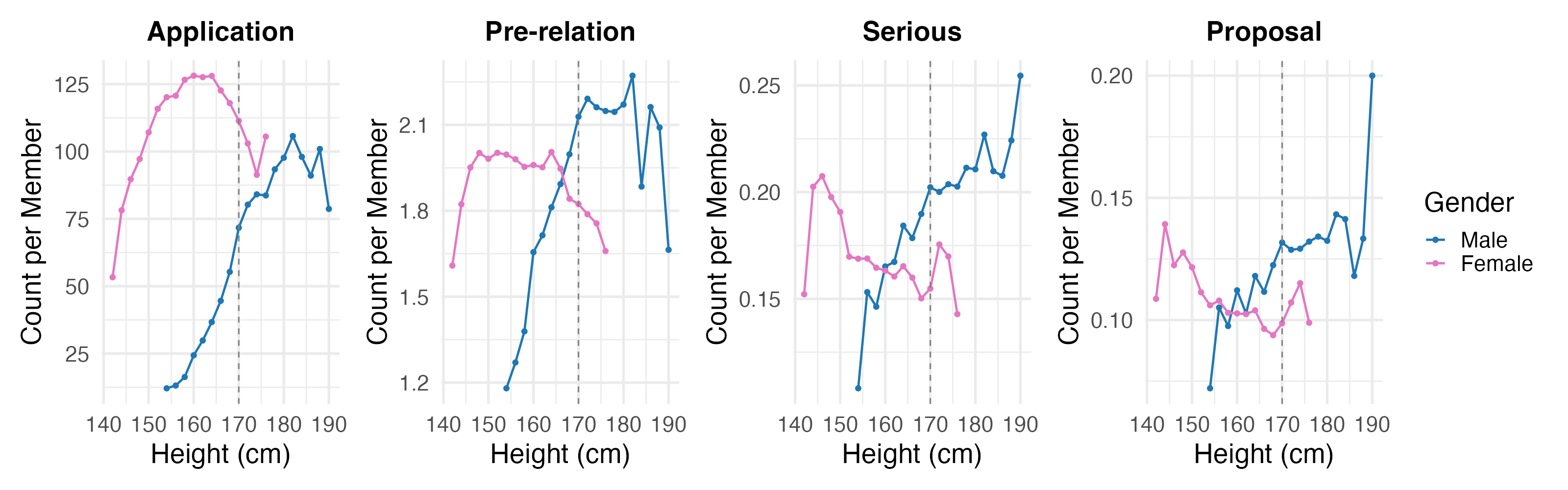}
  \includegraphics[width = 1.0\textwidth]{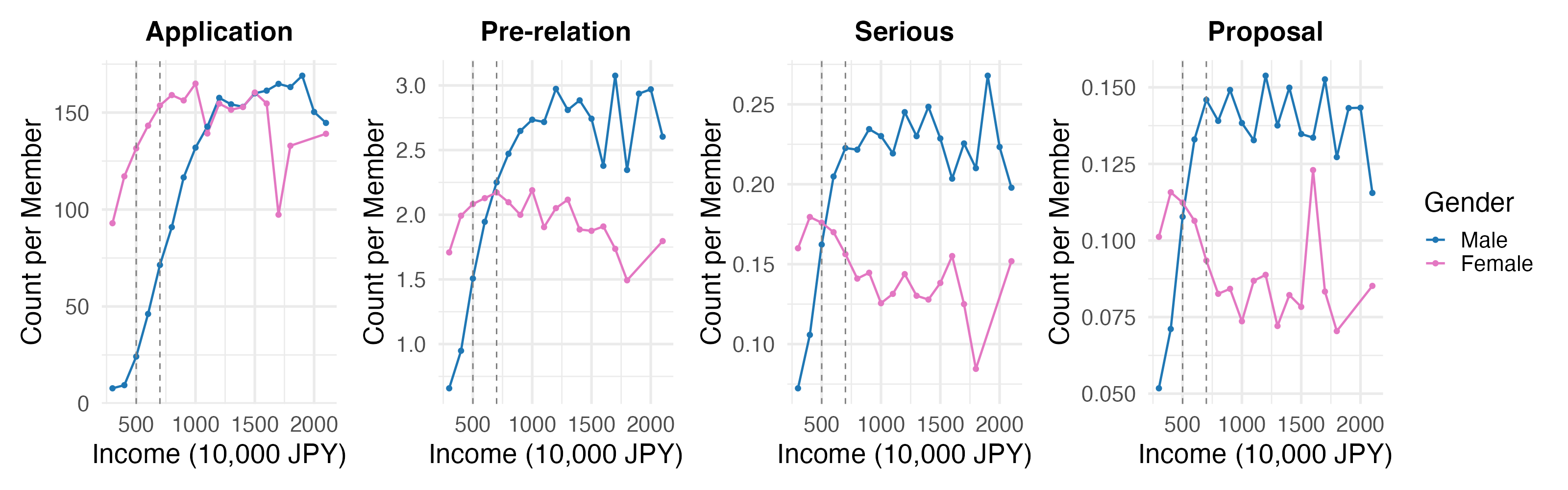}
  \caption{Per-Member Action Counts by Receiver Characteristics}
  \label{fg:action_counts_per_member_by_income}
  \end{center}
  \footnotesize
  Note: The figure plots action intensity (relative counts per member) by observable characteristics. For each gender and each bin or segment (by age, height, or income), the relative count is the total number of actions received by receivers in that bin divided by the number of receivers in the same bin. Dashed vertical lines mark focal thresholds used in the descriptive analysis.
\end{figure}

Three additional descriptive patterns motivate the empirical design. First, per-member action intensity varies sharply with receiver age, income, and height. Figure~\ref{fg:action_counts_per_member_by_income} normalizes action counts by the number of active receivers in each bin, so the plotted series measure per-member action intensity rather than raw market size. For men, per-member action intensity rises strongly with income-category upper edge---men in categories whose lower endpoint is at least 6 million JPY receive more actions at every stage, with an especially pronounced gradient after the application stage---and rises with height while staying relatively stable over central ages. For women, the income gradient is weaker and often reversed at later stages, while action intensity declines sharply after the mid-thirties. These patterns document type-specific differences in received action rates and motivate the proposer-characteristic splits below by age, income, and height.

\begin{figure}[!t]
  \captionsetup{skip=3pt}
  \centering
  \subfloat[Male receivers]{\includegraphics[width = 0.70\textwidth]{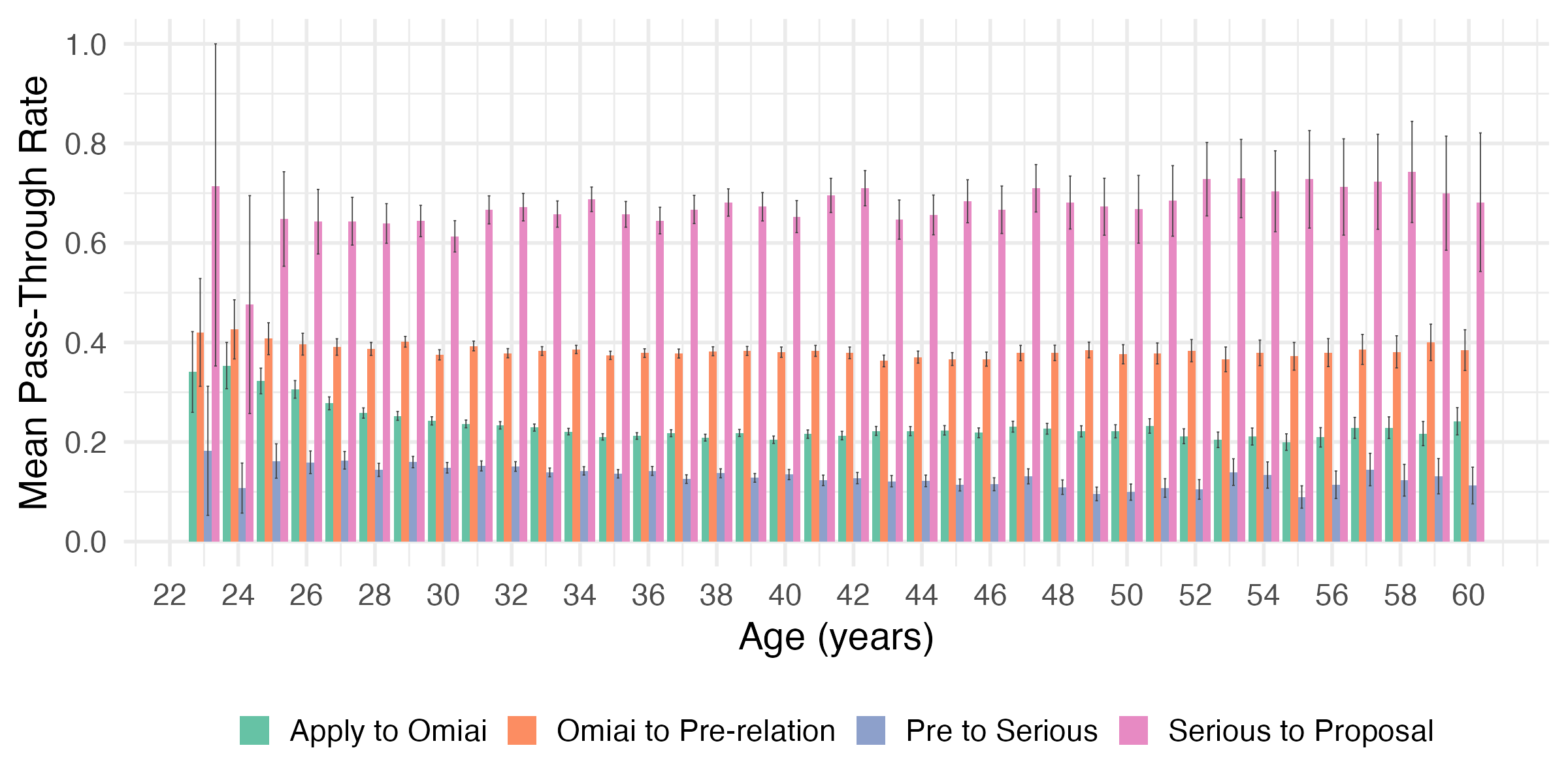}}\\[0.2in]
  \subfloat[Female receivers]{\includegraphics[width = 0.70\textwidth]{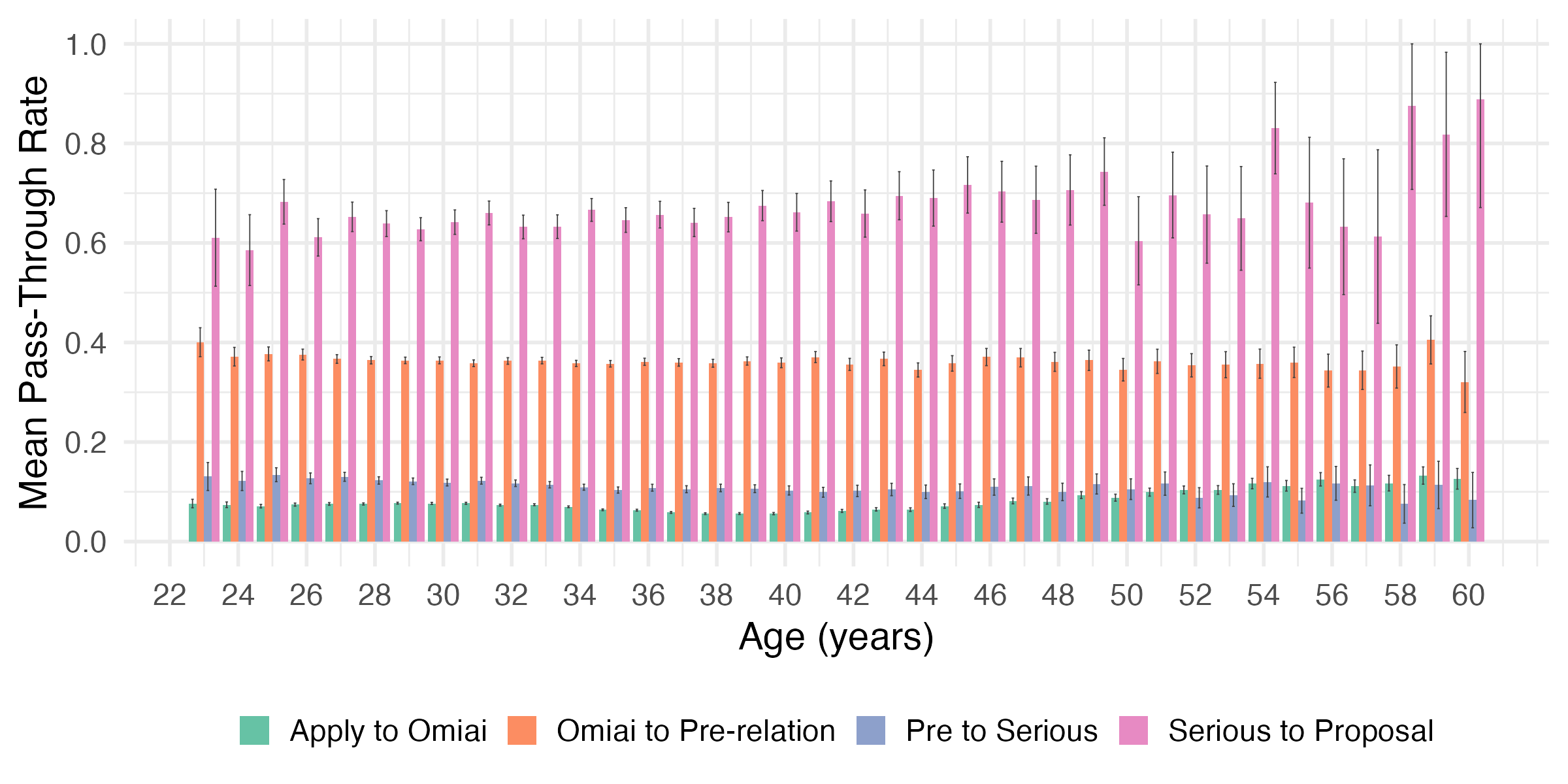}}
  \caption{Stage-to-Stage Pass-Through Rates by Receiver Age}
  \label{fg:pass_through_rate_by_age}
  \par\raggedright\footnotesize
  Note: Bars report conditional transition rates between adjacent stages, with 95\% confidence intervals. \textcolor{black}{Rates are first computed for each receiver and then averaged across receivers with a positive preceding-stage denominator; they are means of receiver-level ratios, not ratios of the aggregate mean counts in Table~\ref{tb:summary_statistics_female_matched}.}
\end{figure}

Second, stage-to-stage pass-through rates are much flatter over receiver age than raw application counts. Figure~\ref{fg:pass_through_rate_by_age} reports the conditional transition rates between adjacent stages: although inbound application volumes differ sharply by age and gender, the pass-through rates are comparatively flat over age for both men and women. One possible explanation is a person-level capacity constraint---each receiver can sustain only a limited number of simultaneous relationships because each relationship from the pre-relationship stage onward requires continued dating, so additional top-of-funnel attention need not translate one-for-one into serious relationships or proposals---although selection into each stage can generate the same pattern. This contrast motivates studying both levels and transition rates and using the application-tracked decomposition in Section~\ref{sec:log_additive_decomp} to separate the pool-size and application-rate channels from conditional transition-rate channels.

A third descriptive fact from members' saved search conditions is documented in Section~\ref{sec:potential_proposers}: age is a near-universal screen, whereas income, height, and education are screened far more by female than by male searchers.\footnote{Together, the interface, summary statistics, and saved-filter data define the empirical object: age is visible before richer information, matching thins rapidly after applications, and the central composition margins are age, income, and height. Section~\ref{sec:potential_proposers} asks whether a birthday changes both application volume and the set of proposers admitted by stated filters.}

\section{Search Criteria, Potential Proposers, and Stated Preferences}\label{sec:potential_proposers}

This section documents the opportunity-set channel that makes a one-year age increment meaningful. A birthday changes a receiver's displayed age discretely but does not mechanically change other characteristics. If proposers use age ranges to decide whom to browse, the birthday can change whether the receiver enters the set of profiles a proposer considers. I first document that age is the near-universal filter. I then use stated age bounds to construct receiver-level opportunity sets, compare those bounds with saved search conditions, and preview the implied birthday event-time patterns.

\subsection{Age as the Dominant Search Filter}\label{subsec:age_filter}

A distinctive feature of the IBJ data is that members configure and save explicit search conditions, so the platform records the partner attributes included in those conditions. These records complement the mechanism evidence in \citet{hwang2026rise}, who use retrospective survey data to show that platform filters and user preferences jointly shape whom online daters meet. Because survey evidence relies on ex post self-reports of dating histories, preferences, and filter settings, the IBJ setting is useful: it records premarital saved search conditions, links them to verified profiles, and follows subsequent relationship outcomes. Whereas Tinder offers only coarse filters and Hinge offers detailed ones, IBJ is firmly a detailed-filter environment: members can screen on age, income, height, weight, education, occupation, marital and child history, and numerous lifestyle attributes.\footnote{The saved search-filter data are a snapshot of users active on the platform as of April 8, 2025. They cover the subset of members who had saved at least one search condition, totaling 133{,}879 conditions across 56{,}371 members. Because the snapshot is not a complete historical record for the 2020--2024 event-time sample, I use it only to document saved-filter settings and to compare them with stated acceptable-age bounds. The main consideration-set analysis instead uses the single stated acceptable-age range.}

\begin{figure}[!t]
  \captionsetup{skip=3pt}
  \centering
  \subfloat[Number of attributes screened per saved condition]{\includegraphics[width = 0.66\textwidth]{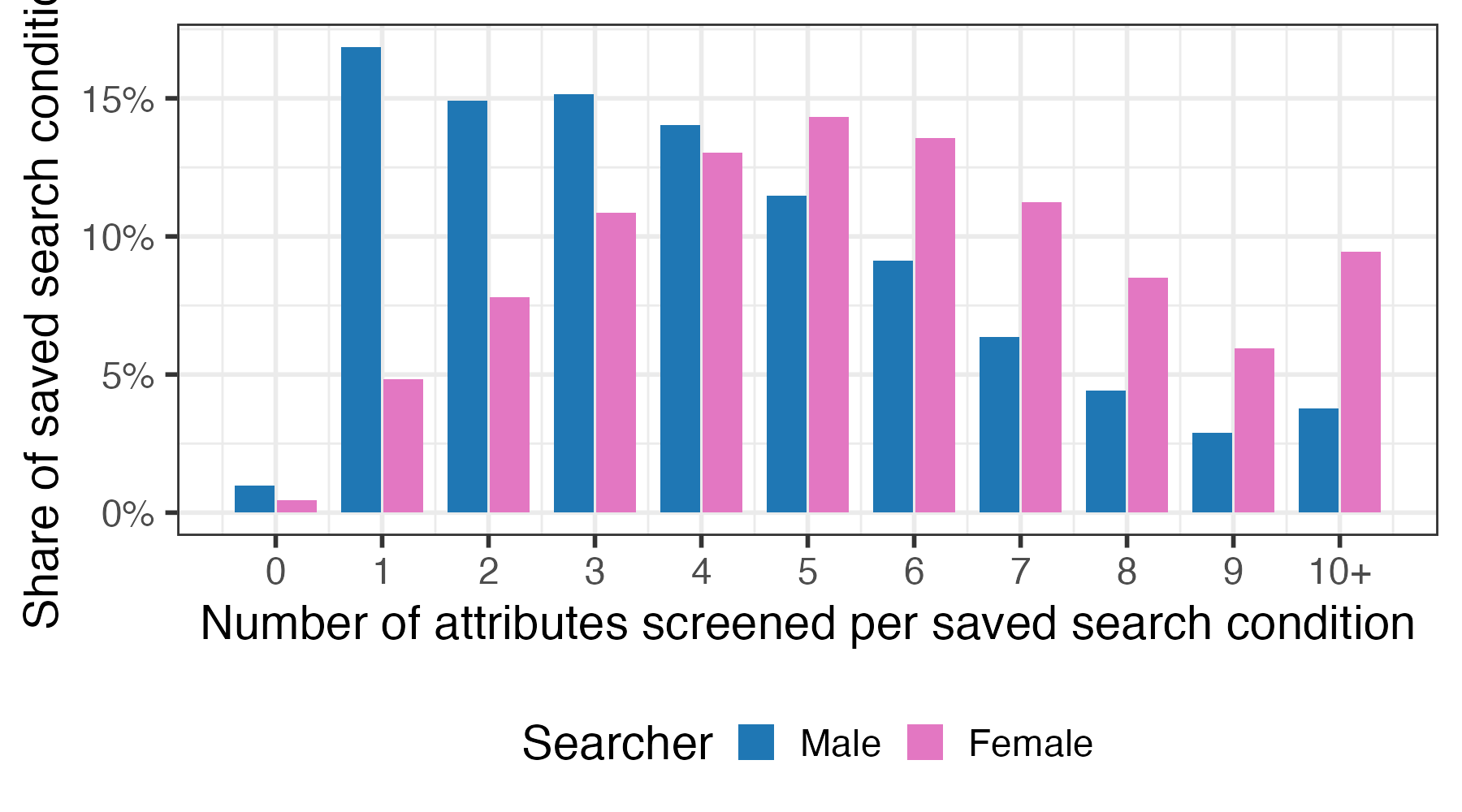}}\\[0.2in]
  \subfloat[Filter usage by attribute and searcher gender]{\includegraphics[width = 0.68\textwidth, height = 0.35\textheight, keepaspectratio]{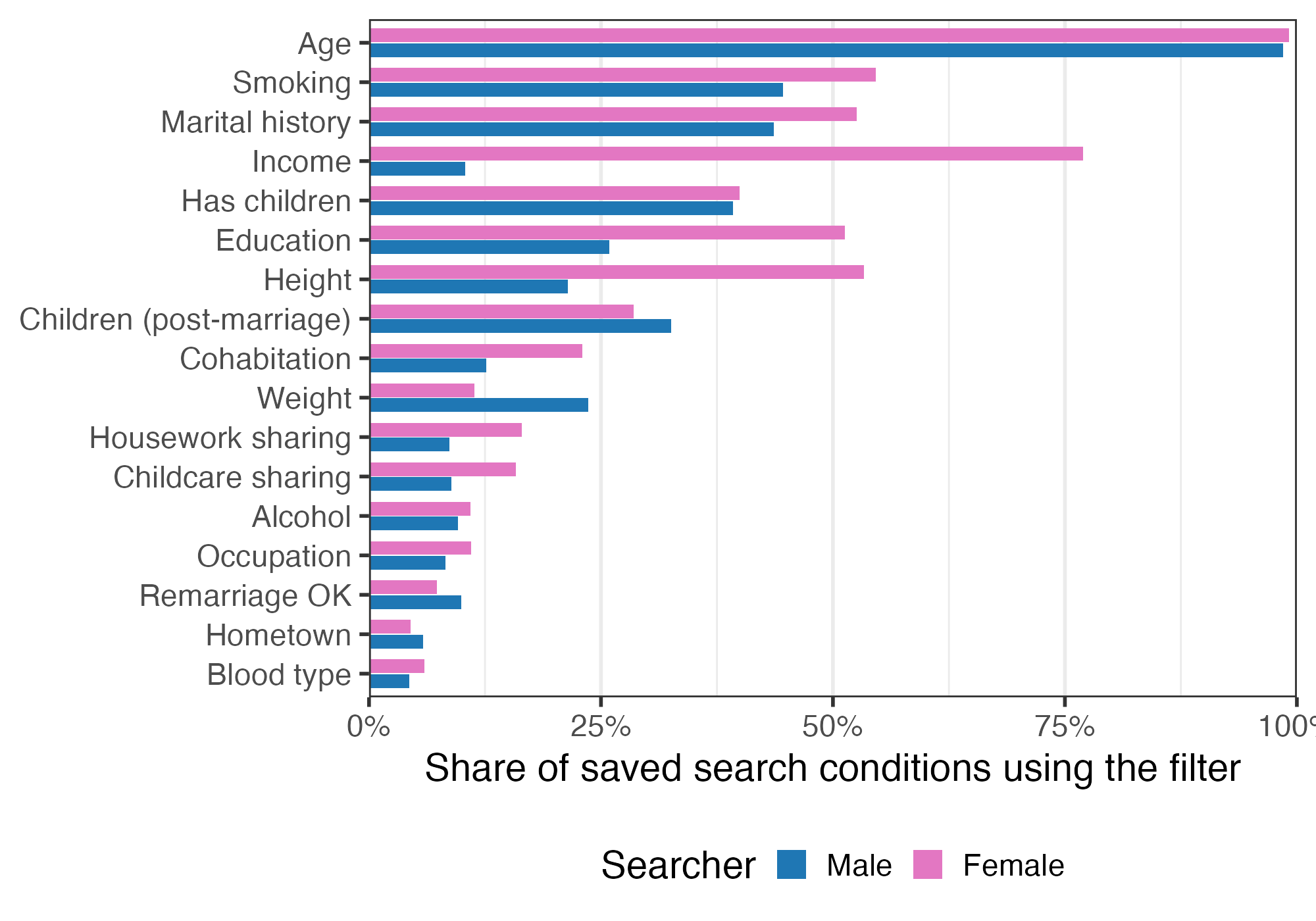}}
  \caption{Filter Usage on the IBJ Platform}
  \label{fg:filter_usage}
  \par\raggedright\footnotesize
  Note: The sample contains 133{,}879 saved conditions from 56{,}371 members. A range filter is active when either bound is set; a categorical filter is active when not all options are selected.
\end{figure}

Figure~\ref{fg:filter_usage} summarizes filter use. Panel~(a) shows that most saved conditions screen on a handful of attributes, with female-authored conditions screening on more attributes. Panel~(b) shows two facts. First, age is near-universal: about 99\% of saved conditions encode an age range, far more than any other attribute. Second, other screens are strongly gendered: female-authored conditions screen more on income (77\% versus 10\%), height (53\% versus 21\%), and education (51\% versus 26\%), whereas male-authored conditions screen somewhat more on weight. Some categorical screens, such as smoking and marital history, are also common, but they do not shift mechanically at a birthday. The empirical focus on age, income, and height is therefore driven by the birthday mechanism, the matching literature, and the descriptive gradients below.

Because age appears in nearly every saved search condition, it is also the margin through which a birthday is most likely to move a receiver into or out of stated eligibility. The rest of this section builds that mechanism around the stated acceptable age ranges, beginning with the size of the potential-proposer pool.

\subsection{The Potential-Proposer Pool}

\begin{figure}[!t]
  \captionsetup{skip=3pt}
  \centering
  \includegraphics[width = 0.55\textwidth]{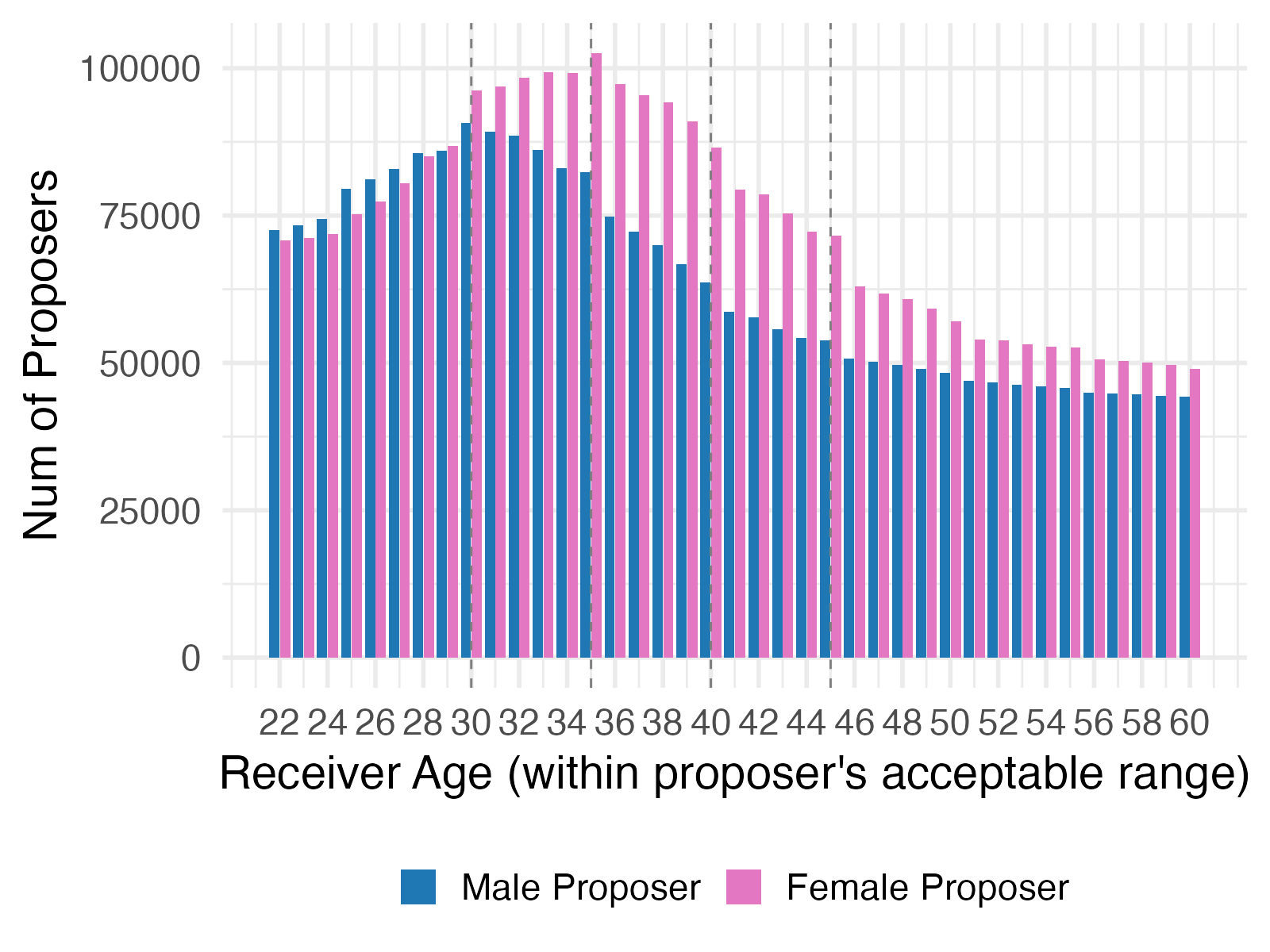}
  \caption{Receiver Ages Included in Proposers' Stated Acceptable Ranges}
  \label{fg:proposer_accepted_age_counts} 
  \par\raggedright\footnotesize
  Note: Counts use proposers' stated lower and upper acceptable-age bounds in the 2020--2024 sample.
\end{figure}

The first building block is the distribution of stated acceptable receiver ages. Figure~\ref{fg:proposer_accepted_age_counts} counts, for each possible receiver age, the number of proposers whose stated acceptable partner-age range includes that age. This is not yet a realized application outcome; it is the size of the stated consideration pool before any pair-specific interaction occurs. The distribution is sharply gendered. Male proposers most often include women in their late twenties and early thirties, and the number who include a given female receiver age declines steeply after the mid-thirties. Female proposers, by contrast, include men over a wider and older age range, with high counts through the thirties and into the early forties. These levels reflect both stated bounds and proposer composition; the adjacent-age changes examined below determine the mechanical birthday effect.

\subsection{Who Is Filtered In and Out}

The second step converts these age-specific counts into receiver-level opportunity sets. For each receiver cohort, I define the \emph{observed-active pool} as opposite-side proposers whose recorded activity windows overlap that cohort's birthday-window dates. A receiver's \emph{filter-implied opportunity set} is the subset of this pool whose stated acceptable age range contains the receiver; this age-based set is the potential-proposer pool used below.

\begin{figure}[!htbp]
  \begin{center}
  \includegraphics[width = 0.98\textwidth]{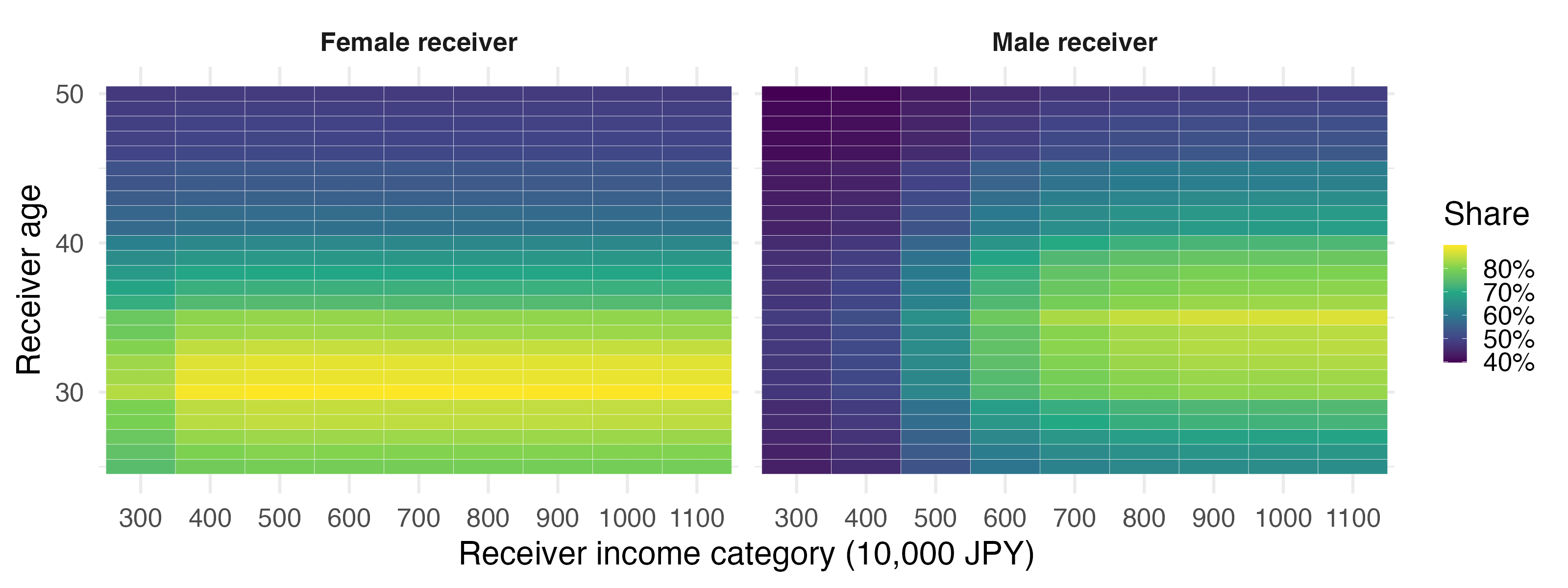}
  \caption{Filter-Implied Opportunity Share by Receiver Age and Income}
  \label{fg:joint_age_income_opportunity}
  \end{center}
  \footnotesize
  Note: Each cell reports the share of observed opposite-sex proposers in the 2020--2024 sample whose stated age and income filters both admit the receiver, by receiver age and income. A missing bound imposes no restriction on that dimension.
\end{figure}

To show how a second stated filter reshapes opportunity, Figure~\ref{fg:joint_age_income_opportunity} instead plots the share of opposite-sex proposers observed in 2020--2024 whose stated age \emph{and} income filters both admit the receiver, by receiver age and income (a missing bound imposes no restriction on that dimension). \textcolor{black}{The income bounds are the stated minimum and maximum acceptable partner-income categories recorded at registration, the same preference record as the stated age bounds.} Income is included because \citet{inoue2026marital} show that, alongside age, it is one of the central observable matching dimensions in this market, especially for male receivers; because the plotted share layers the income filter on top of the age filter, it can vary with receiver income at a fixed age. For male receivers, the heatmap shows a pronounced income gradient: averaging over the plotted ages, the included share rises from about 45\% in the lowest income category to about 70\% in the highest plotted category, while the age gradient is more modest over central ages. For female receivers, the dominant gradient is age: averaging over income, \textcolor{black}{the included share rises from about 79\% at age 25 to a peak of about 89\% near age 30 and then falls to about 48\% at age 50}, while the income gradient is small. Thus, the margin through which stated filters restrict opportunity differs by gender. Men's opportunity sets are more sensitive to income, whereas women's opportunity sets are much more sensitive to age. This difference motivates the gender-specific birthday analysis below.

\subsection{Stated Preferences versus Saved Search Conditions}\label{subsec:actual_filters}

I next compare stated acceptable ages with saved search conditions. For the subset of proposers in the snapshot described in Section~\ref{subsec:age_filter}, Table~\ref{tab:stated_vs_actual} compares the stated range with the median lower and upper age bounds across each proposer's saved conditions. The stated range contains this median saved range in 71.2\% of male-proposer cases and 71.4\% of female-proposer cases; the two ranges are disjoint in only 1.2\% and 0.5\% of cases, respectively. Stated bounds are therefore usually weakly wider than the median saved bounds, so the filter-implied consideration sets tend to be broad. Because the data are a snapshot of saved conditions rather than query-level browsing histories, the comparison does not identify the filter applied in any particular browsing episode.

\begin{table}[!htbp]
  \caption{Stated versus Saved Target Age Range}
\label{tab:stated_vs_actual}
\begin{center}
    \input{figuretable/birthday_project/stated_vs_actual_target_age_containment}
\end{center}
\footnotesize
  Note: Restricted to proposers in the saved-search snapshot. ``Stated $\supseteq$ Saved'' indicates that the stated acceptable age range contains the median lower and upper age bounds across a proposer's saved search conditions. \textcolor{black}{Open or missing endpoints are coded as ages 18 and 65 on both the stated and saved sides before comparison.}
\end{table}

\subsection{Birthday-Relative Mechanism Preview}\label{subsec:reading_panels}

The final step links the stated-filter mechanism to the birthday window.\footnote{Appendix~\ref{app:raw_data_pattern} first shows raw action counts; the application series already displays a break around the birthday.} If age bounds operate through consideration sets, applications should move in the eligibility cells implied by those bounds. The following plots show event-time means after removing member, calendar-year, within-month-position, and registration-tenure fixed effects. I partition inbound applications into four mutually exclusive cells: EXIT if the receiver is eligible before but not after the birthday, ENTER if eligible after but not before, STAY-IN if eligible on both sides, and STAY-OUT if ineligible on both sides.

\begin{figure}[!htbp]
  \begin{center}
  \includegraphics[width = 0.98\textwidth]{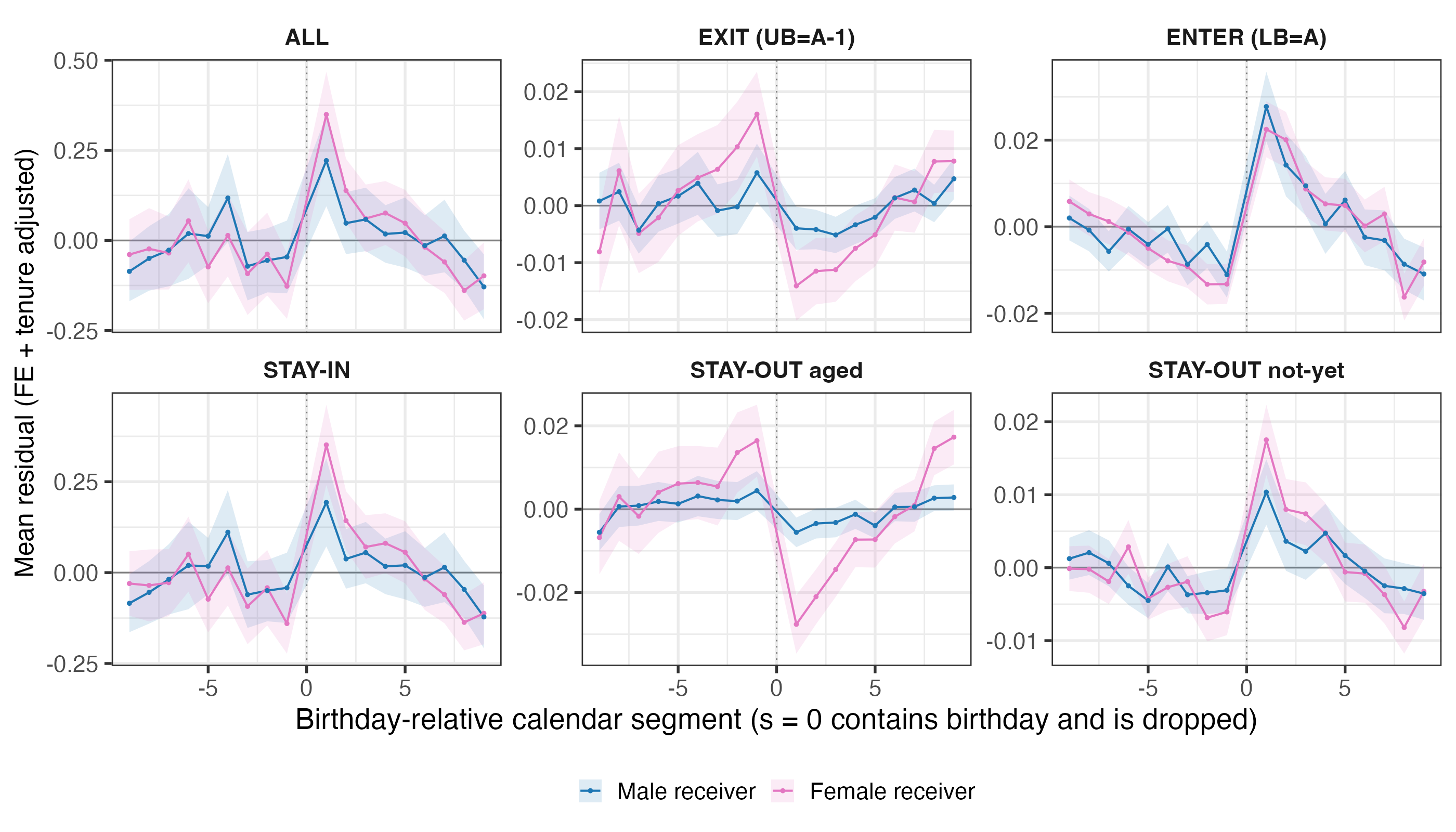}
  \caption{Birthday-Relative Application Patterns by Proposer Eligibility Type}
  \label{fg:event_study_age35_eligibility}
  \end{center}
  \footnotesize
  Note: Application-stage residual path for receiver age 35, with male and female receivers overlaid in each eligibility cell. Points average residual application counts after removing member, calendar-year, within-month-position, and registration-tenure fixed effects; the birthday-containing segment is omitted. EXIT denotes proposers whose stated upper bound is the receiver's pre-birthday age; ENTER denotes proposers whose stated lower bound is the receiver's post-birthday age; STAY-IN denotes proposers for whom the receiver is eligible on both sides; and STAY-OUT denotes proposers for whom the receiver is ineligible on both sides. STAY-OUT is split into proposers already aged out and those not yet reached; ALL contains all proposers. In the panel headings, \(A\) is receiver turning age, UB is the stated upper bound, and LB is the stated lower bound.
\end{figure}

Figure~\ref{fg:event_study_age35_eligibility} illustrates these cells at receiver age 35.\footnote{The decomposition is estimated at every receiver age 25--45. EXIT applications are broadly negative and ENTER applications broadly positive for both genders, although estimates become less precise as cells thin above age 40.} When the receiver turns 35, EXIT proposers have age 34 as their stated upper bound, so their applications should fall; ENTER proposers have age 35 as their stated lower bound, so their applications should rise. The figure shows precisely this pattern. The exercise therefore validates that the stated age bounds have empirical content: they organize the direction of application flows around the birthday in the way the search-interface mechanism predicts. Appendix~\ref{app:illustrative_model} presents an illustrative model that organizes the filter-boundary and within-set application margins.

The more important point is that the pooled response is not exhausted by this mechanical accounting. The pooled series rises, and the STAY-IN series---proposers for whom the receiver remains inside the stated age range on both sides of the birthday---also rises. Because STAY-IN proposers could apply both before and after the cutoff, this movement cannot be explained by a receiver simply entering or exiting a proposer's stated range. It captures an additional interface response that may combine changes in exposure under age-based ordering with proposers' responses to the displayed label.\footnote{STAY-IN is defined by stated bounds, not by every realized search query. The saved-filter snapshot shows that stated ranges contain the median saved range for about 71\% of proposers, but historical query-level filters are unavailable for the event-study period. A STAY-IN response can therefore combine movement across narrower browsing filters, changes in list position or exposure, and a conditional response to the displayed label. The design separates these channels from the broad stated-boundary margin but does not identify them individually.}

\begin{figure}[!htbp]
  \begin{center}
  \includegraphics[width = 0.95\textwidth]{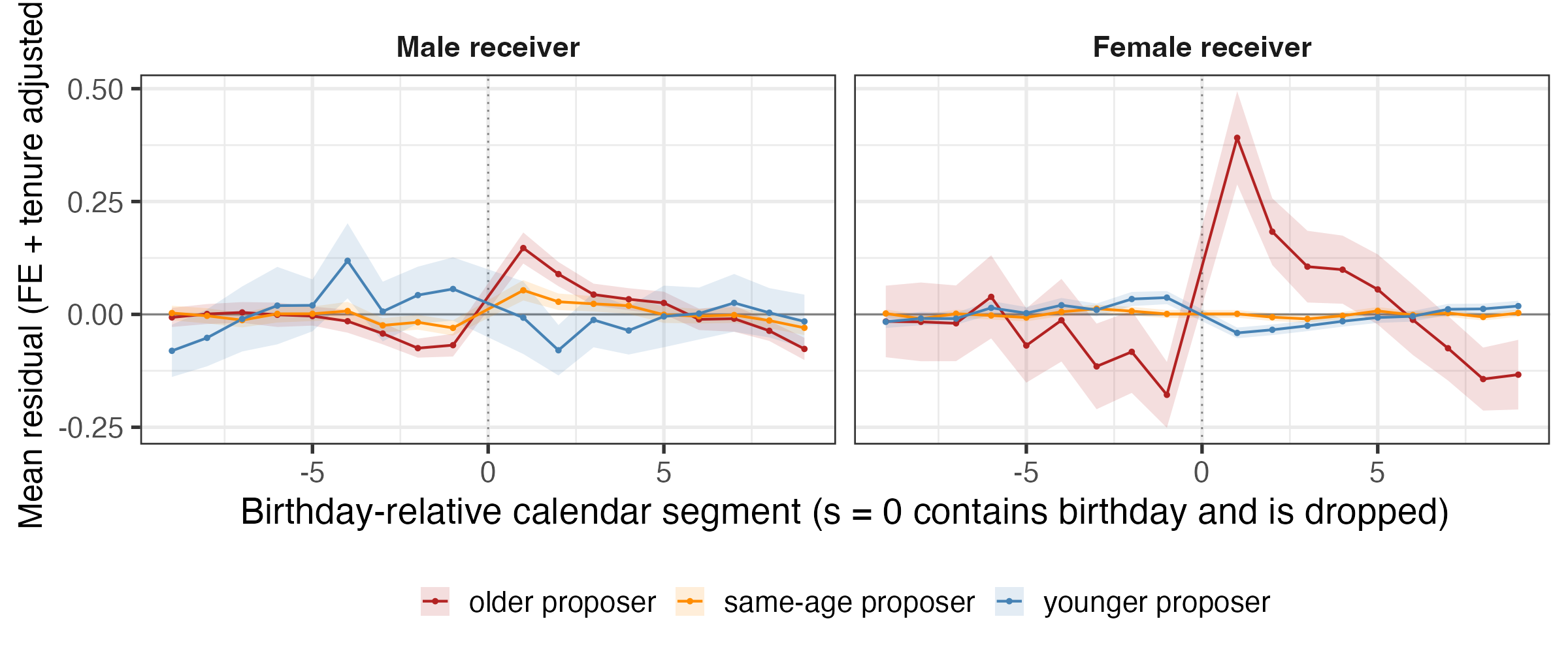}
  \caption{Birthday-Relative Application Patterns within STAY-IN Proposers by Proposer Age}
  \label{fg:stayin_age_es}
  \end{center}
  \footnotesize
  Note: Residual event-time path restricted to STAY-IN proposers and split by proposer age relative to the receiver. The residualization removes member, calendar-year, within-month-position, and registration-tenure fixed effects.
\end{figure}

Figure~\ref{fg:stayin_age_es} shows the relevant composition margin within this always-eligible group: applications from proposers older than the receiver rise at the birthday, while applications from younger proposers are flat or fall. The pattern is consistent with age proximity because the receiver becomes one year closer in age to older proposers and farther from younger proposers. The gender asymmetry is also informative. For male receivers, the response from older proposers is accompanied by a clearer decline from younger proposers; for female receivers, the pattern is dominated by a sharp increase from older proposers. Thus, the birthday re-sorts realized inbound attention even among proposers who remain eligible throughout the window. The design does not, however, distinguish age-gap preferences from age-dependent exposure or salience within the interface. Section~\ref{subsec:proposer_type_composition} estimates these eligibility and re-sorting patterns across receiver ages and proposer types.

Taken together, this evidence supports the mechanism that guides the empirical design. Stated age ranges generate gendered, filter-implied opportunity sets; these ranges usually contain the median bounds in saved search conditions; and the birthday event-time plots show the predicted eligibility movements together with re-sorting among always-eligible proposers. The next section formalizes this evidence in a within-member birthday design for receiver-side cutoffs.

\section{Empirical Strategy: Within-Member Birthday Event Study}\label{sec:empirical_strategy}

The empirical strategy treats the birthday as a sharp, platform-generated event in a within-member event-time design. Three clocks must be kept distinct: calendar time, birthday-relative event time, and platform tenure since registration. The design is informative because birthday dates vary across members, while the label update is mechanically timed and nonmanipulable: displayed age increases by one year, whereas verified attributes such as education, income, occupation, and marital status are not mechanically updated and are assumed to evolve smoothly in the narrow window.

\subsection{Event-Time Panel and Analysis Sample}

I construct an individual-segment panel around the receiver's birthday. Each month is divided into days 1--10, 11--20, and 21--month-end. The unit of observation is receiver \(j\) in birthday-relative calendar segment \(s\): \(s=-1\) immediately precedes the segment containing the birthday, \(s=0\) contains the birthday, and \(s=1\) immediately follows it.\footnote{The platform did not disclose exact birthdays to the author for privacy reasons, so this three-segment monthly partition is the finest event-time unit available.} The birthday segment is dropped as a donut hole, so all identifying comparisons use cleanly pre-birthday or post-birthday observations. The main window is \(s\in\{-9,\ldots,-1,1,\ldots,9\}\), approximately three months on each side.

The main estimation sample is a cleaned subsample of Table~\ref{tb:summary_statistics_female_matched}.\footnote{Appendix~\ref{app:event_time_support} reports sample support by receiver turning age, gender, and funnel stage.} Eligible members contribute the birthday-relative segments in which they are active. The first observed activity segment is excluded, and subsequent active segments with no recorded action are coded as zeros. Segments after exit are not imputed. The estimates therefore condition on remaining active. I analyze activity and exit separately; Appendix~\ref{app:robust_window} shows that exit rates evolve smoothly through the birthday.

\subsection{Estimating Equation}

The design is a within-member birthday event-time design, not a treated-versus-untreated comparison. Every member has a birthday, and the estimand compares the same member's outcomes just before and just after the displayed-age update. Cross-member variation in birthday dates is used to separate this birthday-relative change from calendar time, while variation in registration dates separates it from the platform-tenure profile. Individual fixed effects absorb permanent differences in attractiveness, socioeconomic status, and underlying search intensity. Calendar-year fixed effects absorb market-wide annual shocks, and within-month-position fixed effects absorb the regular within-month activity cycle documented in Appendix~\ref{app:calendar_month_pattern}. Tenure fixed effects absorb the smooth decline in activity after registration, which is pronounced at the application stage.\footnote{This tenure adjustment is central. Within a member, event time and platform tenure move together mechanically. I therefore estimate the post-birthday change while flexibly absorbing the common tenure profile with tenure fixed effects. Appendix~\ref{app:tenure_dispersion} shows that tenure at the birthday is widely dispersed across members, so the post-birthday indicator is separately identified from the tenure profile.}

After dropping the birthday segment \(s=0\), the main estimating equation is
\begin{equation}
Y_{js} = \tau \mathbbm{1}\{s>0\} + \sum_{k}\gamma_k\,\mathbbm{1}\{\text{tenure}_{js}=k\} + \sum_{p}\zeta_p\,\mathbbm{1}\{\text{month-position}_{js}=p\} + \mu_{j} + \delta_{y(j,s)} + \varepsilon_{js}.
\label{eq:rdd}
\end{equation}
Here \(Y_{js}\) is an outcome for receiver \(j\) in segment \(s\), \(k\) indexes tenure segments, \(p\) indexes positions within a calendar month, \(y(j,s)\) is the calendar year, \(\mu_j\) is a receiver fixed effect, and \(\varepsilon_{js}\) is the error term. Birthdays are broadly distributed across calendar segments, and the within-month-position effects directly absorb the regular activity cycle; adding them changes the estimated jump little (Appendix~\ref{app:robust_fe_choice}). The coefficient \(\tau\), which I call the birthday jump, is the average post-birthday change net of member, calendar-year, within-month-position, and tenure effects. It summarizes the average shift over the post-birthday window rather than an instantaneous discontinuity at \(s=0\). The design is closely related to a panel regression discontinuity (RD) in event time, but the preferred implementation is this tenure-adjusted event-time specification because the running variable has only three calendar segments per month and outcomes exhibit strong tenure profiles. Standard errors are clustered at the member level. I estimate equation~\eqref{eq:rdd} separately by receiver gender and, when studying composition, by proposer type.

\paragraph*{Identification.} The identifying assumption is that, absent the displayed-age update, \textcolor{black}{residual outcomes would have the same conditional mean in the pre- and post-birthday windows} after conditioning on member, calendar-year, within-month-position, and tenure effects\textcolor{black}{; smoothness alone is not sufficient, because a remaining birthday-relative trend would load onto the post indicator}. The estimand is the average shift associated with entering the persistent post-birthday age-label regime.\footnote{The within-member comparison follows \citet{list2025experimentalist}: observing the same unit in multiple states removes permanent heterogeneity but shifts attention to time, composition, and sequence effects. Here the relevant concern is temporal stability around the cutoff, not transience after a temporary treatment, because the age-label update is one-way and irreversible.}

The smooth-counterfactual assumption is credible for three institutional reasons. Birthdays are verified from official documents and cannot be manipulated. Proposers observe age in years and birth month but not the exact birthday date, limiting precise strategic timing just before the cutoff. Other verified attributes are not mechanically updated at the birthday and are assumed to evolve smoothly in the narrow window. The remaining concern is residual variation in platform activity rather than manipulation of the cutoff. I address it by absorbing member fixed effects, calendar-year effects, within-month-position effects, and flexible tenure-segment fixed effects, with particular attention to the new-member activity decline. The reported pre-event paths serve as diagnostics for temporal stability, not as proof of the identifying assumption.

A related concern is that the receiver's birthday may change the receiver's own platform activity and hence exposure. The own-birthday exercise shows no general increase in applications or continuation: female application counts decline at ages 35--42, while male effects are mixed (Appendix~\ref{app:proposer_birthday_cutoff}). This evidence is indirect.\footnote{Login histories and profile-update timestamps are not preserved as longitudinal event logs in the research extract. The own-birthday exercise is therefore distinct from the receiver-birthday design: users know their own birthdays and may adjust effort in advance, whereas the main design uses the receiver's birthday as a change in information observed by the opposite side.}

\subsection{Estimator Choices by Outcome}

The pool is a mechanical, filter-implied measure; applications and pre-relationships are dense behavioral counts; serious relationships and proposals are sparse. I therefore match the method to the object. I compute the pool change directly from stated bounds. Applications and pre-relationships use \(\tau\) in equation~\eqref{eq:rdd}, a within-member post-birthday shift. Serious relationships and proposals use descriptive pre--post differences and an application-tracked funnel decomposition. For behavioral outcomes, event time indexes the initiating application. Downstream counts record whether applications initiated in segment \(s\) later reach each stage, rather than the calendar segment in which the transition occurs.

\paragraph*{Potential-proposer pool.} Let \(a\) be receiver \(j\)'s turning age, \(t_j\) the receiver's birthday calendar segment, \(A_i(t_j)\) an indicator that proposer \(i\)'s observed activity window overlaps a 30-day neighborhood of that segment's midpoint, and \([L_i,U_i]\) the proposer's stated acceptable range. Holding the calendar-specific proposer set fixed as the displayed age changes from \(a-1\) to \(a\), the pool change is
\begin{equation*}
\Delta\mathrm{Pool}_{j,a}
=\sum_i A_i(t_j)\mathbbm{1}\{L_i=a,\ U_i\geq a\}
-\sum_i A_i(t_j)\mathbbm{1}\{L_i\leq a-1,\ U_i=a-1\}.
\end{equation*}
The first term is entry through newly satisfied lower bounds; the second is exit through crossed upper bounds.

\paragraph*{Applications and pre-relationships.} I estimate equation~\eqref{eq:rdd} using total counts and cell-specific counts defined by proposer characteristics and stated-age eligibility. To display the path, I regress \(Y_{js}\) on member, calendar-year, within-month-position, and registration-tenure fixed effects without the post-birthday indicator and average the residuals by event time \(s\). These residual paths are descriptive diagnostics. A saturated set of event-time indicators is not separately identified from unrestricted tenure fixed effects and member fixed effects without an additional trend normalization. \textcolor{black}{Because this residualization omits the post indicator, it can project part of a true step onto the fixed effects; within-side slopes depend on the normalization and are not unrestricted pretrend estimates.} The scalar \(\tau\) in equation~\eqref{eq:rdd} remains the estimand and summarizes the average shift over the post-birthday window.

\paragraph*{Serious relationships and proposals.} The same within-member count specification is not the main estimator because these outcomes are sparse at the individual-segment level, as suggested by the summary statistics in Table~\ref{tb:summary_statistics_female_matched}. Estimating many age-by-gender-by-proposer-type cells with member fixed effects yields many zero cells and imprecise estimates for stages reached by very few applications.

These later outcomes also differ conceptually from applications. Once a pair enters the relationship funnel, the interaction is one-to-one rather than a browsing environment governed by filters and sorting rules. The estimates therefore ask whether the birthday response propagates through already-formed matches, not whether the interface changes who is displayed or clicked.

For these stages, I combine descriptive Welch pre--post contrasts over the same \(\pm 9\)-segment window with an application-tracked decomposition. The decomposition follows common application cohorts and separates the pool, application rate, and three downstream transition rates. Because later outcomes are sparse and depend on earlier transitions, these estimates locate propagation rather than identify a separate effect at each stage.

\subsection{Heterogeneity and Eligibility Cells}

The main heterogeneity is by the type of proposer who interacts with the receiver. Motivated by the descriptive gradients in Figure~\ref{fg:action_counts_per_member_by_income}, I split counterparties into higher- and lower-type groups along three observable dimensions: age, income, and height. Age is relative to the focal member: I call an older counterparty an older-cohort proposer and classify that proposer as higher-age; I call a younger counterparty a younger-cohort proposer and classify that proposer as lower-age. For each receiver birthday, I measure the proposer's displayed age in the first clean calendar segment after the birthday-containing segment and hold both that age and the receiver's turning age fixed across the event window. A proposer therefore cannot switch relative-age cells within the window merely because of the proposer's own birthday. Income is split using the income-category upper edge: the category with upper edge 7 million JPY, which corresponds to 6--7 million JPY, and all higher categories are classified as higher-income.\footnote{Income is recorded as banded categories labeled by their exclusive upper edge in 1-million-JPY steps. For example, the category labeled by its 7-million-JPY upper edge covers income of at least 6 million and less than 7 million JPY. Equivalently, in 10{,}000-JPY units, a value of 700 corresponds to the interval 600--699.} Height is split at 170 cm for men and 155 cm for women, two focal thresholds visible in stated minimum-height requirements and in the descriptive attention profiles. These thresholds are always applied to the side whose composition is being studied: in the receiver-birthday analysis, they classify proposers; in proposer-side placebo analyses, they classify receivers. The estimates therefore ask whether the birthday changes not only the number of applications but also the composition of the inbound pool.

Finally, because the mechanism runs through stated age bounds, I decompose inbound applications into mutually exclusive eligibility cells. An EXIT proposer includes the receiver before the birthday but not after it because the receiver crosses the proposer's upper age bound. An ENTER proposer includes the receiver after the birthday but not before it because the receiver reaches the proposer's lower age bound. A STAY-IN proposer includes the receiver on both sides of the cutoff, and a STAY-OUT proposer is ineligible on both sides. Within STAY-IN, I further split applications by the proposer's own age relative to the receiver. This decomposition separates mechanical changes in stated-range membership from re-sorting among proposers whose stated bounds include the receiver on both sides of the birthday.

\section{Empirical Results: Birthday Effects Along the Funnel}\label{sec:rdd_estimation_demean}

This section follows the birthday response through the funnel. I first compute the potential-proposer pool change, then study applications, pre-relationships, and inbound composition, and finally turn to sparse serious-relationship and proposal outcomes.

\subsection{Potential-Proposer Pool}

\begin{figure}[!htbp]
  \begin{center}
  \includegraphics[width = \textwidth]{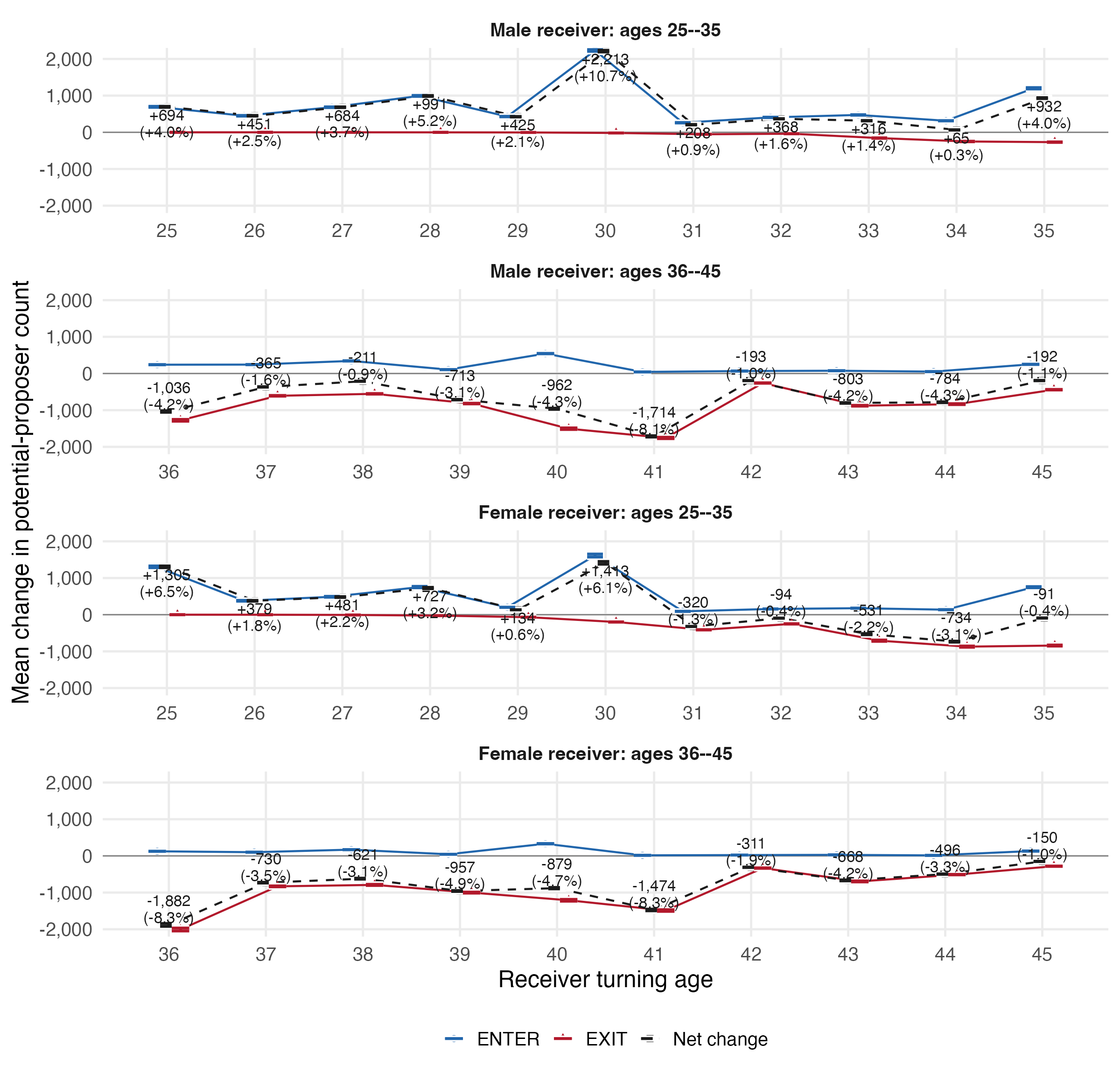}
  \caption{Mechanical Birthday Change in the Number of Stated Age-Eligible Proposers}
  \label{fg:eligibility_margin_pool_change}
  \end{center}
  \footnotesize
  Note: Each point is the receiver-weighted mean count for a gender-age cohort across birthday calendar segments; bars show 95\% confidence intervals with standard errors clustered by absolute birthday segment. The intervals are narrow relative to the count scale and may overlap the plotting symbols. For each receiver's birthday segment, the active proposer set is held fixed while eligibility is evaluated at ages \(a-1\) and \(a\). ENTER is the number gained when a proposer's lower bound equals \(a\). EXIT is the number lost, shown as negative, when a proposer's upper bound equals \(a-1\). Net change is their sum; labels report the rounded count and, in parentheses, its percentage of the pre-birthday pool. The pool contains opposite-sex proposers observed in the event sample whose activity windows overlap the receiver cohort's birthday-window dates.
\end{figure}

Figure~\ref{fg:eligibility_margin_pool_change} plots the numbers entering, exiting, and the resulting net change in the stated-eligible pool. The net change flips sign with age. For male receivers it is positive at every turning age from 25 through 35, peaking at \(+2{,}213\) potential proposers (\(+10.7\%\) of the pre-birthday pool) at age 30, and negative at every age from 36 through 45, reaching \(-1{,}714\) (\(-8.1\%\)) at age 41. For female receivers it is positive through age 30, peaking at \(+1{,}413\) (\(+6.1\%\)) at age 30 \textcolor{black}{in counts and at \(+6.5\%\) (\(+1{,}305\)) at age 25 in relative terms}, and negative from age 31 onward, reaching \(-1{,}882\) (\(-8.3\%\)) at age 36. The net is therefore heterogeneous rather than uniformly negative.

\subsection{Applications and Pre-Relationships}

\begin{figure}[!htbp]
  \begin{center}
  \includegraphics[width = 1.0\textwidth]{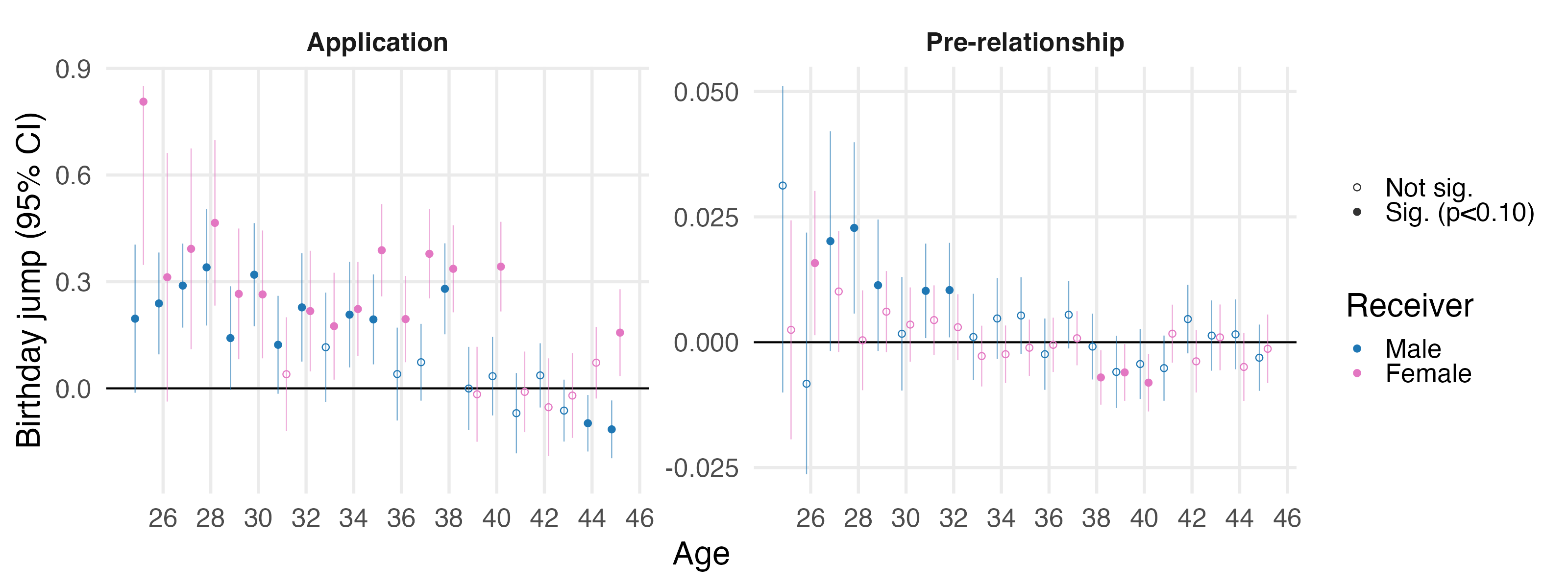}
  \caption{Birthday Jump at Application and Pre-relationship Stages}
  \label{fg:eventstudy_receiver_application_heatmap}
  \end{center}
  \footnotesize
  Note: Each point is the birthday jump, the coefficient on the post-birthday indicator in equation~\eqref{eq:rdd}, estimated on the main sample with zero filling within the active spell, by receiver age. The two panels are the application and pre-relationship stages. Filled markers are statistically significant at the 10\% level; hollow markers are not. Proposer-composition splits (higher- versus lower-type proposers) are reported in Figure~\ref{fg:characteristics_composition} and Appendix Figure~\ref{fg:app_characteristics_prerelation}.
\end{figure}

Figure~\ref{fg:eventstudy_receiver_application_heatmap} reports the birthday jump for applications and pre-relationships. The total application jump is small and, for most receiver ages, slightly positive rather than sharply negative. This matters for interpretation: the birthday does not mainly reduce the volume of inbound attention. Instead, as shown below, it changes which proposers apply. On the other hand, age-specific pre-relationship estimates are mostly small. \textcolor{black}{The pooled estimate for female receivers aged 31--40 is negative (\(-0.0020\) from a \(0.049\) base; Appendix~\ref{app:pooled_inference}), consistent with the recurring decline in this transition in the application-tracked decomposition in Section~\ref{sec:log_additive_decomp}.}\footnote{Appendix~\ref{app:prerelation_mechanism} reports the corresponding mechanism plots.}

\subsection{Composition of Inbound Applications}\label{subsec:proposer_type_composition}

\subsubsection{Eligibility Margins}

\noindent The small average application jump masks offsetting movements across stated age-bound cells. In the main estimation sample, the median jump is \(+0.12\) for male receivers and \(+0.22\) for female receivers. Pooled across receiver ages 25--45 with member-clustered inference, the jump is \(0.21\) for female receivers and \(0.08\) for male receivers (Appendix~\ref{app:pooled_inference}). For female receivers, applications rise for proposers whose stated lower bound is newly satisfied (ENTER, \(+0.02\) on average) and fall for those whose upper bound is crossed (EXIT, \(-0.02\)), while STAY-IN carries most of the total jump (\(+0.25\)). Table~\ref{tab:eventstudy_receiver_application_eligibility_E} reports the corresponding jump by age and eligibility type for female receivers; Appendix~\ref{app:eligibility_male} reports the male-receiver counterpart.

\begin{figure}[!htbp]
  \begin{center}
  \includegraphics[width = 0.98\textwidth]{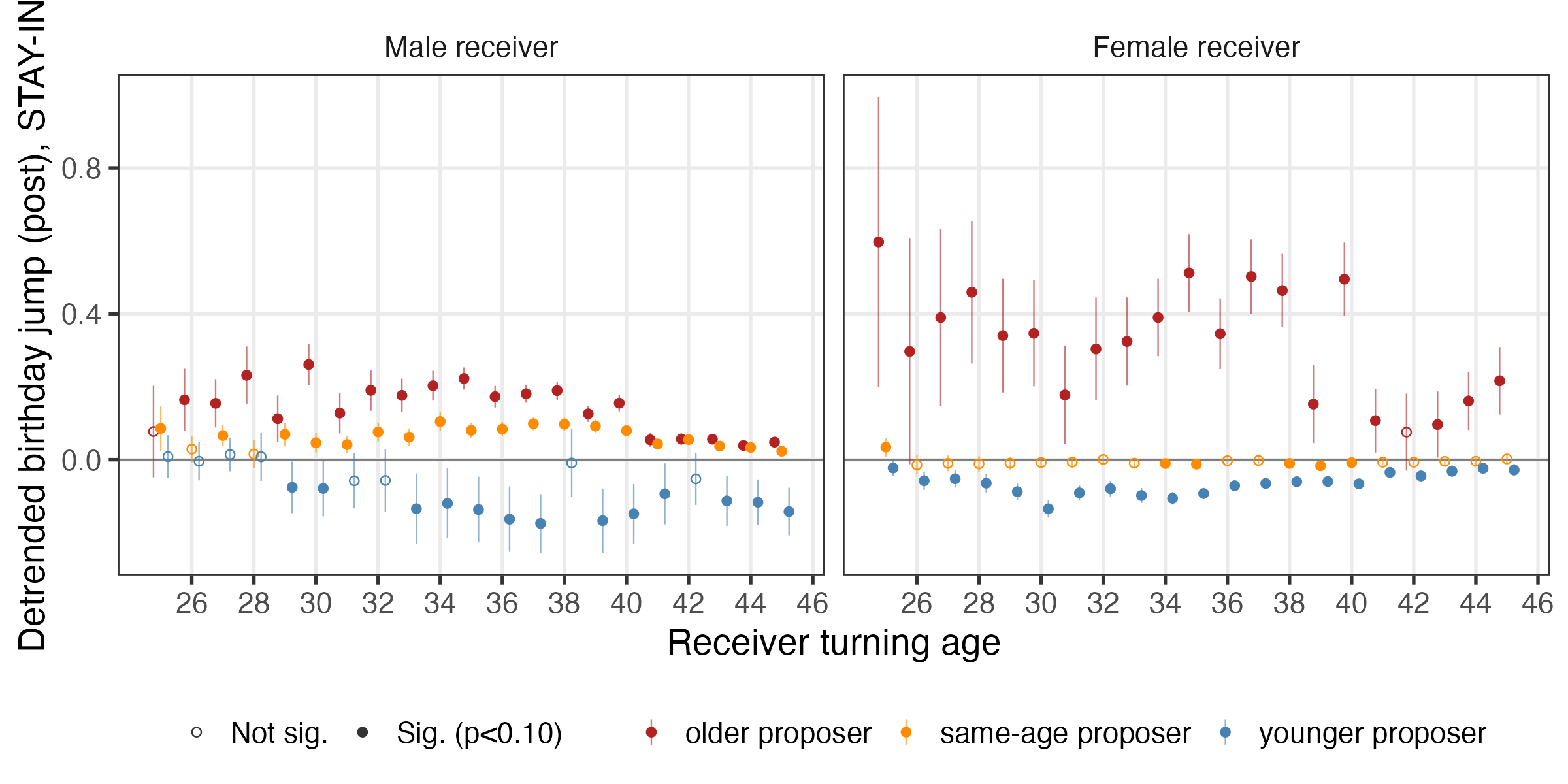}
  \caption{Proposer-Age Re-sorting at the Birthday within STAY-IN}
  \label{fg:resorting_E}
  \end{center}
  \footnotesize
  Note: Birthday jump within STAY-IN proposers, split by the proposer's own age relative to the receiver, by receiver turning age and gender, with 95\% confidence intervals. STAY-IN proposers are eligible to contact the receiver on both sides of the cutoff.
\end{figure}

Within STAY-IN, Figure~\ref{fg:resorting_E} splits the birthday jump by the proposer's own age relative to the receiver. Among receivers turning 35, applications from older-cohort proposers rise by \(0.22\) for men and \(0.51\) for women, while applications from younger-cohort proposers fall by \(0.14\) and \(0.09\), respectively. The same directional re-sorting appears across most central ages. Pooled over ages 25--45, older-cohort applications rise by \(0.33\) for female receivers and \(0.14\) for male receivers, while younger-cohort applications fall by \(0.07\) and \(0.11\), respectively (Appendix~\ref{app:pooled_inference}). Thus, the birthday re-sorts the inbound pool toward older-cohort suitors rather than uniformly reducing attention.

\begin{table}[!t]
  \captionsetup{skip=3pt}
  \caption{Birthday Jump by Eligibility Type, Application Stage, Female Receivers}
  \label{tab:eventstudy_receiver_application_eligibility_E}
  \centering
  \resizebox{\textwidth}{!}{\input{figuretable/birthday_project/eventstudy_receiver_application_eligibility_E_female.tex}}
  \par\vspace{0.1em}
  \begin{minipage}{0.95\textwidth}
  \footnotesize
  Note: Each cell reports the within-member application-stage birthday jump for female receivers, with standard errors in parentheses. STAYIN denotes eligibility on both sides of the cutoff; the STAY-OUT columns split proposers ineligible on both sides into those already aged out and those not yet reached. The five mutually exclusive cells sum to ALL, the all-proposer column. Table~\ref{tab:eventstudy_receiver_application_eligibility_E_male} reports male receivers.
  \end{minipage}
\end{table}

\subsubsection{Observable Proposer Characteristics}

\begin{figure}[!htbp]
  \begin{center}
  \subfloat[Higher-type proposers (older / higher-income / taller)]{\includegraphics[width = 0.90\textwidth, height = 0.45\textheight, keepaspectratio]{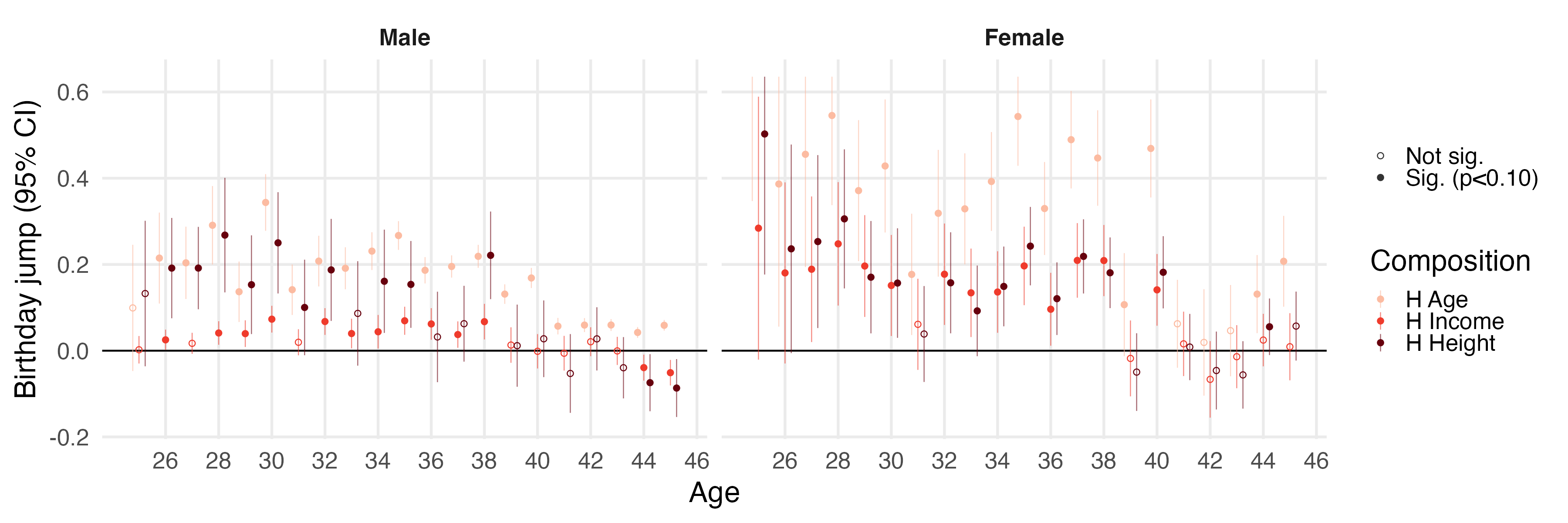}}\\[0.2in]
  \subfloat[Lower-type proposers (younger / lower-income / shorter)]{\includegraphics[width = 0.90\textwidth, height = 0.45\textheight, keepaspectratio]{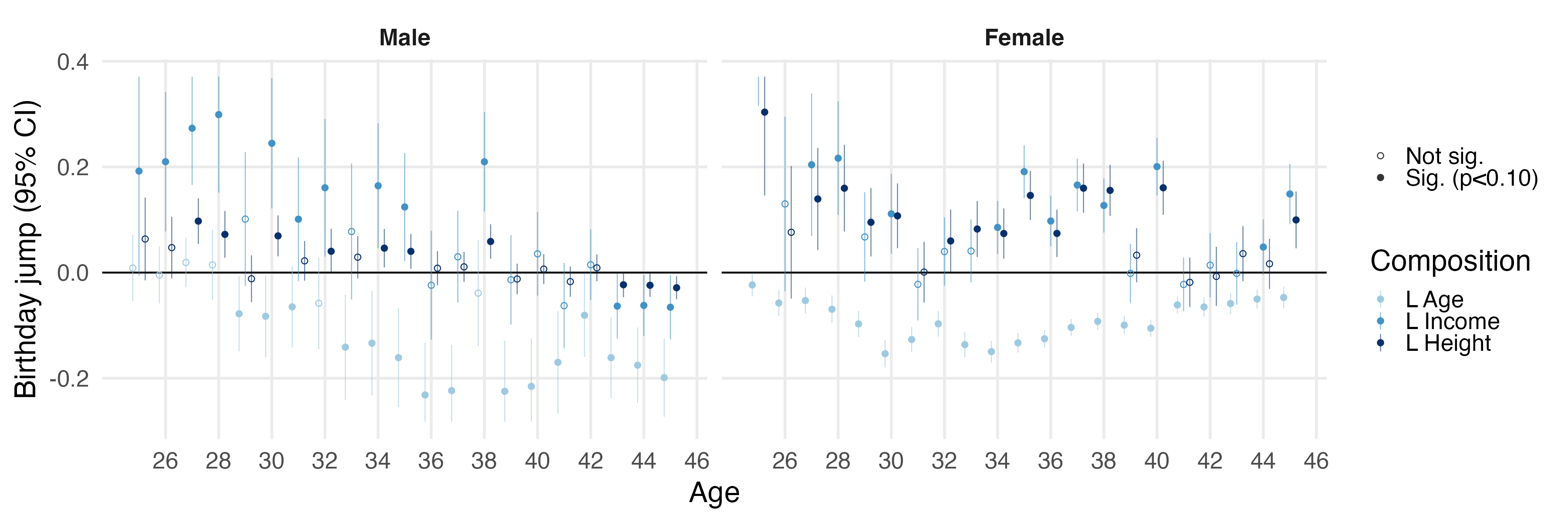}}\\[0.2in]
  \subfloat[By proposer stated child preference (Want / No Preference / Do Not Want)]{\includegraphics[width = 0.90\textwidth, height = 0.45\textheight, keepaspectratio]{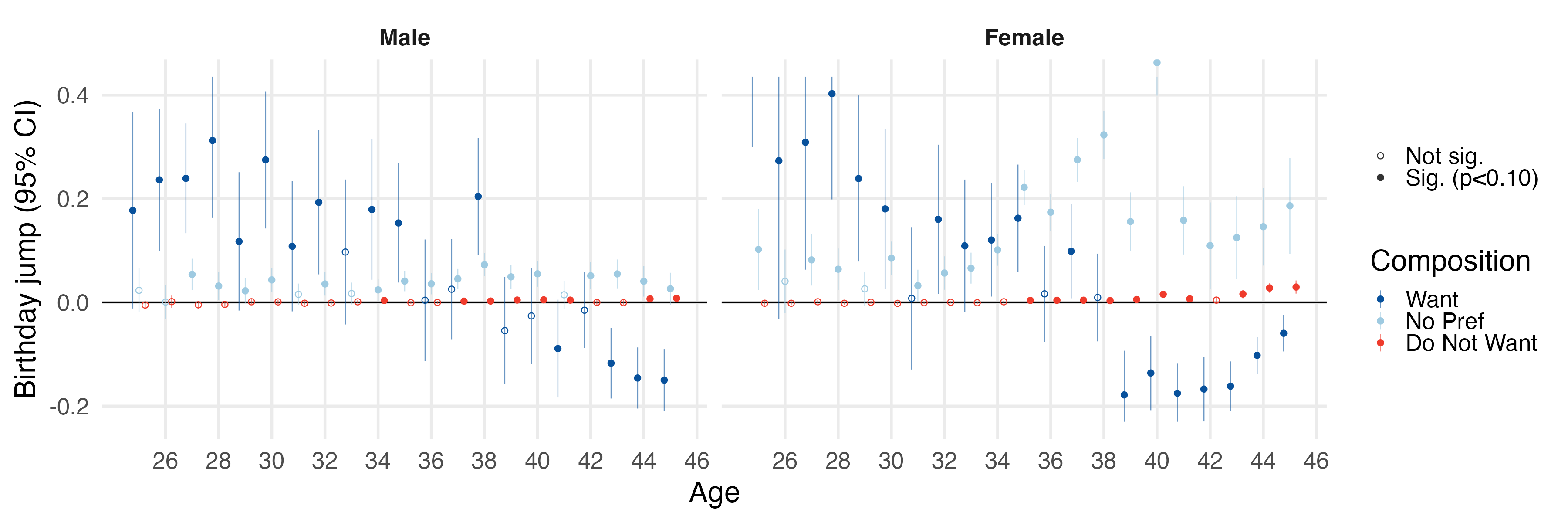}}
  \caption{Application-Stage Birthday Jump by Proposer Characteristics and Child Preference}
  \label{fg:characteristics_composition}
  \end{center}
  \footnotesize
  Note: Application-stage birthday jump by receiver age and gender. H/L Age denotes proposers older/younger than the receiver at the receiver's birthday. H/L Income uses a 7-million-yen income-category upper-edge cutoff. H/L Height uses a cutoff of 170 cm for male proposers and 155 cm for female proposers; H denotes values at or above each cutoff and L denotes values below it.
\end{figure}

\noindent The age-bound decomposition isolates the stated-filter margin. Figure~\ref{fg:characteristics_composition} asks whether the same birthday shock reshapes the composition of inbound applications along observable proposer characteristics: age, income, height, and stated child preference. Panel~(a) shows applications from older, higher-income, and taller proposers; Panel~(b) shows applications from younger, lower-income, and shorter proposers. The contrast is concentrated on age. Applications from older-cohort proposers rise at the birthday while applications from younger-cohort proposers fall, mirroring the STAY-IN pattern. By contrast, income and height splits are smaller, noisier, and do not display a systematic higher-versus-lower tilt at most ages. This pattern is consistent with the institutional shock: the birthday mechanically relabels age, not income or height.

Panel~(c) adds proposer child preference as secondary heterogeneity. For female receivers, the want-child component switches sign with age, positive at younger receiver ages and negative at older ages, while the no-preference component is broadly positive. For male receivers, the want-child component remains positive through the early and mid-thirties and turns negative only later and over a narrower age range. These cells help interpret the female age gradient, but the central composition result remains that applications re-sort primarily by proposer age. Among female receivers aged 31--40, the older-cohort counterparty share rises by 3.00 percentage points at applications and 2.54 points among accepted proposals; the estimates are positive but nonmonotonic across stages and become less precise as the funnel thins. The counterparty age gap rises and the share stating a preference for children falls at every stage (Appendix~\ref{app:realized_match}). Appendix~\ref{app:later_stage_characteristics} reports analogous splits for pre-relationships, serious relationships, and proposals.

\subsection{Serious Relationships and Proposals}

\begin{figure}[!htbp]
  \begin{center}
  \includegraphics[width = 1.0\textwidth]{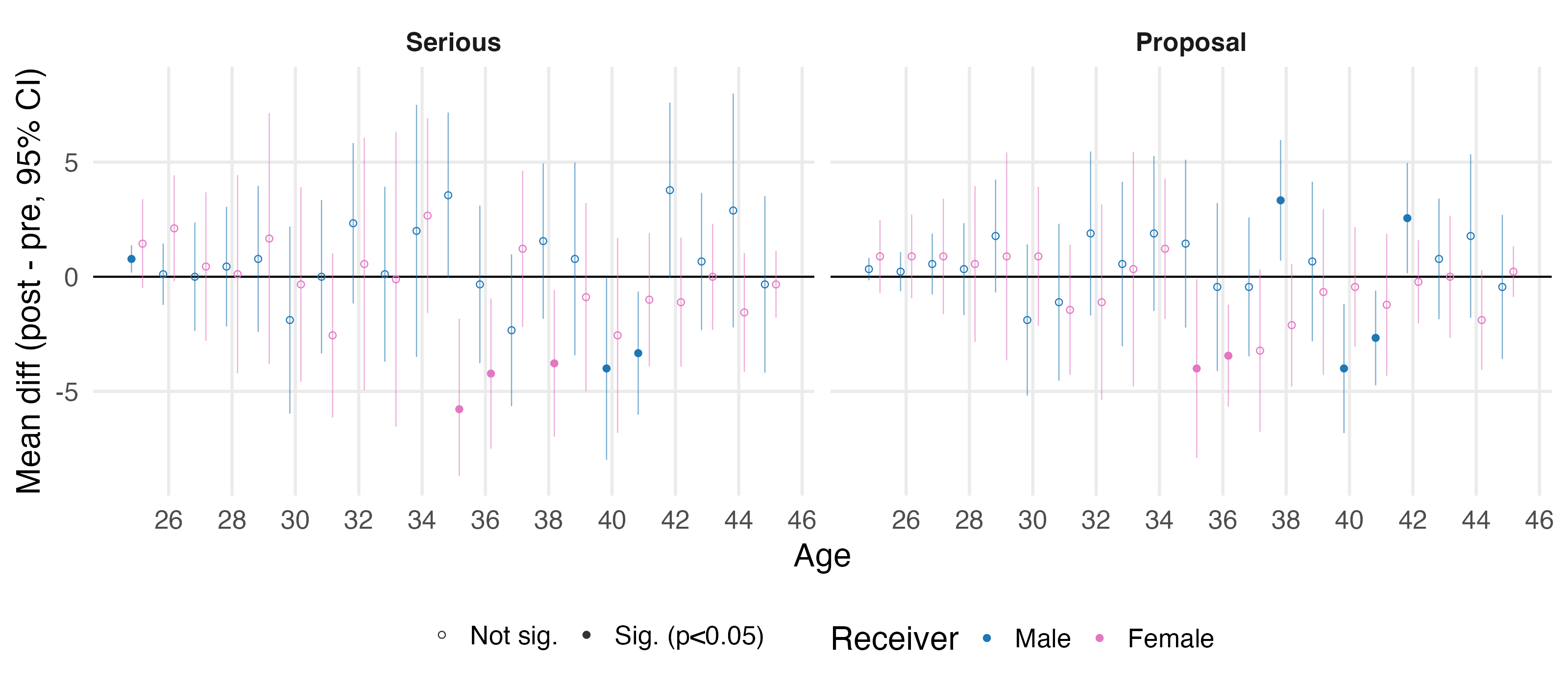}
  \caption{Estimated Pre--Post Differences at Serious Relationship and Proposal Stages}
  \label{fg:ttest_receiver_proposal_heatmap}
  \end{center}
  \footnotesize
  Note: Each point is the difference between the post-birthday and pre-birthday application-cohort segment means by receiver age, estimated using Welch's unequal-variance comparison, with 95\% confidence intervals.
\end{figure}

Serious relationships and proposals are sparse at the individual-segment level, so the within-member specification used for applications and pre-relationships is underpowered by age and proposer type. Accepted proposals also end platform participation, making an active-segment risk set outcome dependent. I therefore track fixed application cohorts by initiating segment, aggregate downstream counts by receiver age and proposer-characteristic cell, and compare the average in the nine pre-birthday segments with the average in the nine post-birthday segments. Figure~\ref{fg:ttest_receiver_proposal_heatmap} reports this Welch pre--post difference; Appendix~\ref{app:trend_check} visualizes the same estimand for receiver age 35.

Figure~\ref{fg:ttest_receiver_proposal_heatmap} reports the full set of descriptive contrasts; the following cells are statistically distinguishable from zero at the 5\% level. For serious relationships, male-receiver differences are positive at age 25 and negative at ages 40 and 41; female-receiver differences are negative at ages 35, 36, and 38. For proposals, male-receiver differences are positive at ages 38 and 42 and negative at ages 40 and 41; female-receiver differences are negative at ages 35 and 36. In the decomposition below, the pre-relationship-rate component is negative and larger in absolute value than the application-rate component at every female receiver age from 31 through 40.

Appendix~\ref{app:later_stage_characteristics} reports corresponding estimates at the pre-relationship, serious-relationship, and proposal stages, split by proposer age, income, height, and stated child preference. Age-related heterogeneity remains the clearest pattern, but the sparse later-stage cells do not support attributing the female proposal decline to a single proposer type.

\subsection{Funnel Decomposition}\label{sec:log_additive_decomp}

I use an application-tracked log-additive identity to describe how the observed pre--post change in proposal counts is distributed across stages of the funnel.

\paragraph*{Construction.}
I index cohort cells by receiver gender \(g\), receiver turning age \(a\), and application segment \(s\), measured relative to the receiver's birthday. Let \(N_{j,s}^{(0)}\) be the potential-proposer pool for receiver \(j\) in segment \(s\), and let \(N_{j,s}^{(k)}\) be the number of applications to receiver \(j\) initiated in segment \(s\) that reach stage \(k\), where \(k=1,2,3,4\) denotes Application, Pre-relationship, Serious relationship, and Proposal. The cohort-segment count \(N^{(k)}_{g,a,s}\) aggregates these individual counts across receivers in the \((g,a)\) cohort:
\begin{equation*}
    N^{(k)}_{g,a,s} \;:=\; \sum_{j \in (g,a)} N_{j,s}^{(k)}.
\end{equation*}
I retain each application record between proposer \(i\) and receiver \(j\) and track whether that application reaches later stages. Direct timestamps count a pre-relationship reached within 60 days, a serious relationship reached within 240 days, and a proposal reached within 420 days. Because each downstream stage institutionally requires all preceding stages, a recorded downstream transition also counts the application as having reached a preceding stage when that stage is not directly observed within its stage-specific window.
Anchoring all stages at the initiating application segment and imposing this institutional ordering ensures \(N^{(k)}_{g,a,s} \leq N^{(k-1)}_{g,a,s}\) \textcolor{black}{for \(k\geq2\)}. Hence \(\pi^{(k\mid k-1)}_{g,a,s} := N^{(k)}_{g,a,s}/N^{(k-1)}_{g,a,s}\) lies in \([0,1]\) \textcolor{black}{for \(k\geq2\)} whenever its denominator is positive\textcolor{black}{; because observed applications include STAY-OUT proposers while \(N^{(0)}\) counts only the stated-eligible pool, \(\pi^{(1\mid 0)}\) is an application-flow rate relative to that pool rather than a nested probability}. The funnel identity
\begin{equation*}
    \log N^{(4)}_{g,a,s} \;=\; \log N^{(0)}_{g,a,s} \;+\; \sum_{k=1}^{4} \log \pi^{(k|k-1)}_{g,a,s}
\end{equation*}
holds exactly at every cohort-segment cell with positive stage counts. For estimation, I avoid dropping zero-count segments by aggregating counts before taking logs. Let \(\mathcal S^- = \{-9,\ldots,-1\}\) and \(\mathcal S^+ = \{1,\ldots,9\}\), and define \(N_{g,a}^{(k),q}:=\sum_{s\in\mathcal S^q}N_{g,a,s}^{(k)}\) for \(q\in\{-,+\}\). Define \(\pi_{g,a}^{(k\mid k-1),q}:=N_{g,a}^{(k),q}/N_{g,a}^{(k-1),q}\), \(\widehat\Delta_0:=\log(N_{g,a}^{(0),+}/N_{g,a}^{(0),-})\), and \(\widehat\Delta_k:=\log(\pi_{g,a}^{(k\mid k-1),+}/\pi_{g,a}^{(k\mid k-1),-})\) for \(k=1,\ldots,4\). Whenever the aggregate counts are positive,

\[
\sum_{k=0}^{4}\widehat\Delta_k
=\widehat\Delta_{\mathrm{total}}
=\log\!\left(\frac{N_{g,a}^{(4),+}}{N_{g,a}^{(4),-}}\right)
\]
holds exactly, without pseudo-counts.

\begin{figure}[!htbp]
  \begin{center}
  \includegraphics[width=0.95\textwidth]{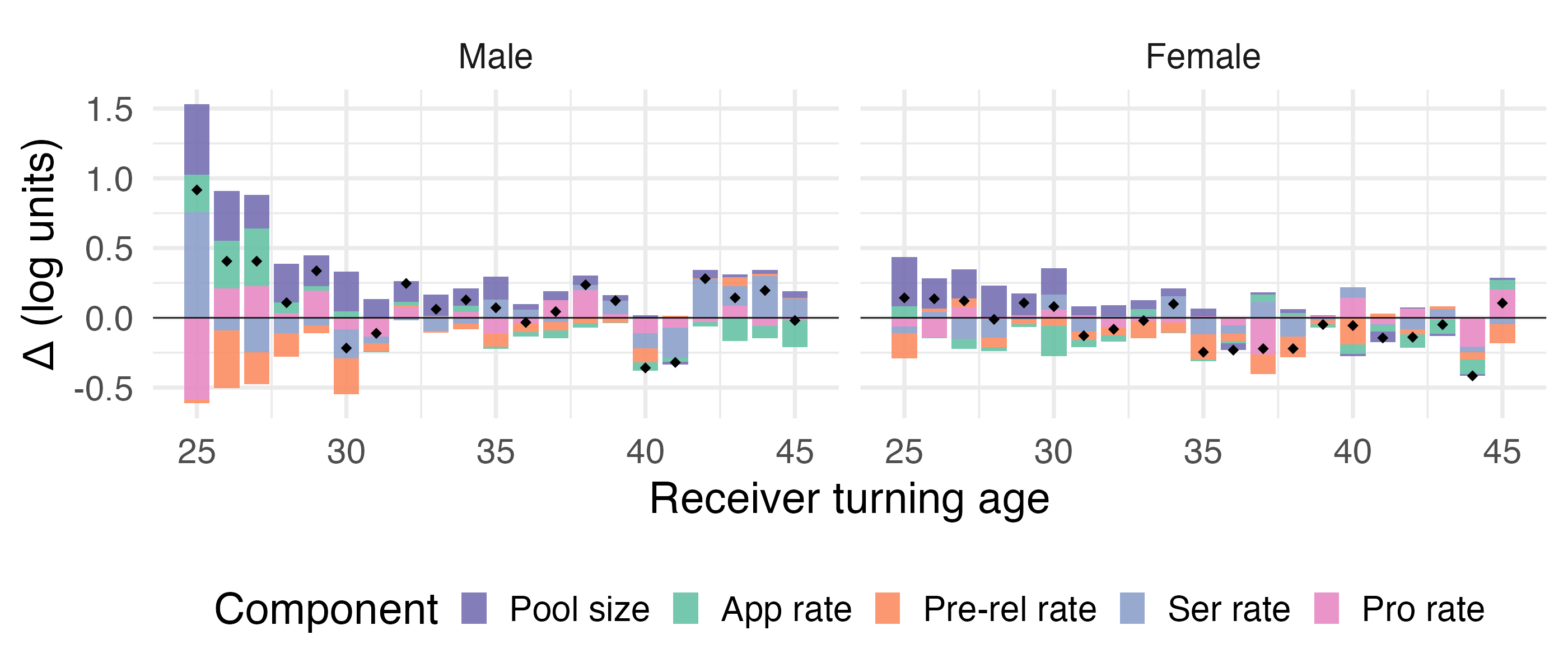}
  \caption{Application-Tracked Log-Additive Decomposition of Birthday-Relative Changes in the Marriage Funnel}
  \label{fg:birthday_decomposition_log_additive}
  \end{center}
  \footnotesize
  Note: Stacked bars decompose the log change in aggregate proposal counts into the contribution of the potential-proposer pool and the contribution of conditional transition rates at the Application, Pre-relationship, Serious, and Proposal stages. Counts are summed separately over nine pre-birthday and nine post-birthday segments before ratios and logs are taken. Black diamonds mark the same total change computed directly from proposal counts. Applications were initiated from June 2020 through November 7, 2023, with 420-day follow-up through December 31, 2024. The reported quantities cover the nine post-birthday segments and are not annualized.
\end{figure}

Figure~\ref{fg:birthday_decomposition_log_additive} reports aggregate pre--post log contrasts in the application-tracked funnel sample, not tenure-adjusted estimates. Three patterns stand out. First, proposal counts decline at nine of the ten female receiver ages from 31 to 40. Five declines exceed \(10\%\), and the largest is about \(22\%\), at age 35. Second, the pre-relationship transition is the recurring negative component, whereas the application-rate component is small for many central-age cohorts. Third, the pool and application-rate contributions vary with age and often offset one another. These contrasts complement rather than replace the tenure-adjusted application estimates. For male receivers aged 31--40, negative contrasts are confined to ages 31, 36, and 40 rather than spread across the age profile.

In Appendix~\ref{app:attribution_passthrough}, two further decompositions describe this margin. A shift-share reweighting associates \(22\%\) of the female decline at ages 31--40 with re-sorting toward older-cohort proposers, whose applications convert at lower rates; for male receivers, composition more than accounts for the decline and is offset by a positive within-type term (Appendix~\ref{app:delta2_reweighting}). Reply data locate the female decline at meeting formation: held meetings per application fall by about 17 percent, while continuation conditional on a held meeting is essentially unchanged (Appendix~\ref{app:who_ended}).

\section{Counterfactuals and Implications}\label{sec:counterfactuals}

I use two accounting counterfactuals over a window of nine calendar segments on each side of the birthday, or about 90 days before and after. Potential-proposer stocks use the exact birthday-segment accounting in Figure~\ref{fg:eligibility_margin_pool_change}. Behavioral stages use the common cohort of applications initiated from June 2020 through November 7, 2023, allowing 420-day follow-up through December 31, 2024. The no-age-bounds accounting (CF1) treats every active opposite-side member as age-eligible at fixed rates. The no-birthday accounting (CF2) retains stated filters and separately removes the mechanical pool change and each stage's fitted birthday change. The reported quantities cover the nine post-birthday segments and are not annualized.

\subsection{Counterfactual Design}

For each gender-age cell, CF1 scales every benchmark stage by the ratio of all observed-active opposite-side members to those whose stated range includes the post-birthday age. It is a fixed-rate all-active benchmark, not a forecast of removing age filters. Age-specific ratios receive different weights across stages.

For the potential-proposer pool, the benchmark evaluates stated eligibility at turning age $A$, while CF2 evaluates the same active proposer set at $A-1$; their difference is the mechanical net change in Figure~\ref{fg:eligibility_margin_pool_change}. From Application onward, the benchmark equals the pre-birthday level plus the estimated birthday change, and CF2 retains that pre-birthday level. Applications and pre-relationships use equation~\eqref{eq:rdd}; serious relationships and proposals use the descriptive Welch differences. Because these fitted changes are removed independently, CF2 reports a separate accounting contrast for each stage and does not chain application counts into later stages.

\subsection{Aggregate Counts Comparison}

Table~\ref{tb:counterfactual_table_female} aggregates receiver ages. Under CF1, the potential-proposer pool is \(32.5\%\) above the benchmark for men and \(33.9\%\) for women; fitted stage counts are \(28.3\)--\(29.4\%\) and \(29.3\)--\(31.5\%\) above it, respectively. Proposal-stage differences are \(28.9\%\) and \(29.9\%\). \textcolor{black}{Because CF1 imposes fixed rates, these downstream differences are scaled benchmark quantities rather than estimated responses to removing age bounds. The directly measured implication is that treating all active counterparts as a user's opportunity set overstates the stated-eligible pool by about one-third, a wedge more than an order of magnitude larger than the aggregate mechanical CF2 pool change.}

\textcolor{black}{The CF2 stage-by-stage accounting differs sharply across receiver channels. Relative to the benchmark, the potential-proposer pool differs by \(-0.1\%\) for male receivers and \(+1.3\%\) for female receivers. Applications, pre-relationships, serious relationships, and proposals are \(4.1\%\), \(3.5\%\), \(3.7\%\), and \(5.2\%\) lower for male receivers. For female receivers, applications are \(7.2\%\) lower, while pre-relationships, serious relationships, and proposals are \(1.3\%\), \(5.7\%\), and \(8.2\%\) higher. The positive female-receiver proposal contrast is estimated independently at the proposal stage, not propagated from the lower application count.}

\begin{table}[H]
\caption{Post-Birthday Benchmark versus Counterfactual Counts}
\label{tb:counterfactual_table_female}
\begin{center}
  \subfloat[Male receivers]{\input{figuretable/birthday_project/counterfactual_table_male}}\\[0.2in]
  \subfloat[Female receivers]{\input{figuretable/birthday_project/counterfactual_table_female}}
\end{center}
\footnotesize
Note: Counts cover the nine post-birthday segments for receiver ages 25--45. Potential-proposer Benchmark, CF1, and CF2 evaluate the same birthday-segment active set using stated eligibility at age $A$, no stated age bounds, and stated eligibility at $A-1$, respectively. CF1 applies the resulting age-specific pool ratio to every behavioral stage; CF2 sets each behavioral-stage birthday change to zero. Applications were initiated from June 2020 through November 7, 2023, with 420-day follow-up through December 31, 2024. The Potential Proposer row sums gender-age-segment stocks; the remaining rows sum fitted pair-level flows and are not directly comparable in levels. CF1 is an all-active scaling benchmark under fixed rates; CF2 is computed independently by stage.

\end{table}

\subsection{Counts Comparison Across Ages}

I next decompose these aggregate results by receiver turning age.

\paragraph{Action counts.}

\begin{figure}[!htbp]
  \begin{center}
  \subfloat[Male receivers]{\includegraphics[width = 0.95\textwidth, height = 0.38\textheight, keepaspectratio]{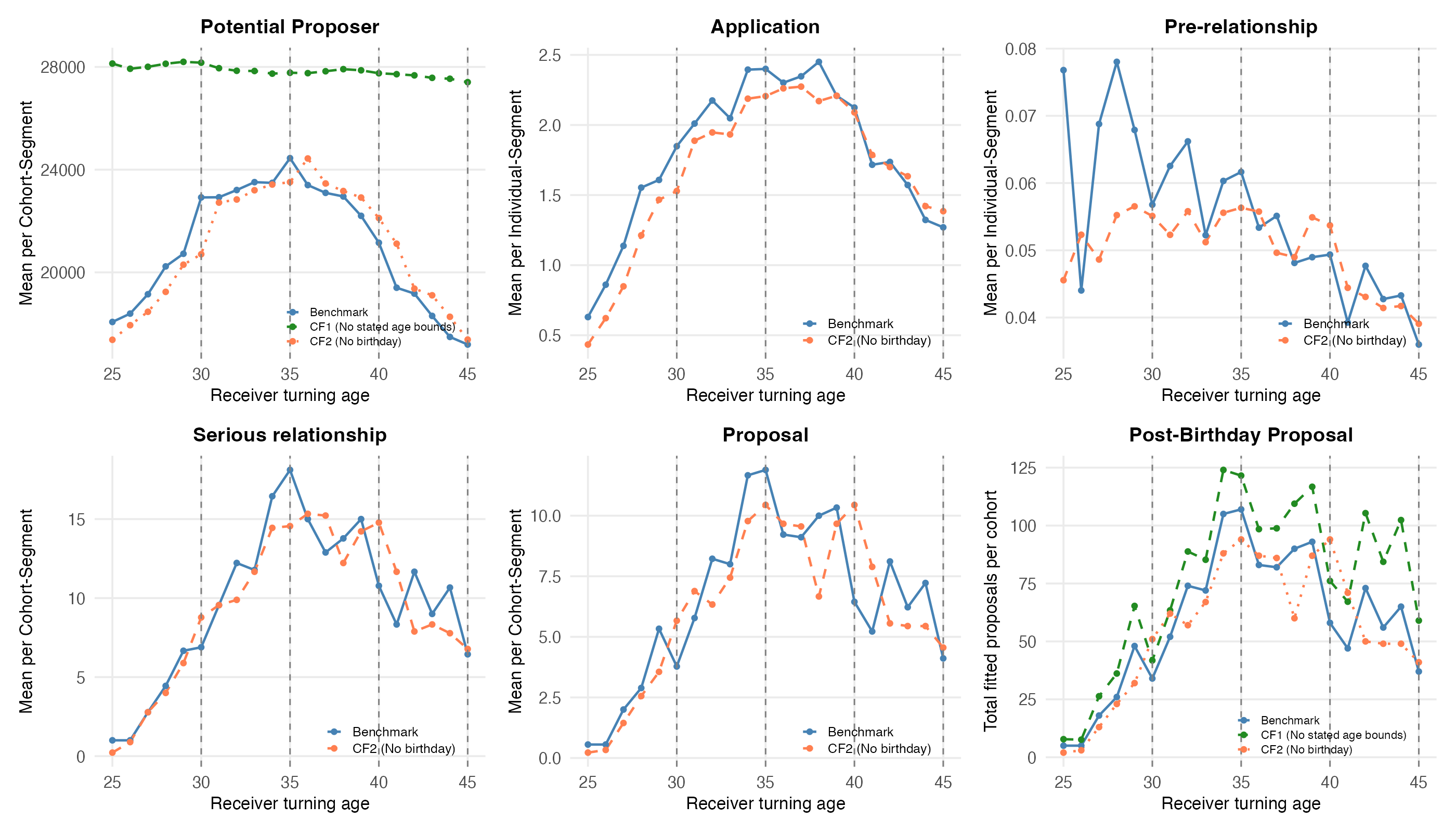}
  }\\[0.2in]
  \subfloat[Female receivers]{\includegraphics[width = 0.95\textwidth, height = 0.38\textheight, keepaspectratio]{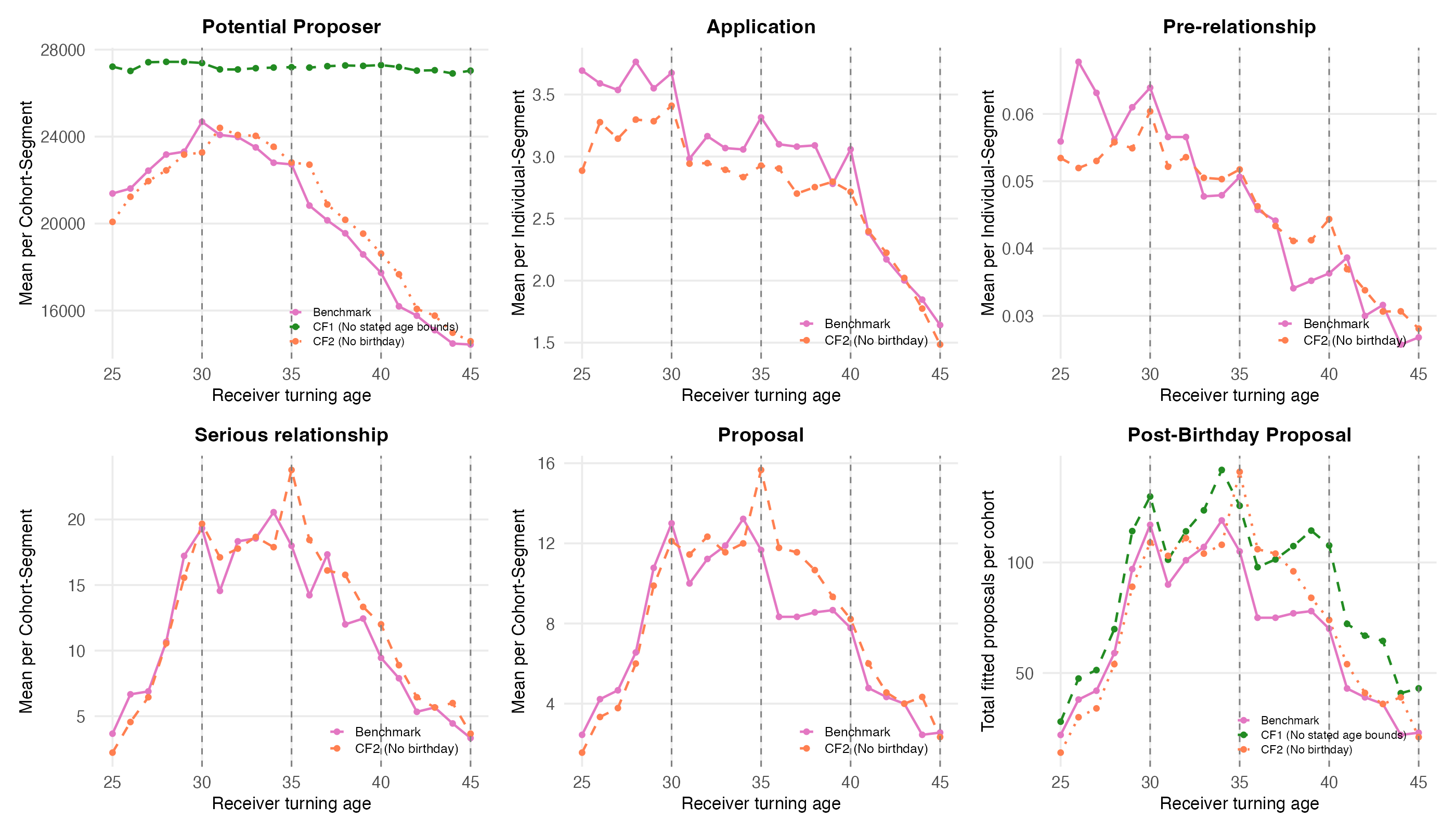}}
  \caption{Post-Birthday Benchmark and Counterfactual Age Profiles}
  \label{fg:counterfactual_female_receiver}
  \end{center}
  \footnotesize
  Note: The Potential Proposer panel evaluates the birthday-segment active set at age $A$ (Benchmark), without stated age bounds (CF1), and at age $A-1$ (CF2). CF1 scales each behavioral stage by the age-specific pool ratio; CF2 sets each behavioral-stage birthday change to zero. Applications were initiated from June 2020 through November 7, 2023, with 420-day follow-up through December 31, 2024. Application and pre-relationship panels report means per receiver-member segment; serious-relationship and proposal panels report means per gender-age segment. The final panel sums fitted proposals over the nine post-birthday segments. The reported quantities cover the nine post-birthday segments and are not annualized.
\end{figure}

Figure~\ref{fg:counterfactual_female_receiver} plots the accounting counts by receiver turning age. CF1 makes potential-proposer pools nearly flat by treating all active opposite-side members as eligible; CF2 removes the mechanical changes in Figure~\ref{fg:eligibility_margin_pool_change}. At many female receiver ages, CF2 applications are lower but fitted proposals are higher; each stage is adjusted independently. \textcolor{black}{For female receivers in their thirties, the no-birthday proposal-stage count is \(12.9\%\) above the benchmark.}

\paragraph{Portfolio composition.}

\begin{figure}[!htbp]
  \begin{center}
  \subfloat[Male receivers]{\includegraphics[width = 0.95\textwidth, height = 0.38\textheight, keepaspectratio]{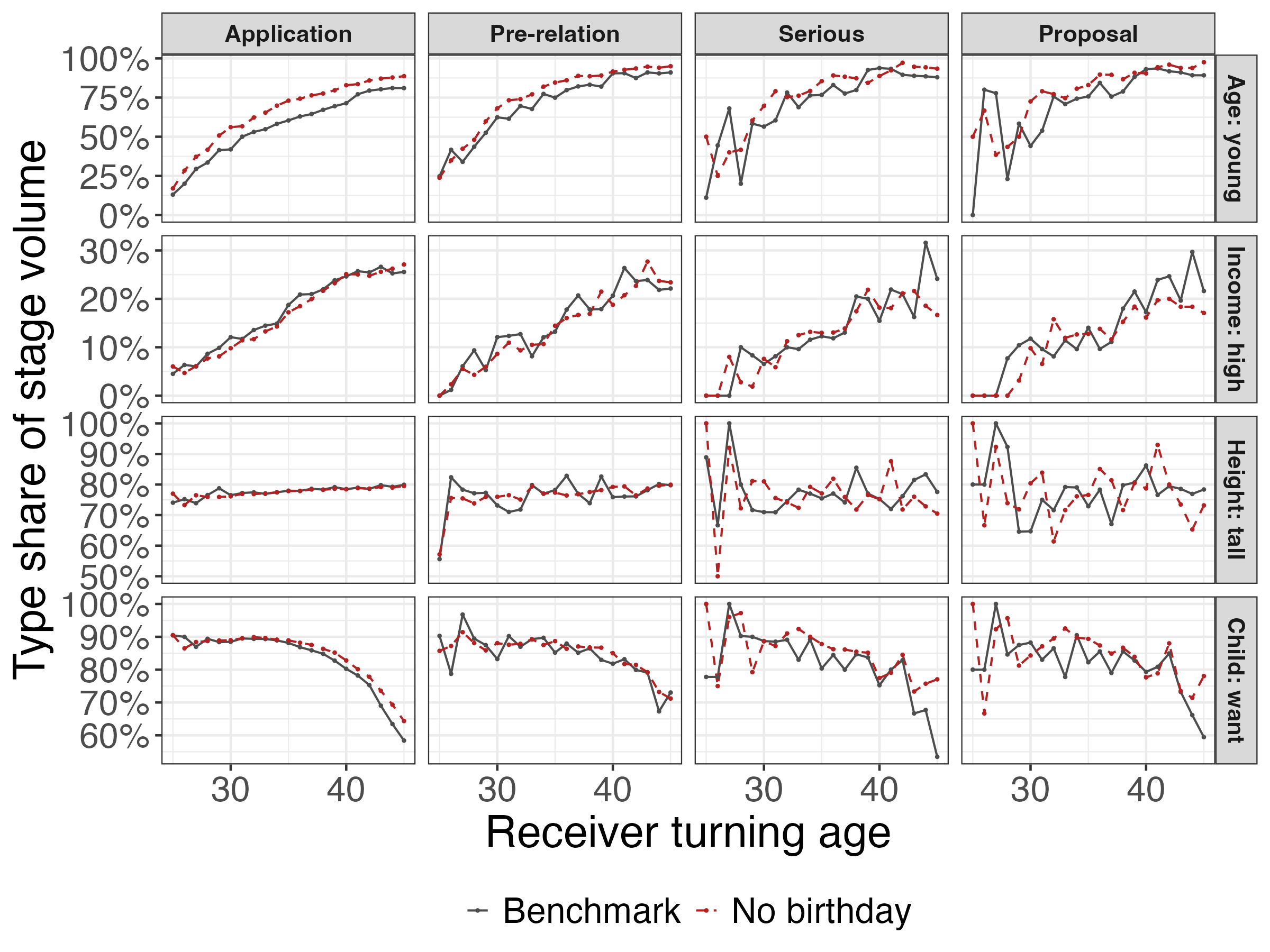}}\\[0.2in]
  \subfloat[Female receivers]{\includegraphics[width = 0.95\textwidth, height = 0.38\textheight, keepaspectratio]{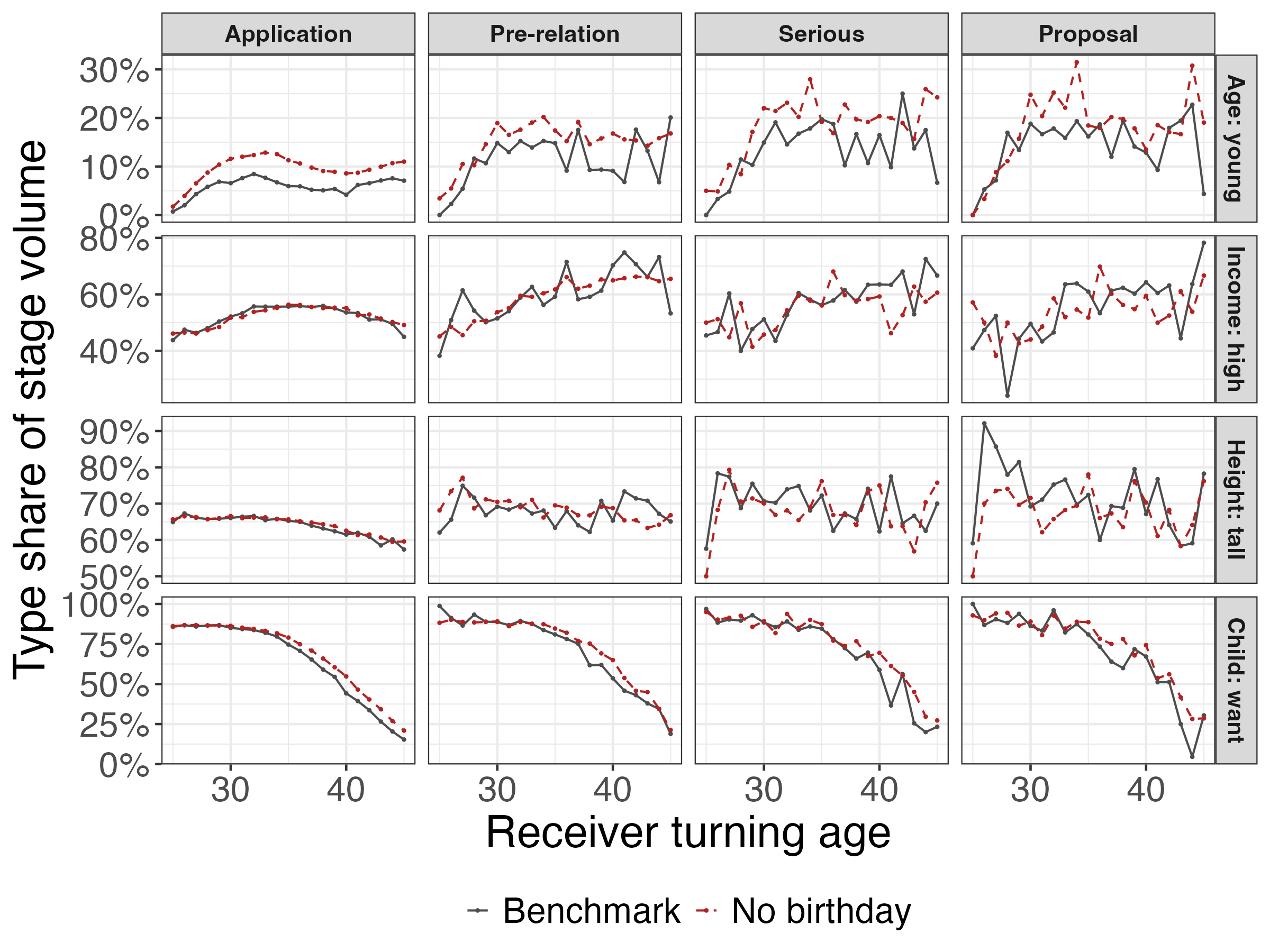}}
  \caption{Counterparty Composition Across Funnel Stages: Benchmark versus CF2}
  \label{fg:counterfactual_portfolio_share}
  \end{center}
  \footnotesize
  Note: Share of fitted stage volume over the nine post-birthday segments accounted for by the indicated proposer type under the benchmark and CF2. The relative-age denominator is total fitted stage volume and therefore includes same-age proposers. Income, height, and child-preference shares use fitted volume with a nonmissing classification on the corresponding dimension.
\end{figure}

Figure~\ref{fg:counterfactual_portfolio_share} compares stage-specific proposer composition under the post-birthday benchmark and CF2. Among all fitted applications, the younger-suitor share rises from \(6.2\%\) to \(10.4\%\) for female receivers and from \(64.6\%\) to \(75.2\%\) for male receivers. The proposal-stage accounting moves in the same direction: \textcolor{black}{the corresponding shares rise from \(15.5\%\) to \(19.7\%\) for female receivers and from \(77.5\%\) to \(84.1\%\) for male receivers}. The benchmark therefore assigns more stage volume to older suitors than the no-birthday accounting at both stages.

The contrast with other characteristics is informative. Child-preference shares also move, consistent with the correlation between proposer age and preferences for children. Income and height application shares differ by about \(0.5\) percentage points or less; some later-stage differences are larger, but they vary in sign and across ages rather than displaying the consistent directional shift seen for relative age.

Female receivers begin with an application pool dominated by older suitors, whereas male receivers begin with one dominated by younger suitors. For both genders, the benchmark is shifted toward older proposers relative to CF2, but from opposite sides of the prevailing age-gap distribution. The same direction at the application and proposal stages shows that the fitted composition difference is not confined to first contact, although CF2 does not trace a common transition vector. Relative age---and correlated child preferences---therefore provides the clearest portfolio contrast.

\paragraph{Summary of Counterfactuals.}

\textcolor{black}{Table~\ref{tb:counterfactual_headline_summary} collects the headline quantities. In the stage-by-stage no-birthday accounting for receivers in their thirties, the proposal-stage count is \(12.9\%\) above the benchmark in the female-receiver channel and \(6.7\%\) below it in the male-receiver channel. The younger-man share in the former rises from \(17.1\%\) to \(21.9\%\), or \(4.8\) percentage points; across male receivers aged 25--45, the younger-woman share rises from \(77.5\%\) to \(84.1\%\). These contrasts describe application-origin channels, not totals by sex, because every accepted proposal pairs a man and a woman.}

\begin{table}[!htbp]
\caption{Headline Counterfactual Quantities: Benchmark versus No Birthday}
\label{tb:counterfactual_headline_summary}
\begin{center}
  {\resizebox{\textwidth}{!}{\input{figuretable/birthday_project/counterfactual_headline_summary}}}
\end{center}
\footnotesize
Note: Aggregate counterparts of the fitted proposal counts in Figure~\ref{fg:counterfactual_female_receiver} and the younger-suitor shares in Figure~\ref{fg:counterfactual_portfolio_share}, summed over the nine post-birthday segments and the indicated receiver turning ages; ``ages 30--39'' denotes receivers in their thirties. Changes report percentage changes for counts and percentage-point (pp) changes for shares.
\end{table}

\subsection{Robustness Check}

Appendix~\ref{app:additional_results} documents sample support, tenure identification, sparse-outcome paths, later-stage heterogeneity, and users' own-birthday behavior. Appendix~\ref{app:attribution_passthrough} reports side attribution, composition reweighting, counterparty composition across stages, and pooled estimates. Appendix~\ref{app:robustness} varies the event window and calendar controls and tests exit, anticipation, and placebo cutoffs. These are targeted diagnostics for the smooth-counterfactual assumption and the downstream interpretation.

\section{Conclusion}

Digital matching platforms make large markets searchable by turning personal traits into visible labels and filters. Do these tools merely reduce search costs, or do they also shape who is considered and which relationships reach engagement? The automatic birthday update on IBJ isolates this question by changing displayed age while leaving other verified attributes unchanged.

\textcolor{black}{The answer is that age-based platform design primarily reshapes composition and continuation rather than aggregate first contact. Entry at lower age bounds often offsets exit at upper bounds, leaving modest net changes in stated eligibility and applications; within the eligible pool, however, applications shift from younger toward older suitors. The application-tracked decomposition locates the recurring female-receiver proposal loss mainly in the transition into early dating. Treating all active counterparts as the opportunity set overstates the stated-eligible pool by about one-third; downstream differences of similar size in CF1 follow from its fixed-rate scaling. In the stage-specific no-birthday accounting for receivers in their thirties, the proposal count is \(12.9\%\) above the benchmark in the female-receiver channel and \(6.7\%\) below it in the male-receiver channel, while the former's younger-man share is \(4.8\) percentage points higher. Age labels and filters therefore reshape observed opportunity sets, initiation-channel proposal counts, and accepted-pair composition.}

\bibliographystyle{apalike}
\setlength{\bibsep}{5pt}
\bibliography{ibj_project}

\newpage
\appendix
\numberwithin{figure}{section}
\numberwithin{table}{section}

\section*{Online Appendix}

\section{Additional Data Details}\label{app:data_details}

\subsection{IBJ Platform and Marriage Agency}\label{app:marriage_agency_platform}

\begin{figure}[!htbp]
\begin{center}
\begin{tikzpicture}[
  node distance=0.6cm and 0.6cm,
  every node/.style={draw, rectangle, rounded corners, minimum width=1.2cm, minimum height=1cm, align=center}
]

\node (ibj) {IBJ platform};

\node (agency1) [below left=of ibj] {Marriage agency};
\node (agency2) [below right=of ibj] {Marriage agency};

\node (user1a) [below left=of agency1] {User};
\node (user1b) [below=of agency1] {User};

\node (user2a) [below=of agency2] {User};
\node (user2b) [below right=of agency2] {User};

\draw[-] (ibj) -- (agency1);
\draw[-] (ibj) -- (agency2);

\draw[-] (agency1) -- (user1a);
\draw[-] (agency1) -- (user1b);

\draw[-] (agency2) -- (user2a);
\draw[-] (agency2) -- (user2b);

\end{tikzpicture}
\end{center}
\caption{Structure of the IBJ Platform: Agencies and Users}
\label{fig:ibj_structure}
\end{figure}
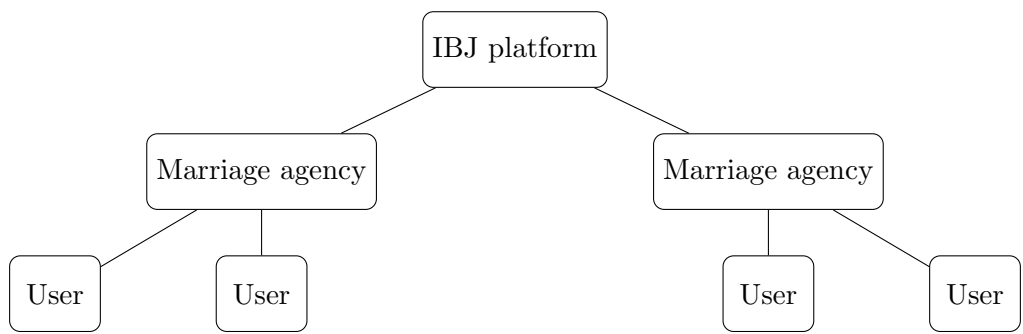

Figure~\ref{fig:ibj_structure} illustrates the matching environment operated by the IBJ platform. The main players are the IBJ platform operator, affiliated marriage agencies, and users seeking partners.

\paragraph*{Platform}

The IBJ platform provides affiliated marriage agencies with a nationwide matchmaking infrastructure. As of June 2026, the network included approximately 4,800 marriage agencies and more than 110,000 active registered members. It operates a shared database of prospective marriage partners and a proprietary arranged-meeting system that links agencies and clients. Through this system, counselors can search member profiles, communicate across agencies, and manage client interactions in real time. IBJ also uses algorithmic match suggestions and facilitates inter-agency coordination so that counselors can guide clients from initial introduction to engagement. The platform is designed for marriage-oriented matching rather than casual dating.

\paragraph*{Marriage Agency}
To join the IBJ platform, an independent marriage consulting agency must pass IBJ's admission process and agree to its operating standards. New agencies sign a contract and pay a one-time membership fee to the IBJ federation. Agency owners also complete IBJ's initial training program, which covers platform rules, matchmaking procedures, and operational practices. Once on the platform, member agencies gain access to IBJ's nationwide pool of vetted singles and its matchmaking system, as well as ongoing support such as marketing tools, professional development seminars, and inter-agency communication channels. Affiliation with IBJ therefore allows small agencies to operate within a standardized national network.

\paragraph*{User}

Individuals seeking a spouse join one of the local marriage agencies in the IBJ network, through which they become part of IBJ's matchmaking system. Enrollment involves an orientation or interview with a counselor and the submission of documents verifying identity and eligibility for marriage, including government-issued identification, proof of single status, residence, income, education, and relevant professional qualifications. After screening, the individual signs a membership agreement and pays the agency's fees. These typically include an initial registration or enrollment fee, monthly membership dues, and a success fee payable upon engagement, although fee levels vary by agency. Once affiliated with an IBJ agency, the client gains access to the IBJ network and works with a marriage counselor. The counselor helps prepare the profile, discusses the client's values and preferences, uses the IBJ system to filter and recommend suitable matches, and coordinates with other agencies when arranging official introductions. After an omiai meeting, the counselor continues to mediate and advise the client through trial dating, serious relationship formation, and formal engagement.


\subsection{Platform Volume and Coverage}\label{app:platform_volume_coverage}

\begin{figure}[!htbp]
  \begin{center}
  \subfloat[Active users and market tightness]{\includegraphics[width = 0.33\textwidth]{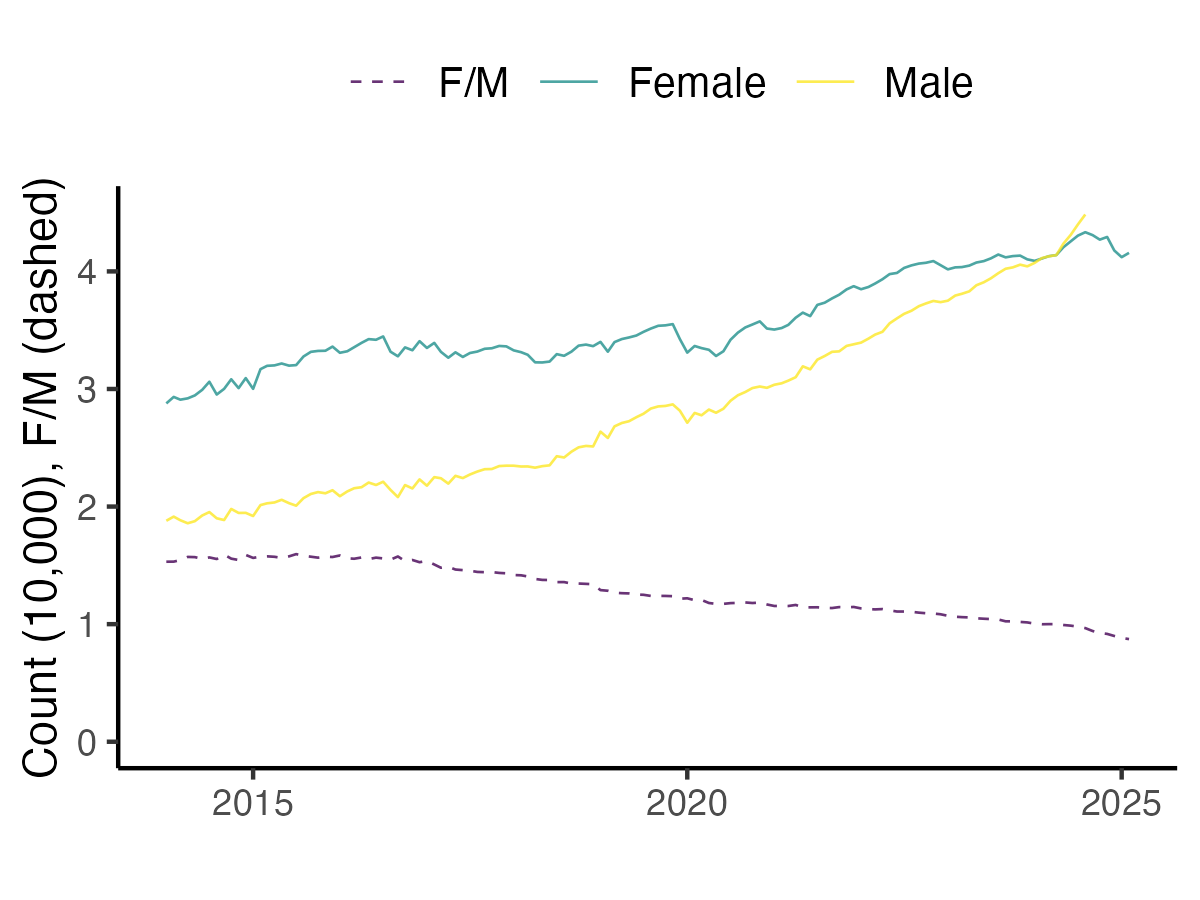}}
  \subfloat[Engagements]{\includegraphics[width = 0.33\textwidth]{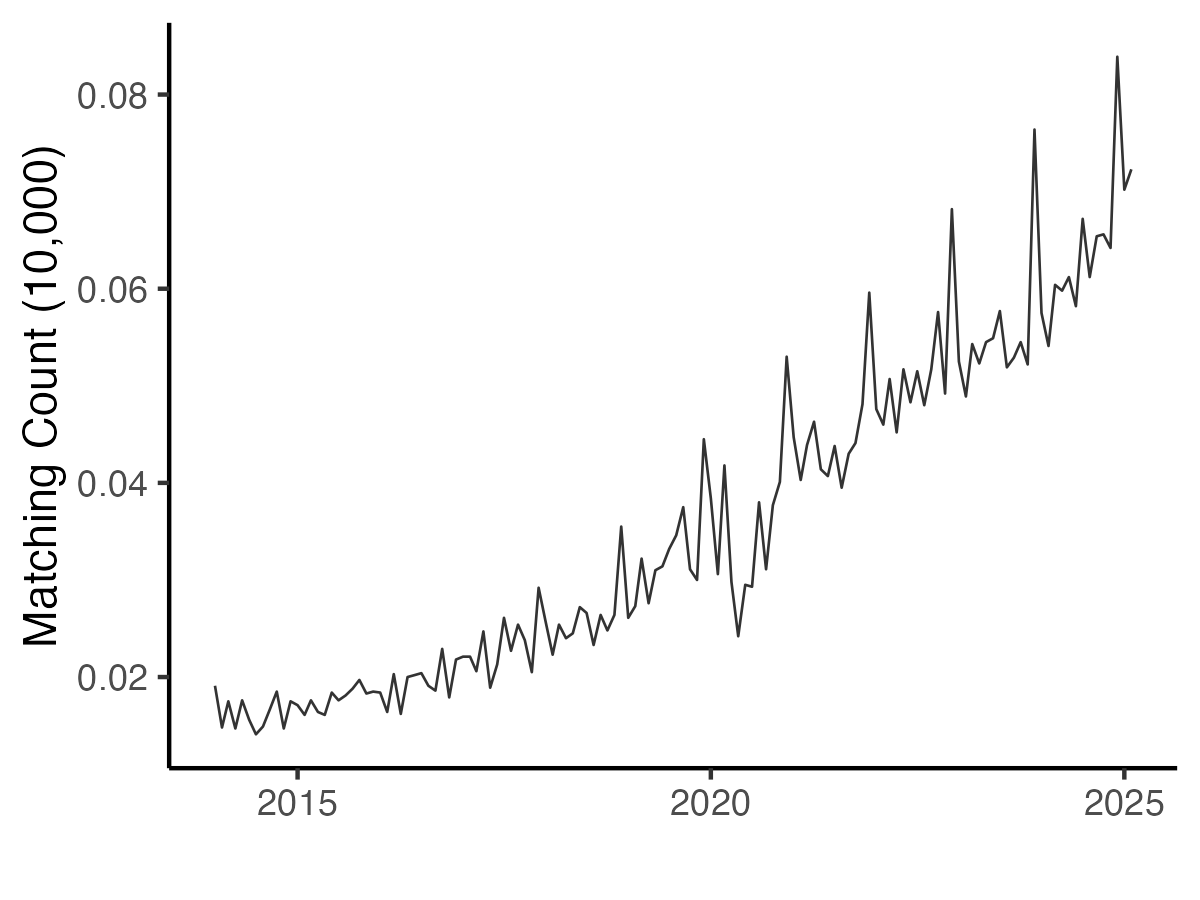}}
  \subfloat[Partner-finding rates]{\includegraphics[width = 0.33\textwidth]{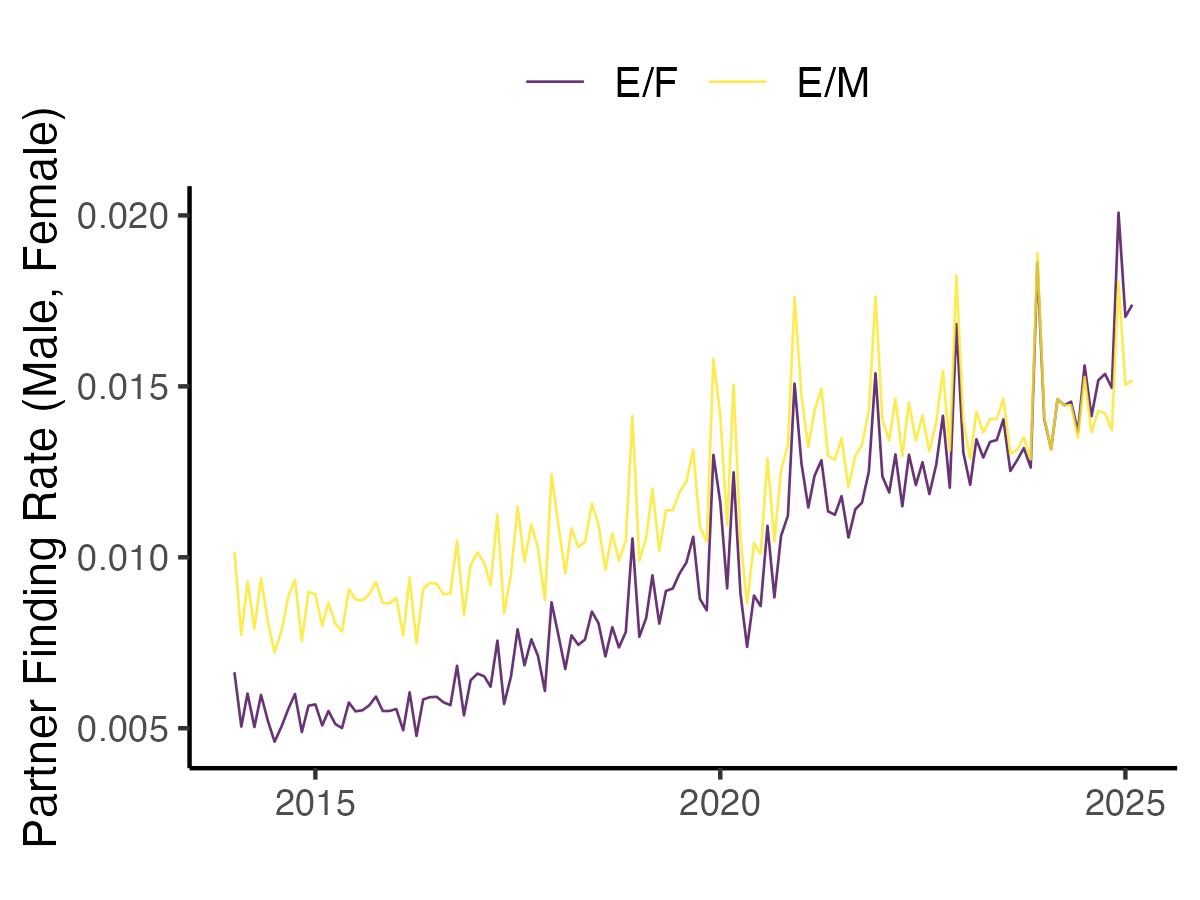}}
  \caption{Platform Volume, Market Tightness, and Engagements, 2014--2025}
  \label{fg:platform_volume}
  \end{center}
  \footnotesize
  Note: Panel (a) plots active female users, active male users, and market tightness measured by the female-to-male ratio. Panel (b) plots confirmed engagements. Panel (c) plots engagement counts relative to the number of active women and active men. See \citet{otani2025nonparametric}.
\end{figure}

Figure~\ref{fg:platform_volume} summarizes the scale of the IBJ platform over time. Active participation expanded steadily from 2014 to 2025, with faster growth on the male side after 2020 and a corresponding decline in the female-to-male ratio. Within-platform engagements in the confidential analytical extract also rose sharply, totaling 7,587 in 2024. Across its services, IBJ reported 16,398 engagements that year, about 3.4\% of Japanese marriages. Thus, although IBJ is a selected, high-verification platform, it represents a sizable segment of contemporary partner search \citep{otani2025nonparametric}.

\begin{figure}[!htbp]
  \begin{center}
  \subfloat[Age (Proportion)]{\includegraphics[width = 0.9\textwidth, height = 0.28\textheight, keepaspectratio]{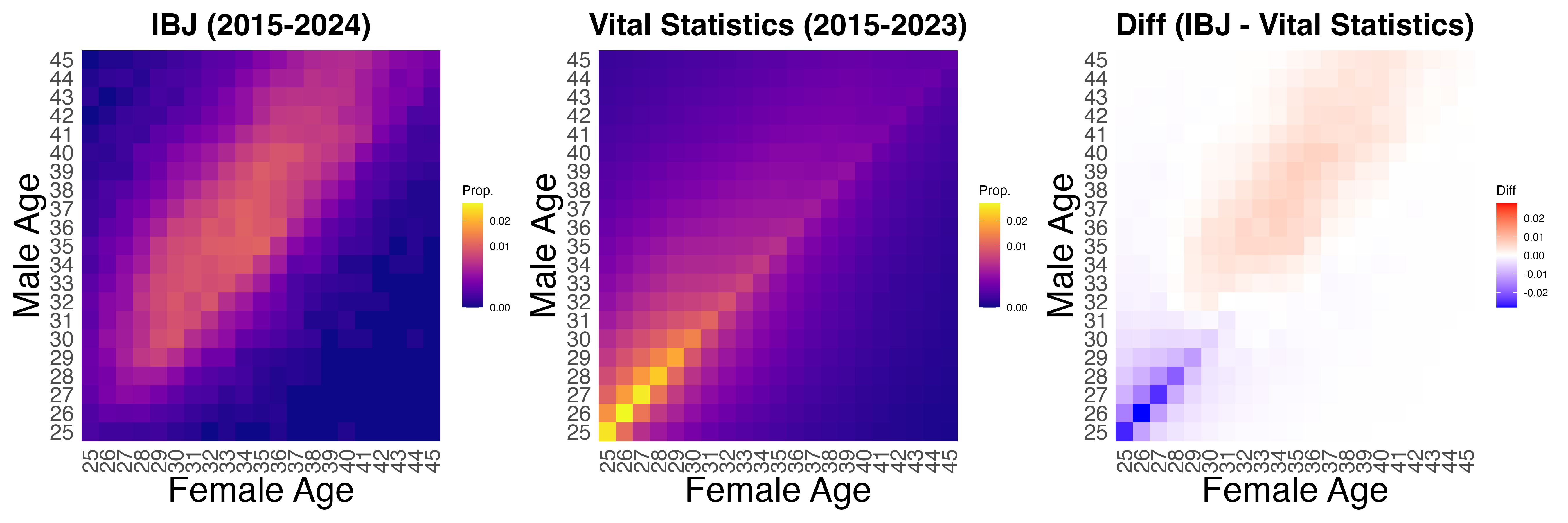}}\\[0.2in]
  \subfloat[Education (Proportion)]{\includegraphics[width = 0.9\textwidth, height = 0.28\textheight, keepaspectratio]{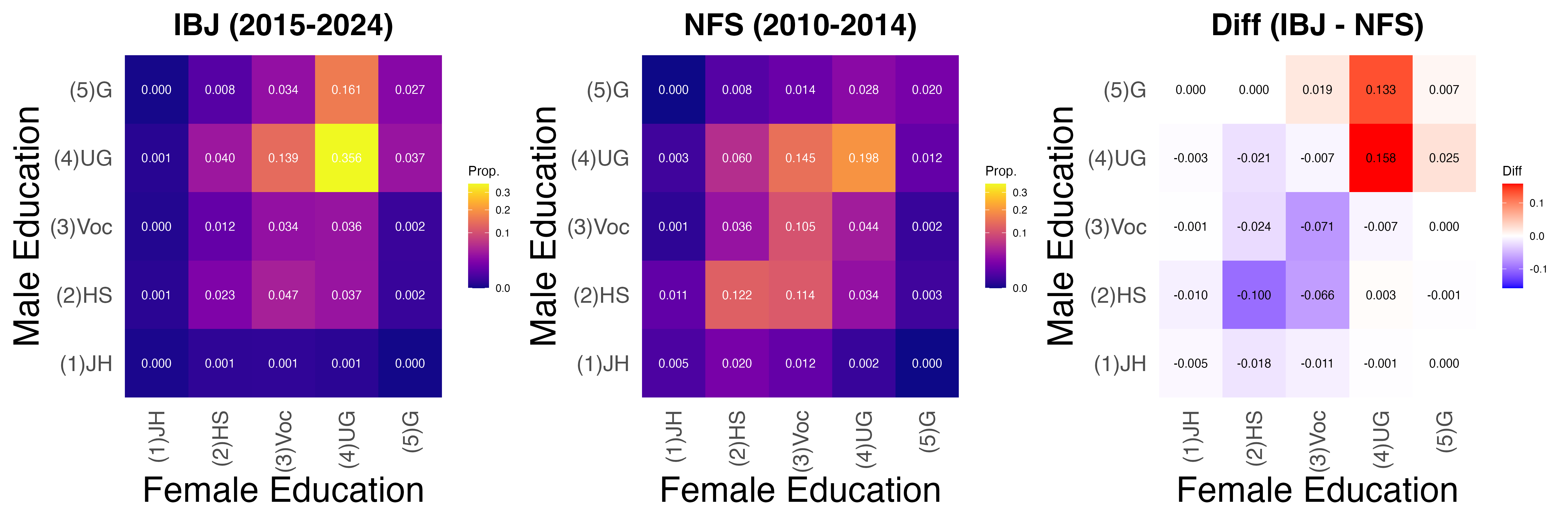}}\\[0.2in]
  \subfloat[Income (Proportion)]{\includegraphics[width = 0.9\textwidth, height = 0.28\textheight, keepaspectratio]{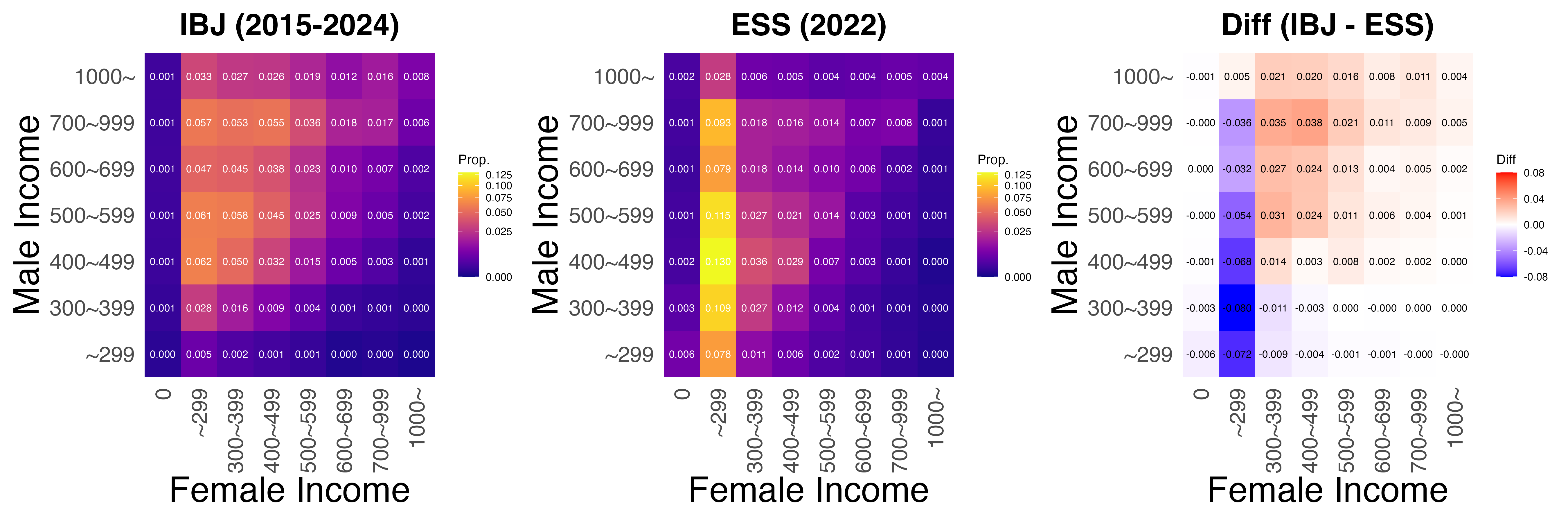}}
  \caption{IBJ Market Coverage Compared with National Statistics: Joint Proportions}
  \label{fg:coverage_prop}
  \end{center}
  \footnotesize
  Note: The joint proportion is $P(i,j)=n_{ij}/N$, where $n_{ij}$ is the count in cell $(i,j)$ and $N$ is the total count. The age comparison uses IBJ data for 2015--2024 and Vital Statistics for 2015--2023; education uses the 15th National Fertility Survey for the 2010--2014 marriage cohort; income uses the 2022 Employment Status Survey for husbands aged 49 or younger. See \citet{inoue2026marital} for definitions and zero-income coding.
\end{figure}

\begin{figure}[!htbp]
  \begin{center}
  \subfloat[Age (Likelihood Ratio)]{\includegraphics[width = 0.9\textwidth, height = 0.28\textheight, keepaspectratio]{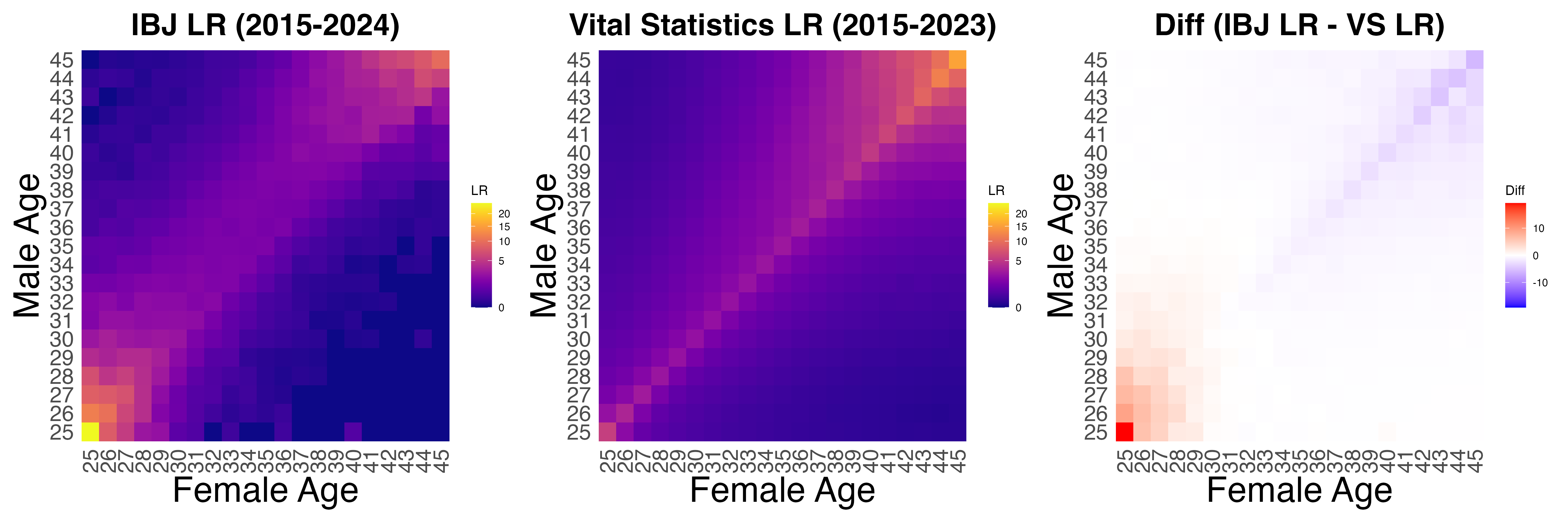}}\\[0.2in]
  \subfloat[Education (Likelihood Ratio)]{\includegraphics[width = 0.9\textwidth, height = 0.28\textheight, keepaspectratio]{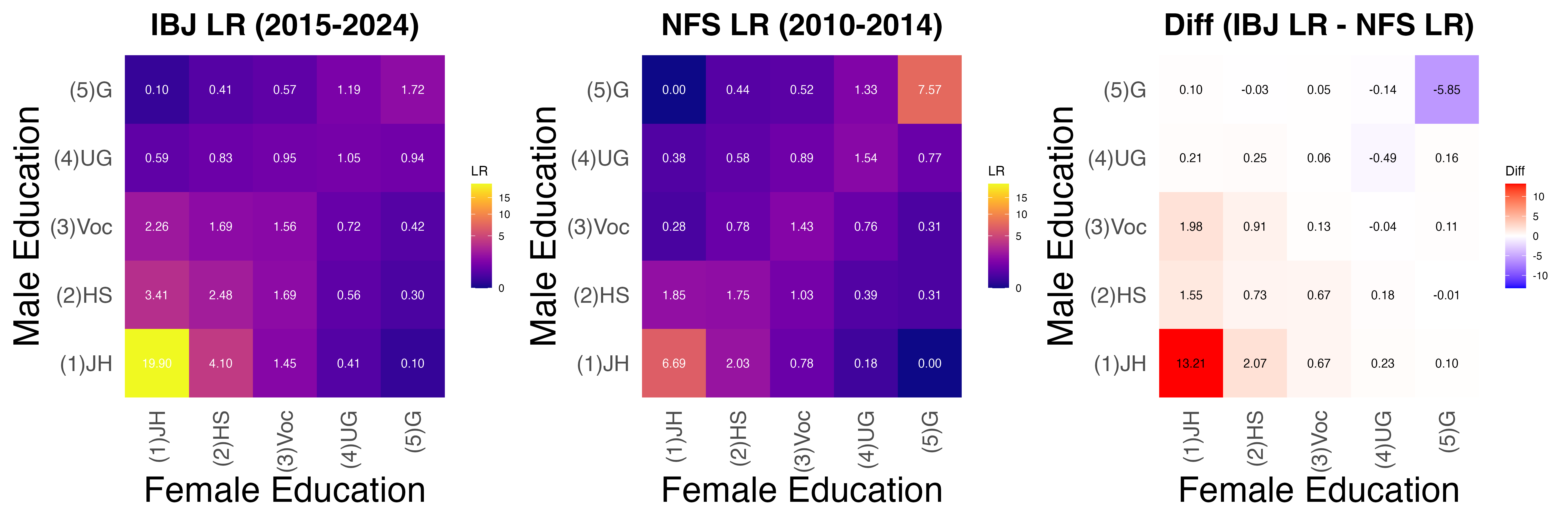}}\\[0.2in]
  \subfloat[Income (Likelihood Ratio)]{\includegraphics[width = 0.9\textwidth, height = 0.28\textheight, keepaspectratio]{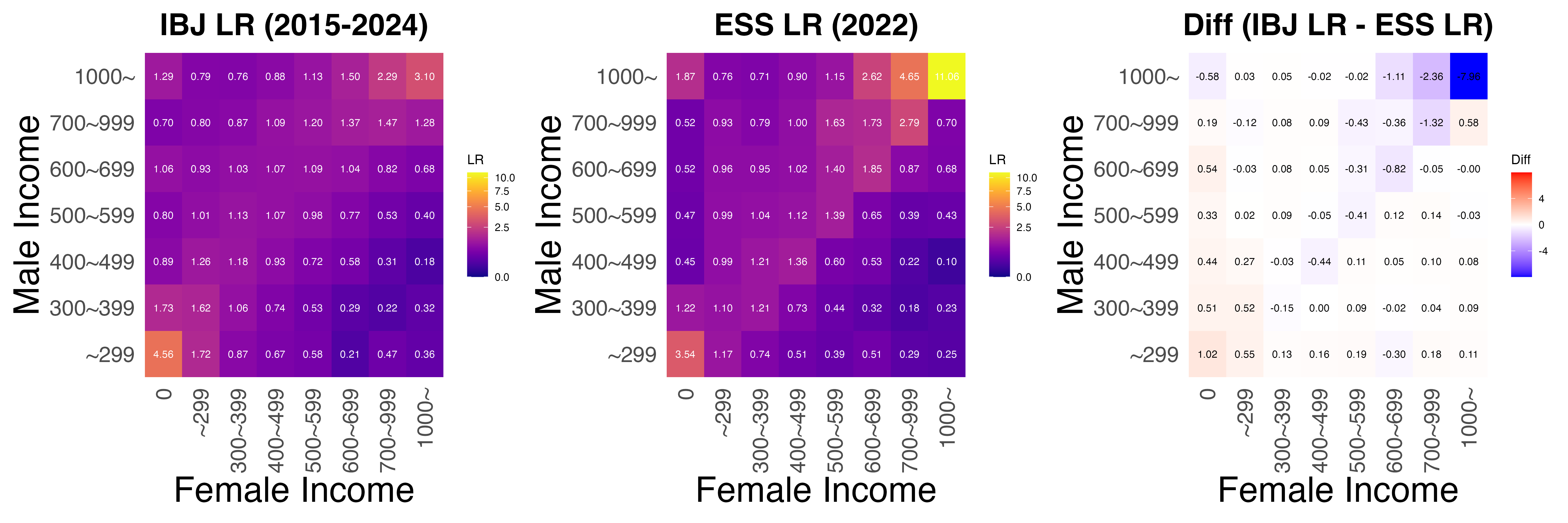}}
  \caption{IBJ Market Coverage Compared with National Statistics: Likelihood Ratios}
  \label{fg:coverage_lr}
  \end{center}
  \footnotesize
  Note: The likelihood ratio $\mathrm{LR}_{ij}=P(i,j)/[P_{\text{male}}(i)\,P_{\text{female}}(j)]$ divides the joint proportion by the male and female marginal proportions. A value above (below) one indicates over- (under-) representation relative to random matching. See \citet{inoue2026marital}.
\end{figure}

This appendix compares the joint distribution of matched IBJ couples with national benchmarks; \citet{inoue2026marital} provides the full analysis. Figure~\ref{fg:coverage_prop} reports raw joint proportions and Figure~\ref{fg:coverage_lr} reports likelihood ratios, which net out marginal-distribution differences to isolate assortative-matching strength. The proportion panels show that IBJ engagements are concentrated among older and more educated couples than the national benchmarks, while the likelihood-ratio panels show that the diagonal (positive assortative) structure for age and education is broadly comparable across IBJ and representative data once composition is controlled for. The income panels additionally reflect that IBJ records pre-marital income whereas the Employment Status Survey records post-marital income, \textcolor{black}{generating the large mass of wives in the lowest income category in the latter}.

\subsection{Raw Data Pattern}\label{app:raw_data_pattern}

\begin{figure}[!htbp]
  \begin{center}
  \includegraphics[width = 0.95\textwidth]{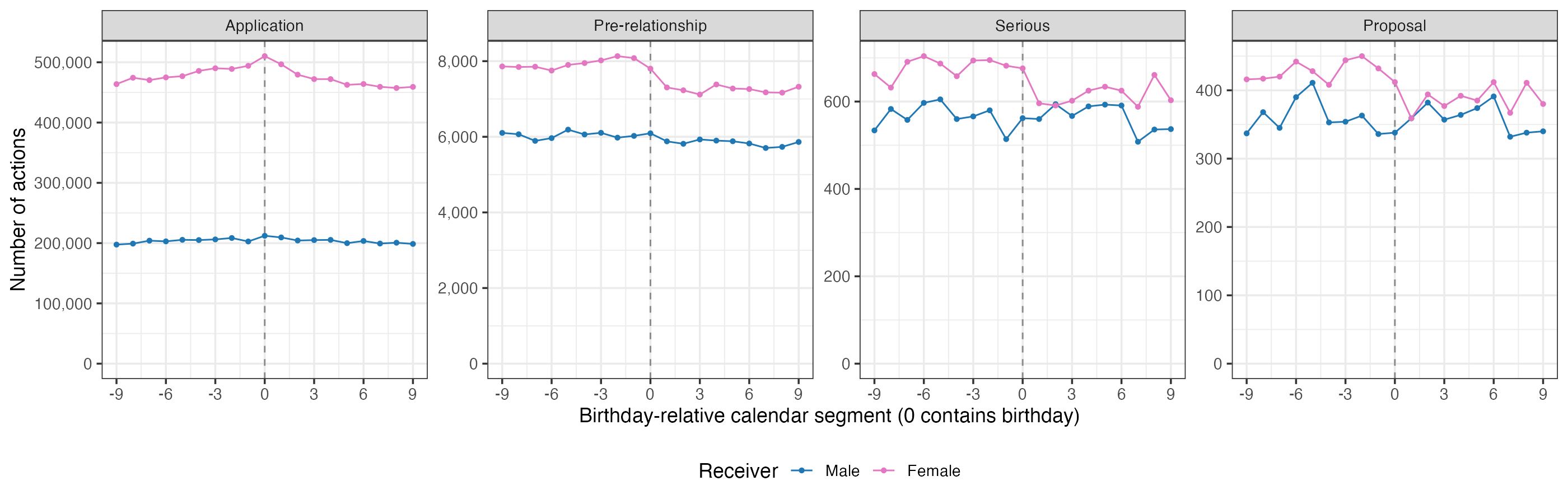}
  \caption{Raw Received Action Counts by Birthday-Relative Segment}
  \label{fg:birthday_segment_action_counts}
  \end{center}
  \footnotesize
  Note: Raw counts of received actions by receiver gender, stage, and birthday-relative calendar segment. Segment 0 contains the receiver's birthday.
\end{figure}

Figure~\ref{fg:birthday_segment_action_counts} plots raw received-action totals by birthday-relative calendar segment and receiver gender. Female application counts rise into the birthday-containing segment and decline afterward; male application counts are flatter. Lower-frequency outcomes are noisier. These unadjusted totals motivate the within-member analysis, which accounts for member, calendar-year, within-month-position, and platform-tenure effects.

\subsection{Calendar-Month Data Pattern}\label{app:calendar_month_pattern}

\begin{figure}[!htbp]
  \begin{center}
  \includegraphics[width = 0.95\textwidth]{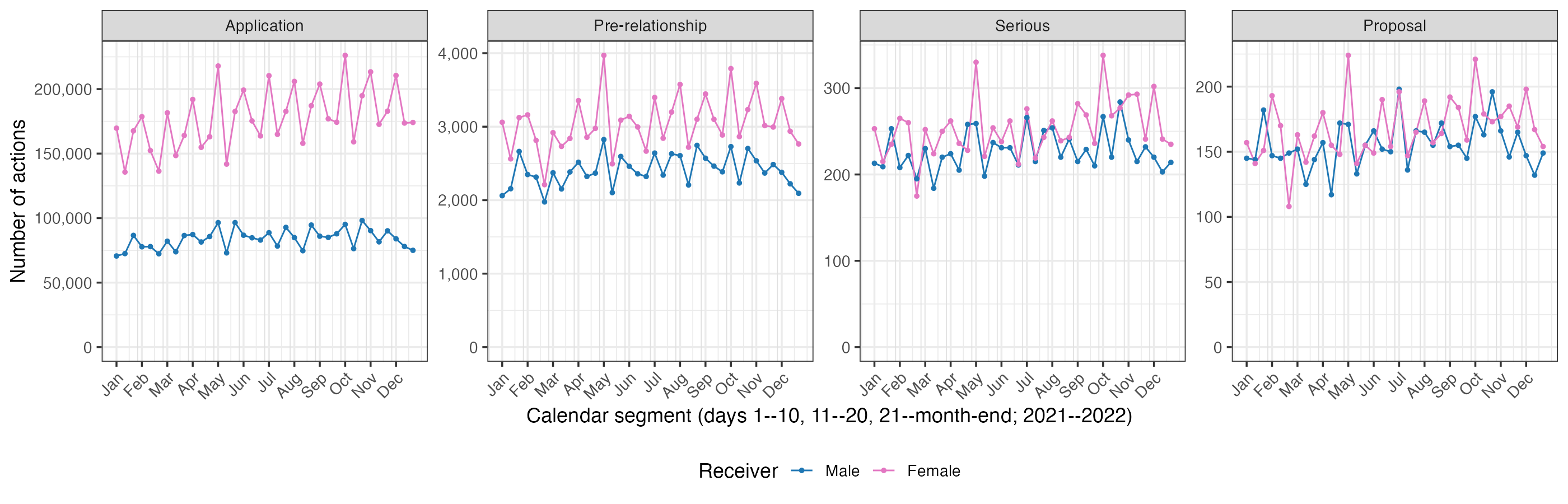}
  \caption{Raw Received Action Counts by Calendar Segment}
  \label{fg:calendar_segment_action_counts}
  \end{center}
  \footnotesize
  Note: The figure reports raw received-action counts by receiver gender, stage, and calendar segment. Each month is divided into days 1--10, days 11--20, and days 21--month-end. \textcolor{black}{Initiating applications are restricted to 2021--2022, so all stages have at least 420 days of follow-up through December 31, 2024.}
\end{figure}

Figure~\ref{fg:calendar_segment_action_counts} plots the same action counts by calendar segment rather than birthday-relative segment. \textcolor{black}{To give all four stages at least 420 days of follow-up, it restricts initiating applications to 2021--2022.} Applications and pre-relationships are typically high in days 1--10, lower in days 11--20, and higher again in the final segment. Under some agency plans, members receive a monthly allotment of meeting requests and may use remaining capacity near month-end. The first two bins each contain 10 days, whereas the final bin contains 8--11 days; because the raw counts are not normalized by bin length, the final-bin level may partly reflect this difference. The main specification absorbs the observed within-month pattern without relying on a behavioral interpretation.

\begin{figure}[!htbp]
  \begin{center}
  \includegraphics[width = 0.90\textwidth]{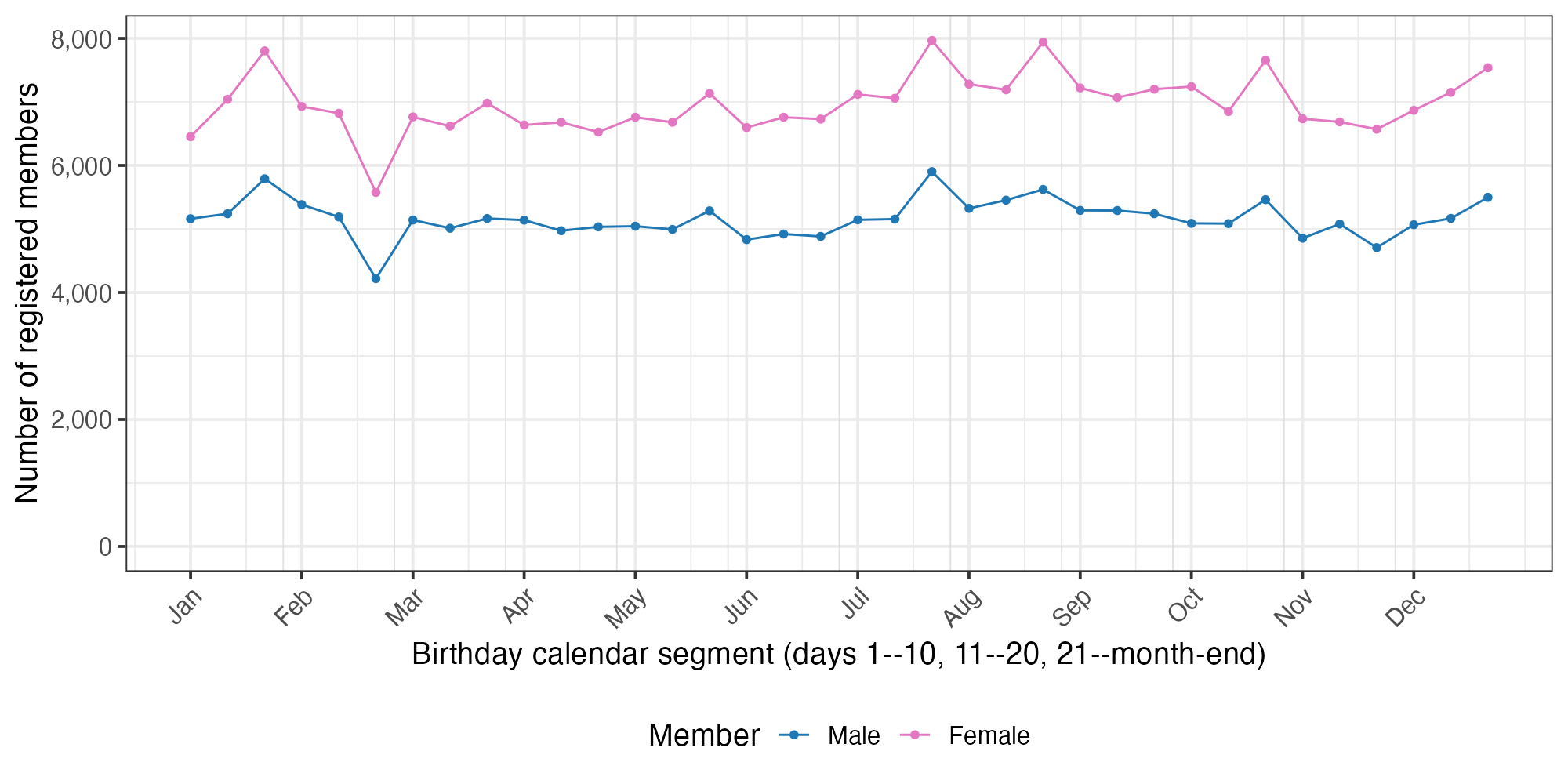}
  \caption{Registered Members by Birthday Calendar Segment}
  \label{fg:calendar_segment_registered_users}
  \end{center}
  \footnotesize
  Note: The figure reports registered members by gender and birthday calendar segment. Each month is divided into days 1--10, days 11--20, and days 21--month-end.
\end{figure}

Figure~\ref{fg:calendar_segment_registered_users} provides a complementary diagnostic on birthday timing. Birth segments are broadly distributed over the calendar year, with no sharp concentration in particular calendar segments. This dispersion reduces concern that a few calendar cells drive the birthday-relative estimates, but it does not by itself establish orthogonality. The main specification controls directly for within-month position, and Appendix~\ref{app:robust_fe_choice} adds month-of-year and specific year-month fixed effects.

\clearpage
\section{Theoretical Model Details}\label{app:theoretical_model}

\subsection{An Illustrative Model for Interpreting the Results}\label{app:illustrative_model}

This appendix presents an illustrative model for interpreting the top-of-funnel results. Although applications are measured as a receiver-side outcome, the proposer is the decision maker; I therefore model applications as proposer-side demand for receiver profiles. The model combines an age-filter-based consideration set with logit errors in proposer utility. Its utility index follows the profile-contact formulation used in online dating by \citet{hitsch2010makes}. The model organizes the empirical margins and ends at the application stage; it is not structurally estimated.

\subsubsection{Consideration-Set Formation}

Let $\mathcal I_t$ and $\mathcal J_t$ be the active proposer and receiver profiles available at time $t$, respectively. Each proposer $i\in\mathcal I_t$ has a stated acceptable partner-age range $[L_i,U_i]$. Proposer $i$'s filter-implied consideration set is
\begin{equation}
    \mathcal C_{it}
    =\left\{j\in\mathcal J_t:L_i\leq A_{jt}\leq U_i\right\},
    \label{eq:model_consideration_set}
\end{equation}
where $A_{jt}$ is receiver $j$'s displayed age. The empirical construction treats the single observed stated range as fixed over the birthday window. Other profile attributes and search restrictions are held fixed and suppressed from the notation. When proposer $i$ applies these stated bounds in condition search, a profile outside $\mathcal C_{it}$ cannot be selected through that search, while a profile inside the set competes with the other considered profiles and an outside option of sending no application.

Equation~\eqref{eq:model_consideration_set} is an empirical, filter-implied set rather than a claim that every eligible profile is viewed. Historical query-level filters, profile rankings, and exposure are not observed over the event-study period. Empirically, STAY-OUT applications capture other browsing routes and differences between stated and realized filters. The model abstracts from these distinctions; Section~\ref{subsec:actual_filters} instead documents how stated bounds compare with saved search conditions.

\subsubsection{Choice within the Consideration Set}

On each application opportunity, proposer $i$ chooses one receiver from $\mathcal C_{it}$ or the outside option. The deterministic utility index for receiver $j$ is
\begin{equation}
    v_{ijt}
    =\alpha_j+X_j'\beta_{g(i)}
    +h_{g(i)}\!\left(A_{jt}-A_{it}\right),
    \label{eq:model_application_utility}
\end{equation}
where $\alpha_j$ is a receiver-specific utility component, $X_j$ contains receiver attributes other than age, $\beta_{g(i)}$ is a proposer-gender-specific coefficient vector, $A_{it}$ is proposer $i$'s displayed age, $g(i)$ denotes proposer gender, and $h_{g(i)}(\cdot)$ captures preferences over the receiver--proposer age difference. With i.i.d.\ type-I extreme-value utility shocks and the outside-option utility normalized to zero, the probability that proposer $i$ selects receiver $j$ is
\begin{equation}
    P_{ijt}
    =
    \begin{cases}
    \displaystyle
    \frac{\exp(v_{ijt})}
    {1+\sum_{\ell\in\mathcal C_{it}}\exp(v_{i\ell t})},
    & j\in\mathcal C_{it},\\[1.2em]
    0, & j\notin\mathcal C_{it}.
    \end{cases}
    \label{eq:model_application_probability}
\end{equation}
Members can submit multiple applications. I represent this as repeated choice opportunities and let $M_{it}$ denote proposer $i$'s number of opportunities at time $t$ and $Y_{jt}$ the resulting number of applications received by receiver $j$. Conditional on $M_{it}$, expected applications are
\begin{equation}
    \mathbb E[Y_{jt}]
    =\sum_{i\in\mathcal I_t} M_{it}P_{ijt}.
    \label{eq:model_expected_applications}
\end{equation}
The model conditions on search intensity. The comparative statics hold $M_{it}$ and non-age profile characteristics fixed.

\subsubsection{A One-Year Change in Displayed Age}

Consider receiver $j$ turning age $a$, so displayed age changes from $A_j^-=a-1$ to $A_j^+=a$. Partition active proposers into
\begin{align*}
    \mathcal E_j(a)
    &:=\left\{i\in\mathcal I_t:L_i=a,\ U_i\geq a\right\},
    &&\text{(ENTER)},\\
    \mathcal X_j(a-1)
    &:=\left\{i\in\mathcal I_t:L_i\leq a-1,\ U_i=a-1\right\},
    &&\text{(EXIT)},\\
    \mathcal S_j(a)
    &:=\left\{i\in\mathcal I_t:L_i\leq a-1,\ U_i\geq a\right\},
    &&\text{(STAY-IN)},\\
    \mathcal O_j(a)
    &:=\left\{i\in\mathcal I_t:a-1\notin[L_i,U_i],\ a\notin[L_i,U_i]\right\},
    &&\text{(STAY-OUT)}.
\end{align*}
These four sets are exhaustive. The birthday adds the receiver to the consideration sets of ENTER proposers, removes the receiver from those of EXIT proposers, and leaves membership unchanged for STAY-IN and STAY-OUT proposers. Under equation~\eqref{eq:model_application_probability}, STAY-OUT proposers have zero application probability on both sides. Let $N_j^-$ and $N_j^+$ denote the sizes of receiver $j$'s filter-implied potential-proposer pool before and after the birthday. Their change is the identity
\begin{equation}
    N_j^+-N_j^-
    =\left|\mathcal E_j(a)\right|-\left|\mathcal X_j(a-1)\right|.
    \label{eq:model_pool_decomposition}
\end{equation}
The net pool change can therefore be small even when both boundary flows are large.

Suppressing time subscripts throughout, let $P_{ij}^-$ and $P_{ij}^+$ denote the choice probabilities evaluated just before and after the displayed-age update, and let $M_i$ be proposer $i$'s fixed number of choice opportunities over this comparison. The corresponding application effect decomposes exactly as
\begin{align}
    \Delta\mathbb E[Y_j]
    ={}&
    \underbrace{
    \sum_{i\in\mathcal E_j(a)}M_iP_{ij}^+
    -\sum_{i\in\mathcal X_j(a-1)}M_iP_{ij}^-
    }_{\text{consideration-set effect}}
    \nonumber\\
    &+\underbrace{
    \sum_{i\in\mathcal S_j(a)}M_i
    \left(P_{ij}^+-P_{ij}^-\right)
    }_{\text{choice-within-set effect}}.
    \label{eq:model_application_decomposition}
\end{align}
The first term gives the mechanically signed ENTER and EXIT components. The second term captures the response among proposers for whom the receiver remains eligible on both sides of the birthday. This is the model counterpart of the empirical eligibility decomposition.

\subsubsection{Age-Difference Comparative Statics}

A transparent special case of equation~\eqref{eq:model_application_utility}, consistent with the asymmetric age-difference specification in \citet{hitsch2010matching}, is
\begin{equation}
    h_g(d)
    =-\kappa_g^+[d]_+^2-\kappa_g^-[-d]_+^2,
    \qquad \kappa_g^+,\kappa_g^->0,
    \label{eq:model_age_utility}
\end{equation}
where $d=A_j-A_i$ and $[x]_+:=\max\{x,0\}$. The two coefficients allow the utility cost of an older receiver to differ from that of a younger receiver.

For a STAY-IN proposer, hold the other profiles in the choice set fixed and define
\[
    D_{i,-j}
    :=1+\sum_{\ell\in\mathcal C_i,\,\ell\neq j}\exp(v_{i\ell}).
\]
Then the one-year change in receiver $j$'s choice probability is
\begin{equation}
    P_{ij}^+-P_{ij}^-
    =
    \frac{D_{i,-j}\left[\exp(v_{ij}^+)-\exp(v_{ij}^-)\right]}
    {\left[D_{i,-j}+\exp(v_{ij}^+)\right]
     \left[D_{i,-j}+\exp(v_{ij}^-)\right]}.
    \label{eq:model_probability_change}
\end{equation}
Its sign is therefore the sign of $v_{ij}^+-v_{ij}^-$. Classify proposer age relative to receiver $j$'s turning age $a$, as in the empirical analysis. If proposer $i$ is $r\geq1$ years older than $a$, equation~\eqref{eq:model_age_utility} gives
\begin{equation}
    v_{ij}^+-v_{ij}^-
    =\kappa_{g(i)}^-(2r+1)>0.
    \label{eq:model_older_proposer}
\end{equation}
If proposer $i$ is $r\geq1$ years younger than $a$, it gives
\begin{equation}
    v_{ij}^+-v_{ij}^-
    =-\kappa_{g(i)}^+(2r-1)<0.
    \label{eq:model_younger_proposer}
\end{equation}
Thus, among STAY-IN proposers, the displayed-age update raises applications from strictly older proposers and lowers applications from strictly younger proposers. These are the main empirical re-sorting margins; the same-age boundary cell is not used to discipline this illustrative specification. Together with offsetting ENTER and EXIT flows, the model explains how total applications can change little while their age composition shifts. Because STAY-IN uses broad stated bounds, the empirical response can also reflect narrower filters, age-based ranking or exposure, and label salience; the model does not separate these channels.

\section{Additional Results}\label{app:additional_results}

\subsection{Support by Receiver Age, Gender, and Funnel Stage}\label{app:event_time_support}

Table~\ref{tb:event_time_support} reports distinct receivers with at least one application reaching each stage, by turning age and gender. Applications and pre-relationships involve thousands of receivers at central ages, although the youngest male cells are smaller. Serious relationships and proposals typically involve only a few hundred receivers per age, motivating the aggregate pre--post estimator.

\begin{table}[!htbp]
  \caption{Receivers Reaching Each Funnel Stage by Age and Gender}
  \label{tb:event_time_support}
  \begin{center}
  \small
  \input{figuretable/birthday_project/event_time_support_by_age_stage_gender.tex}
  \par\vspace{0.4em}
  \begin{minipage}{0.92\textwidth}
  \footnotesize
  Note: Each cell is the number of distinct receivers of the given turning age and gender with at least one application initiated in the main nine-segment window on either side of the birthday that reaches the indicated stage. The application and pre-relationship samples require 0 and 60 days of follow-up, respectively; serious relationships and proposals use the common 420-day follow-up cohort. Downstream stages are institutionally nested. A receiver may contribute at more than one turning age; the final row sums the age-specific cells.
  \end{minipage}
  \end{center}
\end{table}

\subsection{Event Study without Tenure Controls and the Tenure Profile}\label{app:tenure_diagnostic}

\begin{figure}[!htbp]
  \begin{center}
  \subfloat[Event-time path without tenure controls]{\includegraphics[width = 0.95\textwidth, height = 0.34\textheight, keepaspectratio]{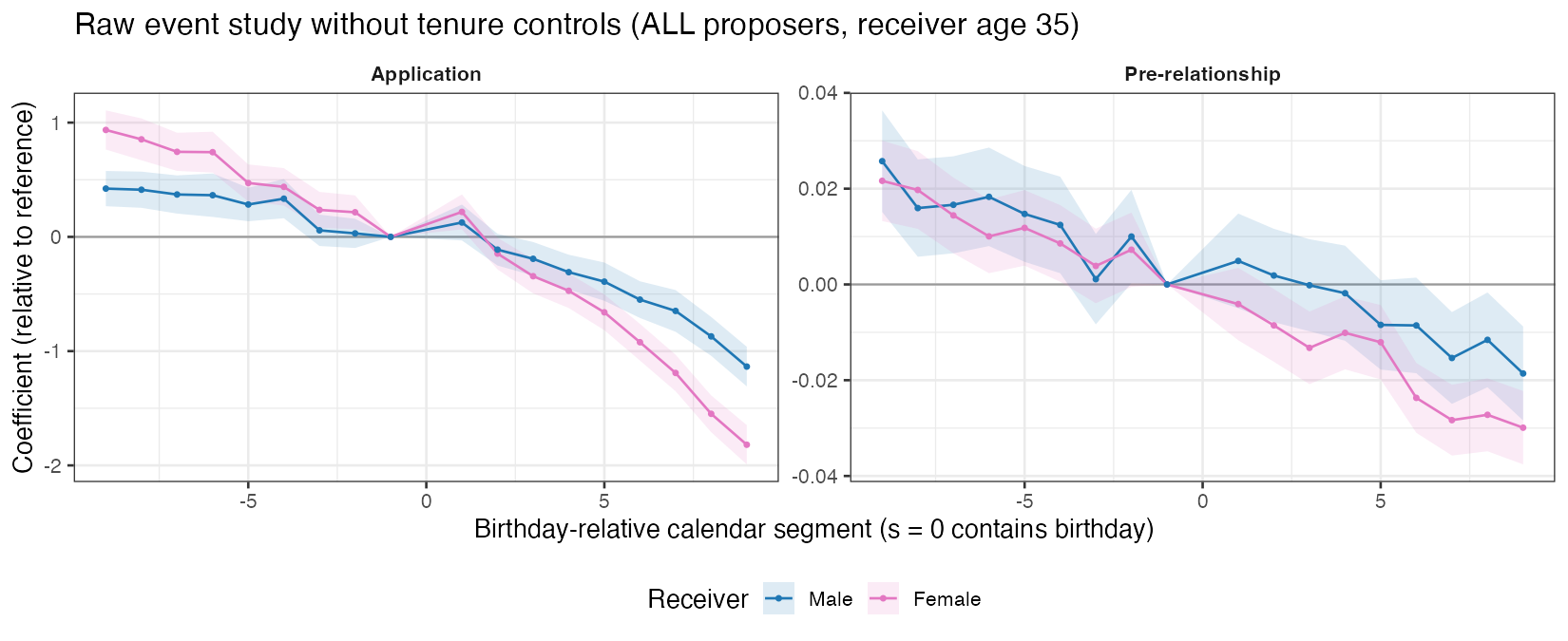}}\\[0.2in]
  \subfloat[Estimated tenure profile]{\includegraphics[width = 0.95\textwidth, height = 0.34\textheight, keepaspectratio]{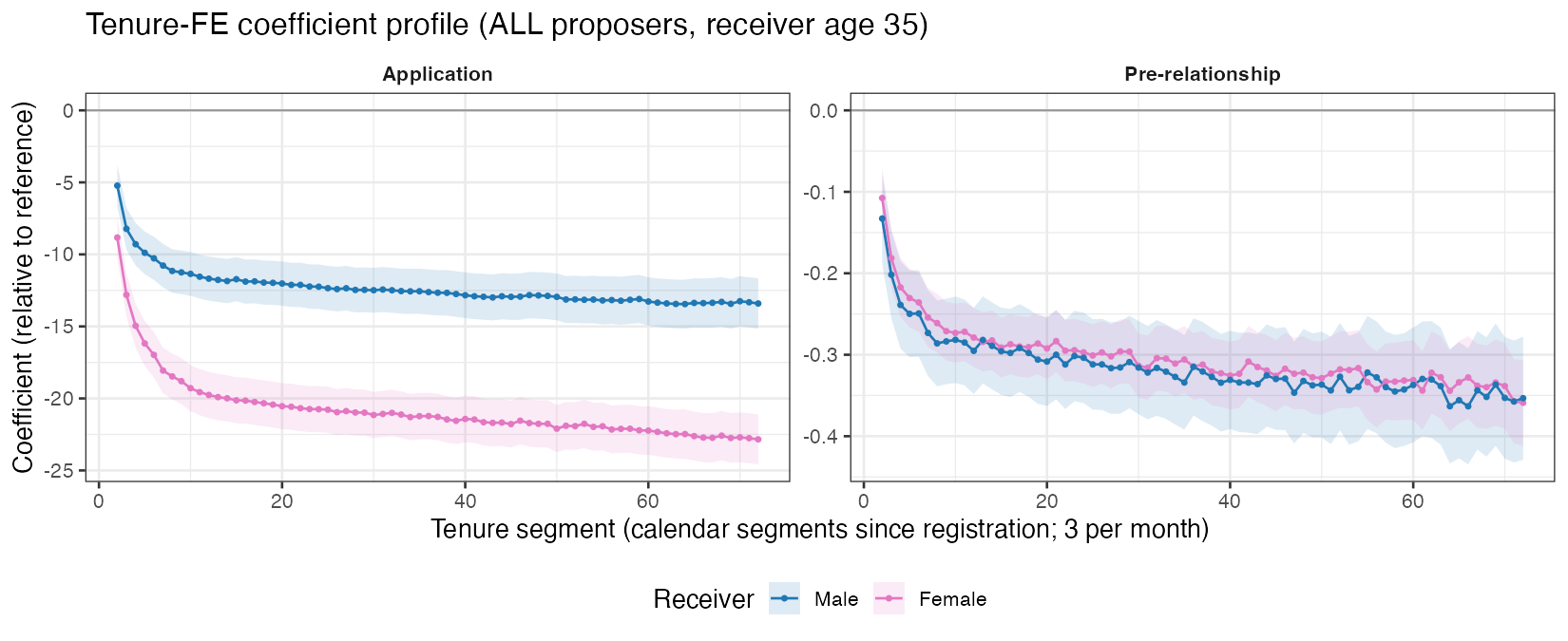}}
  \caption{Event Study without Tenure Controls and the Estimated Tenure Profile}
  \label{fg:tenure_fe_vs_eventstudy}
  \end{center}
  \footnotesize
  Note: All proposers, receiver age 35, application (left) and pre-relationship (right) stages. Panel (a) plots event-time coefficients after accounting for member, calendar-year, and within-month-position fixed effects but without tenure controls. Panel (b) plots tenure-segment coefficients from the preferred specification, capped at 72 segments because later coefficients are flat and imprecise. Bands are 95\% confidence intervals; coefficients are relative to each model's omitted level. The common downward pattern is consistent with contamination of the uncontrolled event-time path by the tenure decline.
\end{figure}

The within-member design must separate the birthday response from the steep decline in activity as platform tenure grows. Because event time and tenure move together within a birthday window, an event study without tenure controls can load the tenure decline onto event time. Figure~\ref{fg:tenure_fe_vs_eventstudy} illustrates this for all-proposer applications and pre-relationships at receiver age 35. Panel~(a), which omits tenure controls, declines smoothly across the cutoff rather than displaying a discrete post-birthday shift. Panel~(b) shows that the estimated tenure profile falls sharply at low tenure and then flattens. The panels use different horizontal axes and are not pointwise comparable, but their common downward pattern shows why the preferred specification absorbs the tenure profile before estimating the birthday jump.

\subsection{Tenure Distribution at the Birthday Segment}\label{app:tenure_dispersion}

Because the specification absorbs both a flexible tenure profile and a post-birthday indicator, it is worth checking that these two are separately identified. Within a member, tenure moves one-for-one with birthday-relative event time, so tenure at segment \(s\) equals the tenure at the birthday plus \(s\), whereas the post-birthday indicator depends on \(s\) alone. The two are therefore distinguished only by cross-member variation in tenure-at-birthday: if every member reached the birthday at nearly the same tenure, post-birthday observations of some members would never share a tenure cell with pre-birthday observations of others, and the post-birthday jump would be weakly identified relative to the tenure profile.

\begin{figure}[!htbp]
  \begin{center}
  \includegraphics[width = 0.85\textwidth, height = 0.42\textheight, keepaspectratio]{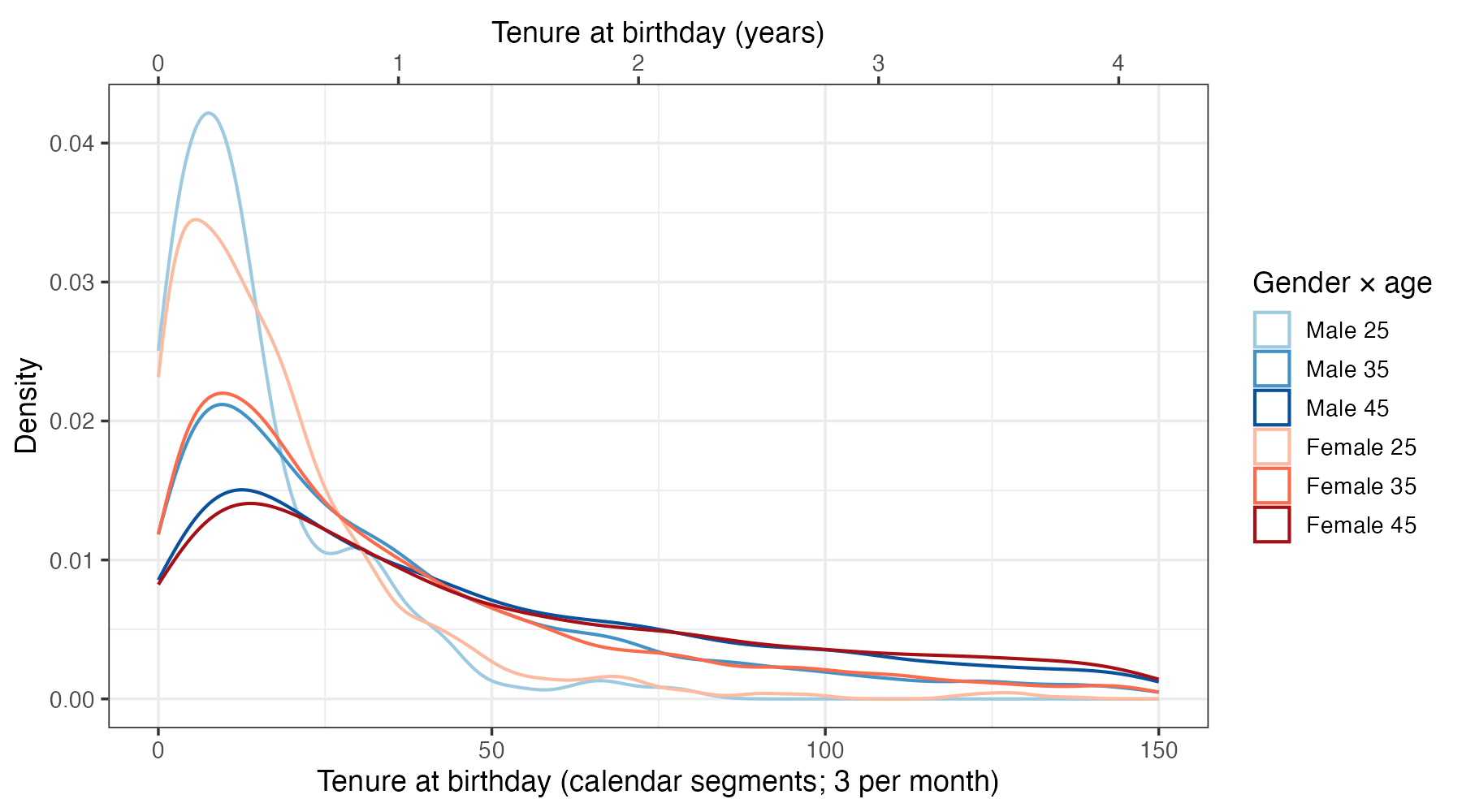}
  \caption{Tenure-at-Birthday Distribution by Receiver Age and Gender}
  \label{fg:tenure_at_birthday}
  \end{center}
  \footnotesize
  Note: Kernel densities of platform tenure at the birthday segment, measured from registration in calendar segments (three per month); the secondary axis shows years. Curves are shown by receiver age and gender at ages 25, 35, and 45.
\end{figure}

Figure~\ref{fg:tenure_at_birthday} plots tenure at the birthday, measured from registration in calendar segments, for representative receiver ages 25, 35, and 45 by gender. In the pooled sample, its standard deviation is about 63 segments, the distribution spans more than 380 distinct values, and about 6\% of observations lie within one segment of the mode. This dispersion is large relative to the \(\pm 9\)-segment event window and supplies the cross-member variation that separates the post-birthday indicator from the common tenure profile.

\subsection{Eligibility Decomposition for Male Receivers}\label{app:eligibility_male}

Table~\ref{tab:eventstudy_receiver_application_eligibility_E_male} reports the application-stage birthday jump by eligibility cell for male receivers, the counterpart to Table~\ref{tab:eventstudy_receiver_application_eligibility_E} in the main text. The eligibility margins are smaller and noisier than on the female side, consistent with age screens binding less tightly for male receivers.

\begin{table}[!htbp]
  \caption{Birthday Jump by Eligibility Type, Application Stage, Male Receivers}
  \label{tab:eventstudy_receiver_application_eligibility_E_male}
  \begin{center}
  \resizebox{\textwidth}{!}{\input{figuretable/birthday_project/eventstudy_receiver_application_eligibility_E_male.tex}}
  \par\vspace{0.4em}
  \begin{minipage}{0.95\textwidth}
  \footnotesize
  Note: Each cell reports the within-member application-stage birthday jump by receiver age and eligibility cell for male receivers, with standard errors in parentheses. STAYIN denotes STAY-IN proposers, who are eligible on both sides of the birthday cutoff; the two STAY-OUT columns are proposers ineligible on both sides, split into those already aged out (stated upper bound below the pre-birthday age) and those not yet reached (stated lower bound above the post-birthday age). The five mutually exclusive cells are additive, so EXIT, ENTER, STAYIN, and the two STAY-OUT columns sum to ALL, the all-proposer column, in each row. NA denotes a cell with no observations.
  \end{minipage}
  \end{center}
\end{table}

\subsection{Pre-relationship-Stage Mechanism Preview}\label{app:prerelation_mechanism}

\begin{figure}[!htbp]
  \begin{center}
  \includegraphics[width = 0.98\textwidth]{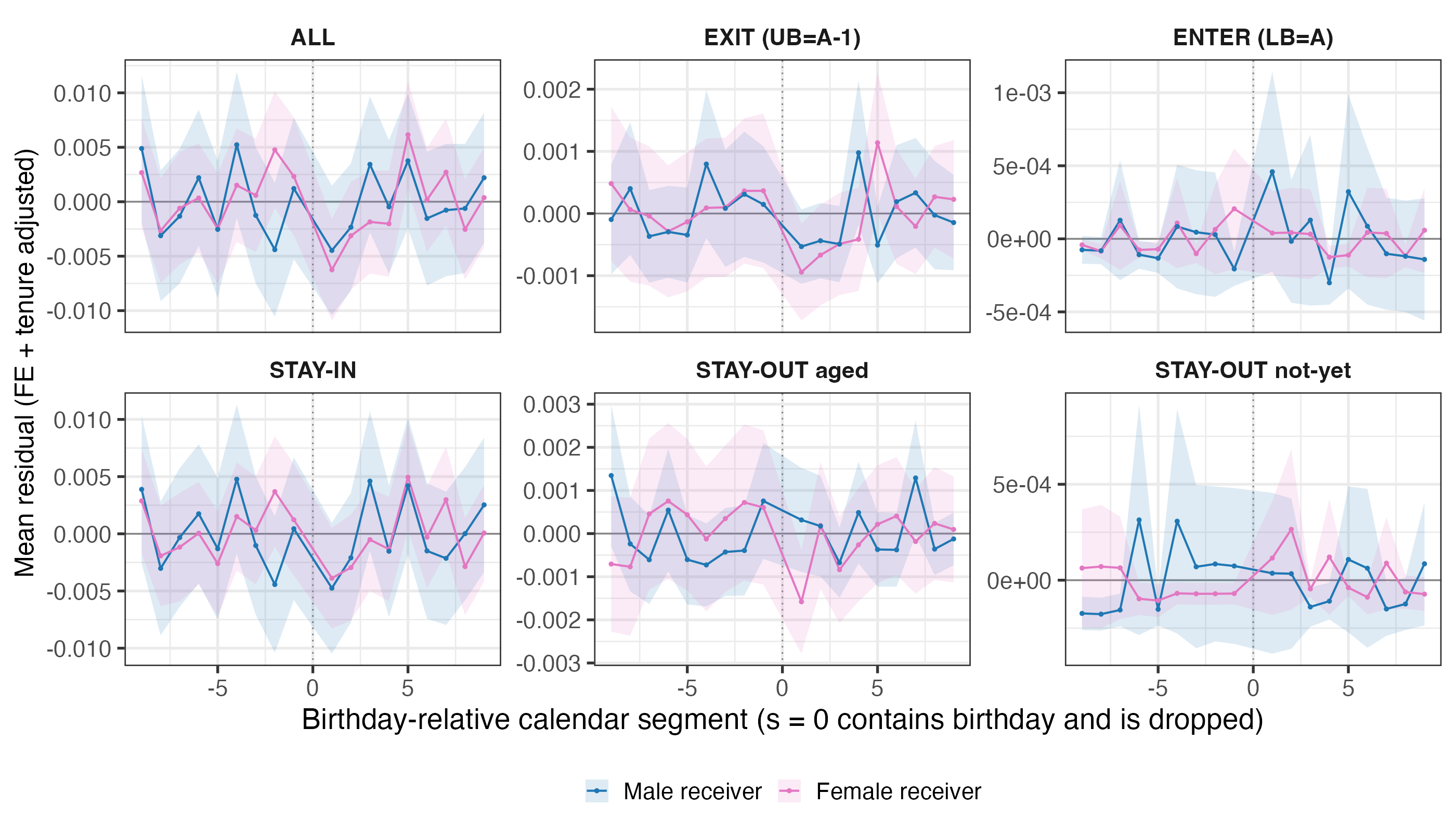}
  \caption{Birthday-Relative Pre-relationship Patterns by Proposer Eligibility Type (Receiver Age 38)}
  \label{fg:app_event_study_age38_eligibility_prerel}
  \end{center}
  \footnotesize
  Note: Pre-relationship-stage counterpart to Figure~\ref{fg:event_study_age35_eligibility}, shown at receiver age 38 (where the total female pre-relationship birthday jump is significant; see Figure~\ref{fg:eventstudy_receiver_application_heatmap}). After accounting for member fixed effects, calendar-year fixed effects, the within-month calendar position, and the platform-tenure profile, the figure plots pre-relationship counts by birthday-relative segment and eligibility cell. Each panel has its own vertical scale.
\end{figure}

\begin{figure}[!htbp]
  \begin{center}
  \includegraphics[width = 0.95\textwidth]{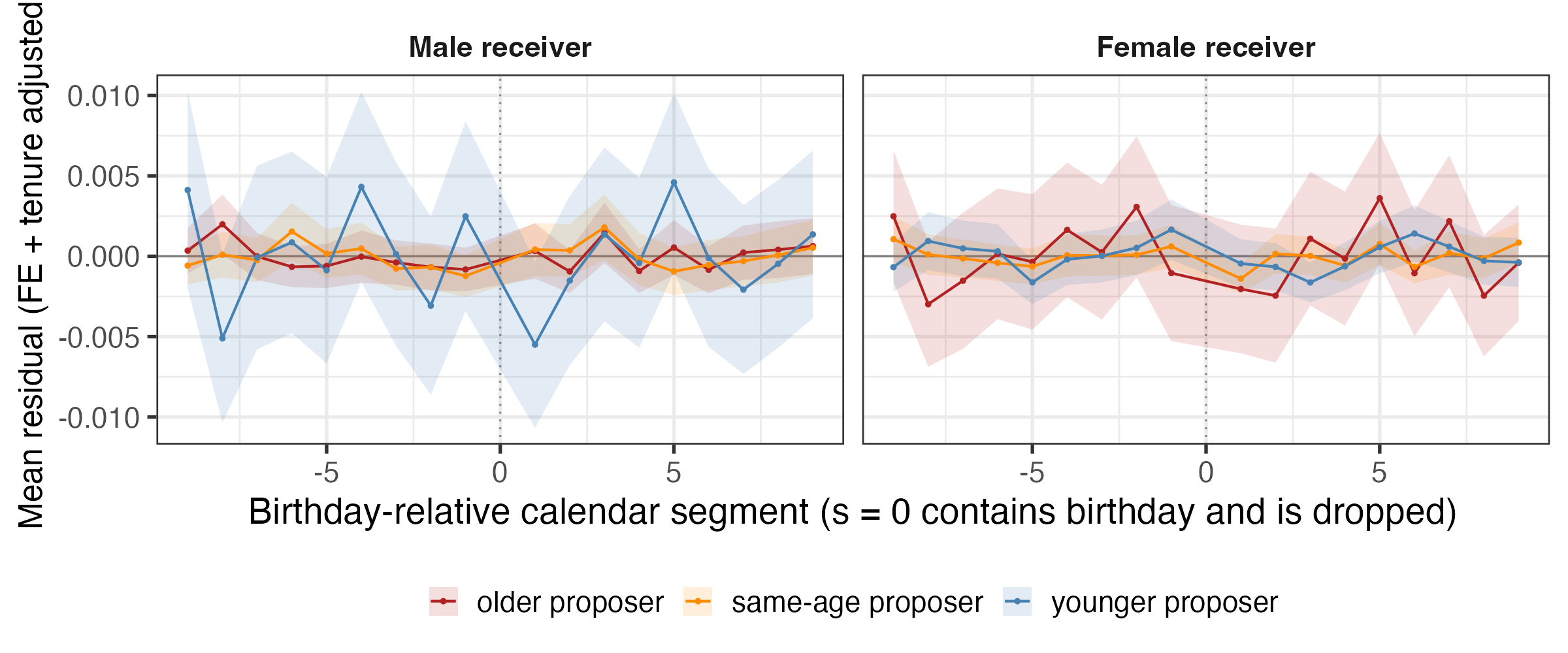}
  \caption{Birthday-Relative Pre-relationship Patterns within STAY-IN Proposers by Proposer Age (Receiver Age 38)}
  \label{fg:app_stayin_age_es_prerel}
  \end{center}
  \footnotesize
  Note: Pre-relationship-stage counterpart to Figure~\ref{fg:stayin_age_es}, shown at receiver age 38. The plot uses the same tenure-adjusted event-time series and is restricted to STAY-IN proposers, split by the proposer's own age relative to the receiver (older, same-age, younger). Panels are male (left) and female (right) receivers.
\end{figure}

Figures~\ref{fg:event_study_age35_eligibility} and~\ref{fg:stayin_age_es} in the main text preview the eligibility and proposer-age re-sorting margins at the application stage for receiver age 35. Figures~\ref{fg:app_event_study_age38_eligibility_prerel} and~\ref{fg:app_stayin_age_es_prerel} report the same tenure-adjusted event studies at the pre-relationship stage. They use receiver age 38, where the total female pre-relationship jump is \(-0.0070\) (\(p=0.011\)); at age 35 it is \(-0.0011\) (\(p=0.702\)). At age 38, most of the female decline occurs within STAY-IN (\(-0.0043\)) rather than at the ENTER or EXIT boundaries. The male age-38 estimate is \(-0.0009\) (\(p=0.796\)), and the proposer-age split is more muted than at the application stage.

\subsection{Segment-Level Visualization of the Pre--Post Difference}\label{app:trend_check}

\begin{figure}[!htbp]
  \begin{center}
  \subfloat[Male receivers]{\includegraphics[width = 0.90\textwidth, height = 0.45\textheight, keepaspectratio]{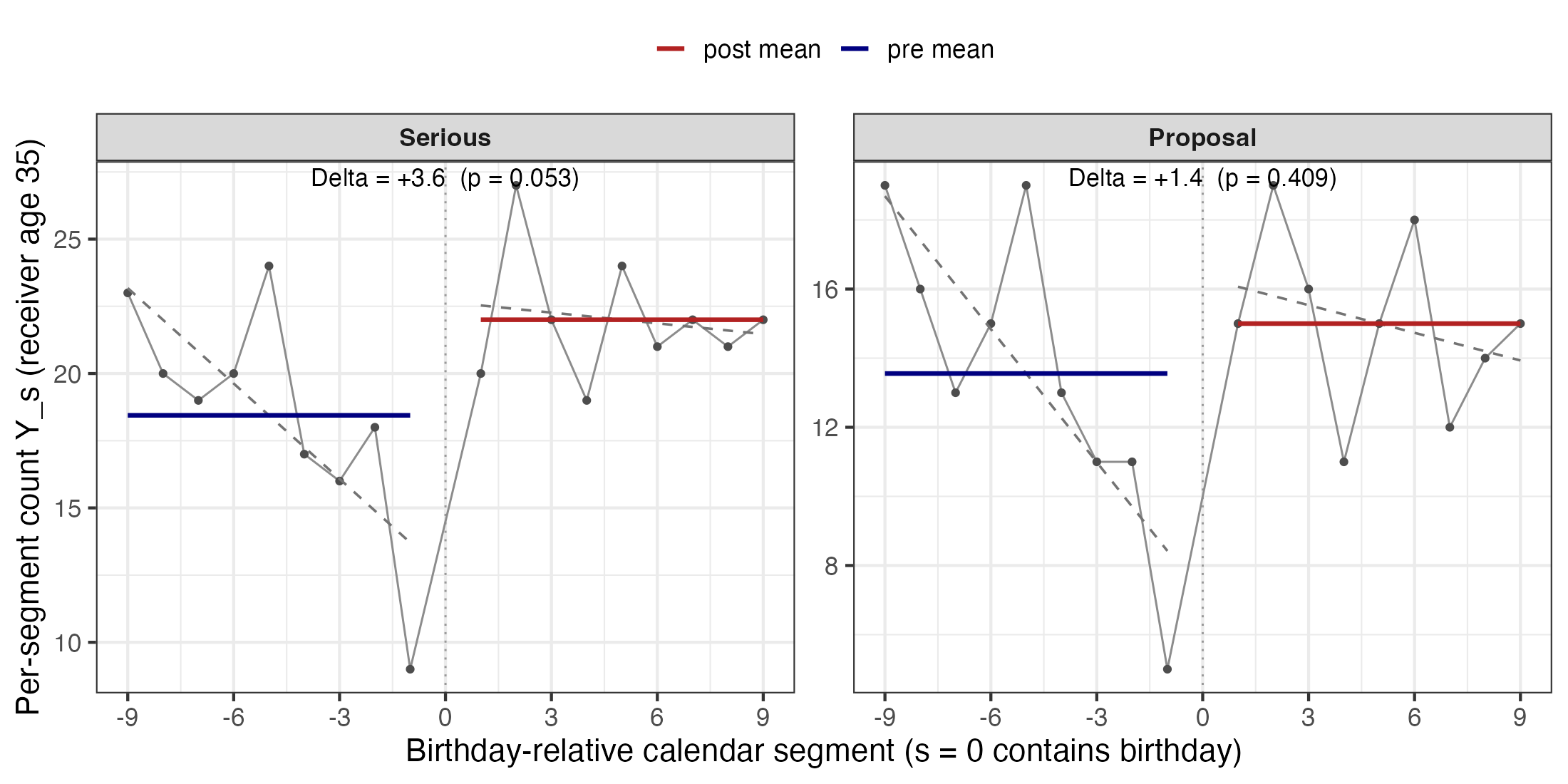}}\\[0.2in]
  \subfloat[Female receivers]{\includegraphics[width = 0.90\textwidth, height = 0.45\textheight, keepaspectratio]{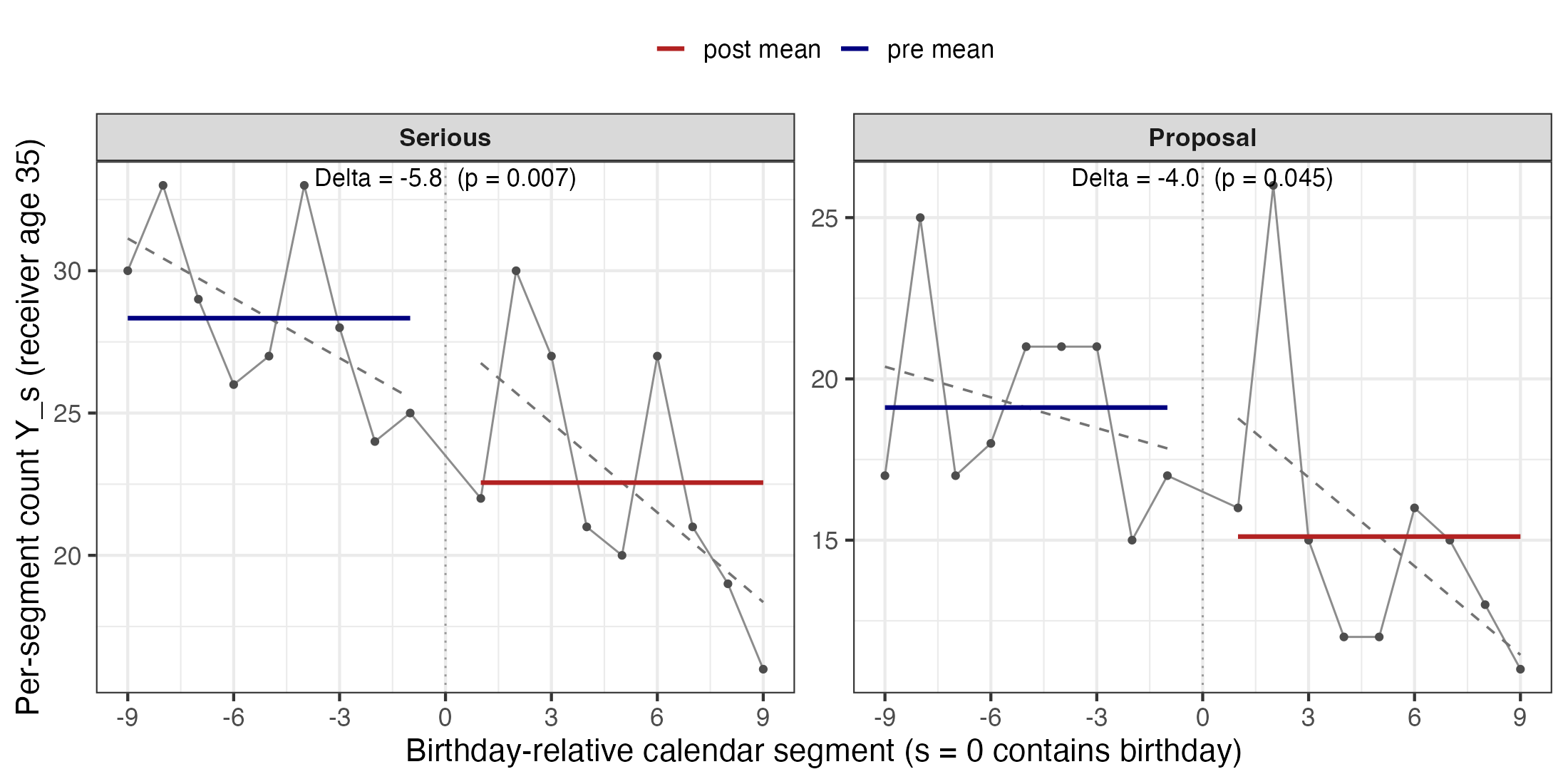}}
  \caption{Segment-Level Pre--Post Difference, Visualized for Receiver Age 35}
  \label{fg:trend_check_ttest_consistent}
  \end{center}
  \footnotesize
  Note: The figure plots total outcome counts by birthday-relative segment and the pre- and post-birthday means. Their gap equals the Welch pre--post difference in Figure~\ref{fg:ttest_receiver_proposal_heatmap}; dashed lines show separate linear fits on each side.
\end{figure}

Figure~\ref{fg:trend_check_ttest_consistent} plots the segment-level totals underlying the Welch pre--post contrasts for receiver age 35. The gap between the pre- and post-birthday means equals the difference reported in Figure~\ref{fg:ttest_receiver_proposal_heatmap}. The visible pre-birthday drift, especially for female receivers, reinforces that these contrasts are descriptive rather than clean discontinuity estimates.

\subsection{Receiver Birthday Effects by Proposer Characteristics at Later Stages}\label{app:later_stage_characteristics}

\begin{figure}[!htbp]
  \begin{center}
  \subfloat[Higher-type proposers (older / higher-income / taller)]{\includegraphics[width = 0.90\textwidth, height = 0.25\textheight, keepaspectratio]{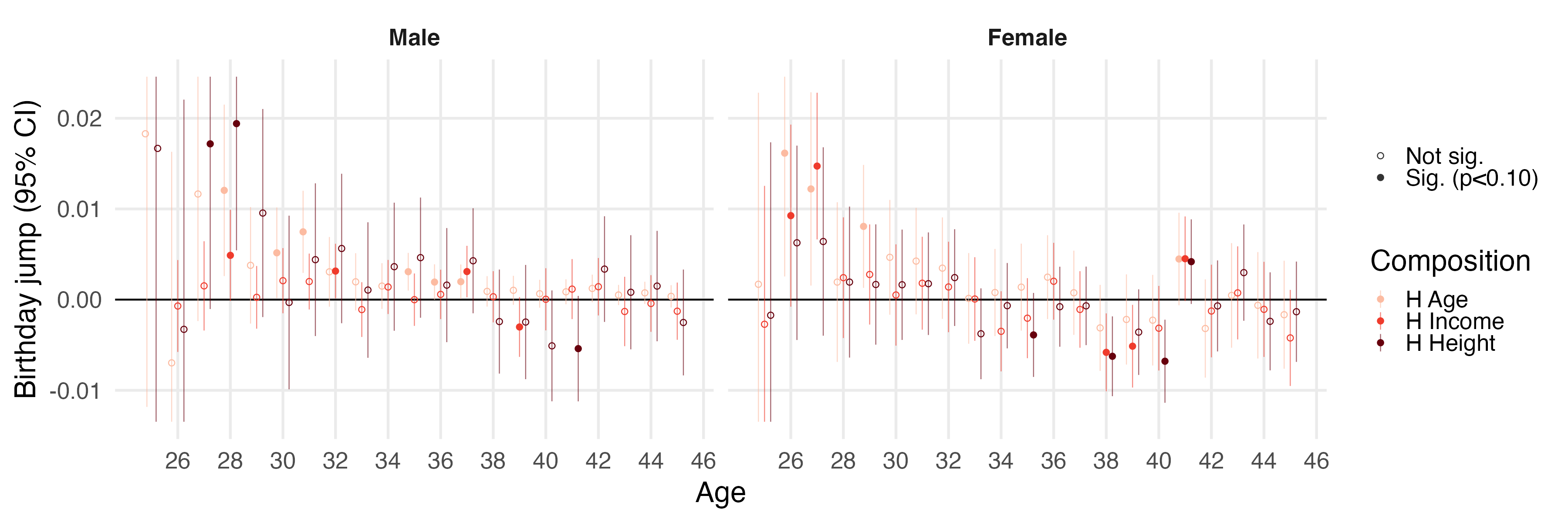}}\\[0.2in]
  \subfloat[Lower-type proposers (younger / lower-income / shorter)]{\includegraphics[width = 0.90\textwidth, height = 0.25\textheight, keepaspectratio]{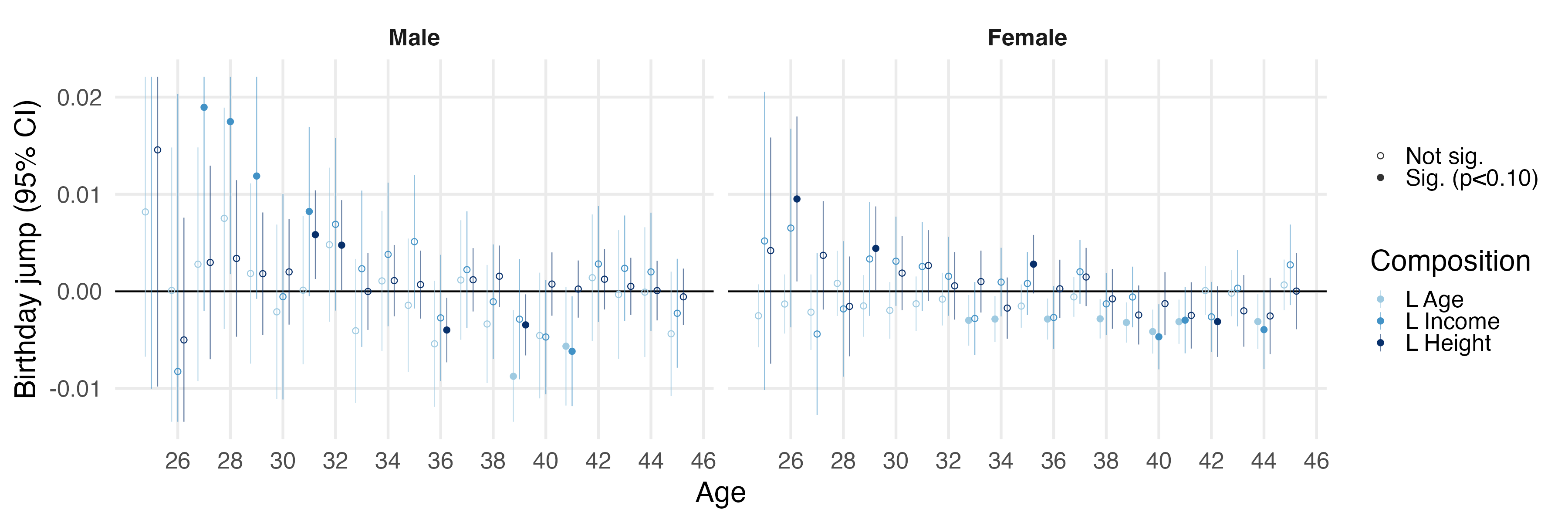}}\\[0.2in]
  \subfloat[By proposer stated child preference (Want / No Preference / Do Not Want)]{\includegraphics[width = 0.90\textwidth, height = 0.25\textheight, keepaspectratio]{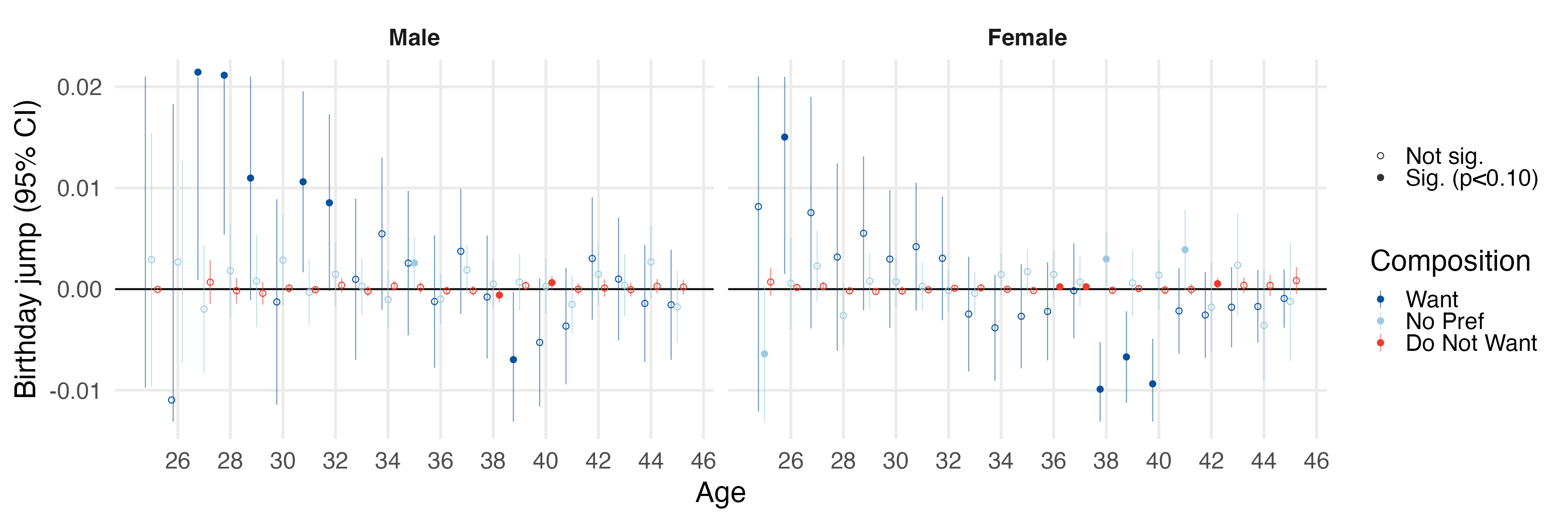}}
  \caption{Pre-relationship-Stage Birthday Jump by Proposer Characteristics and Child Preference}
  \label{fg:app_characteristics_prerelation}
  \end{center}
  \footnotesize
  Note: Pre-relationship-stage birthday jump by receiver age and gender. Panel (a) shows higher-type proposers; panel (b) shows lower-type proposers; panel (c) splits by the proposer's stated child preference.
\end{figure}

\begin{figure}[!htbp]
  \begin{center}
  \subfloat[Higher-type proposers (older / higher-income / taller)]{\includegraphics[width = 0.90\textwidth, height = 0.25\textheight, keepaspectratio]{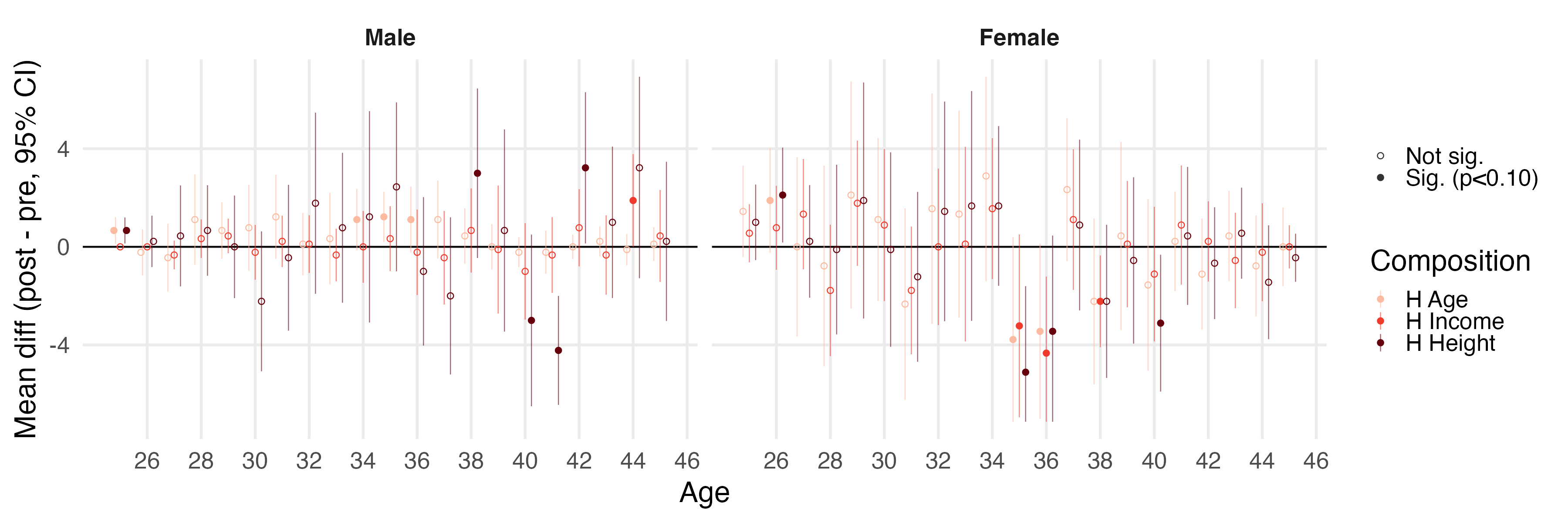}}\\[0.2in]
  \subfloat[Lower-type proposers (younger / lower-income / shorter)]{\includegraphics[width = 0.90\textwidth, height = 0.25\textheight, keepaspectratio]{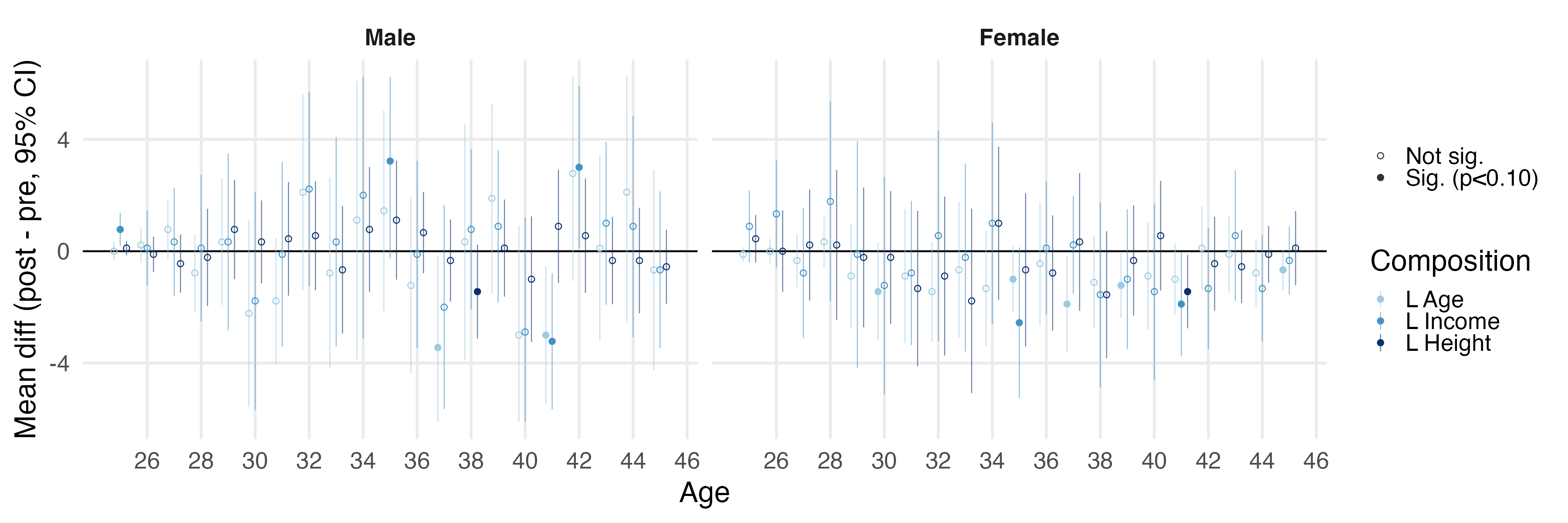}}\\[0.2in]
  \subfloat[By proposer stated child preference (Want / No Preference / Do Not Want)]{\includegraphics[width = 0.90\textwidth, height = 0.25\textheight, keepaspectratio]{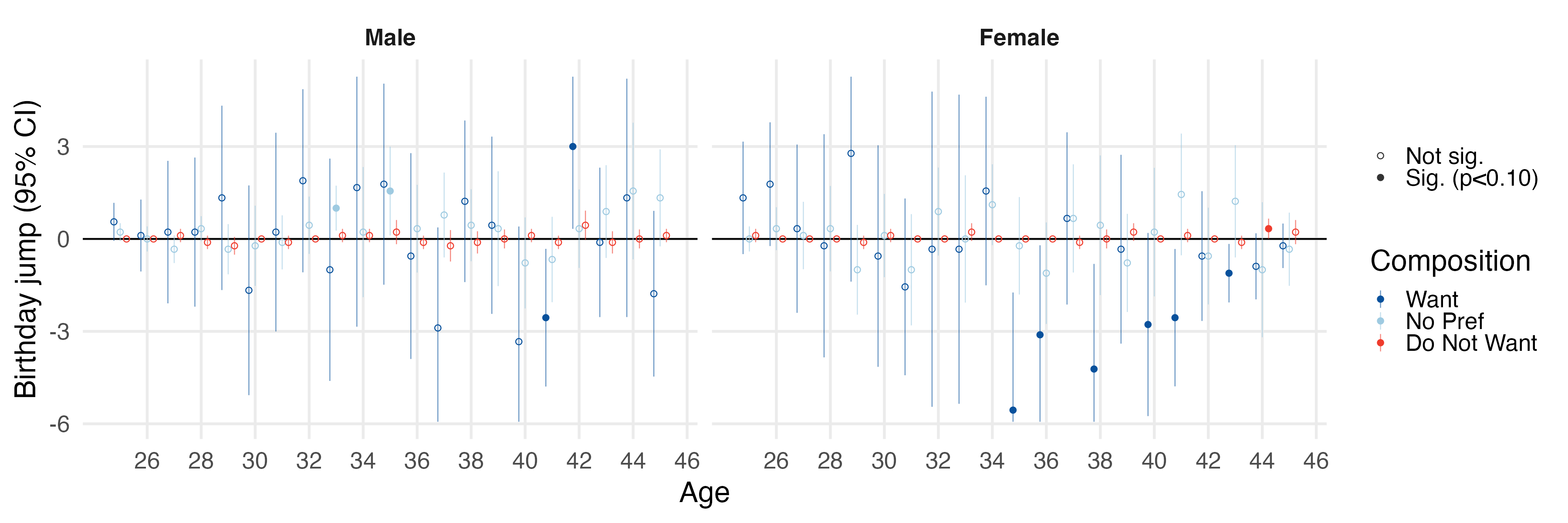}}
  \caption{Serious-Relationship-Stage Pre--Post Difference by Proposer Characteristics and Child Preference}
  \label{fg:app_characteristics_serious}
  \end{center}
  \footnotesize
  Note: Segment-level Welch pre--post difference at the serious-relationship stage by receiver age and gender. Panel (a) shows higher-type proposers; panel (b) shows lower-type proposers; panel (c) splits by the proposer's stated child preference.
\end{figure}

\begin{figure}[!htbp]
  \begin{center}
  \subfloat[Higher-type proposers (older / higher-income / taller)]{\includegraphics[width = 0.90\textwidth, height = 0.25\textheight, keepaspectratio]{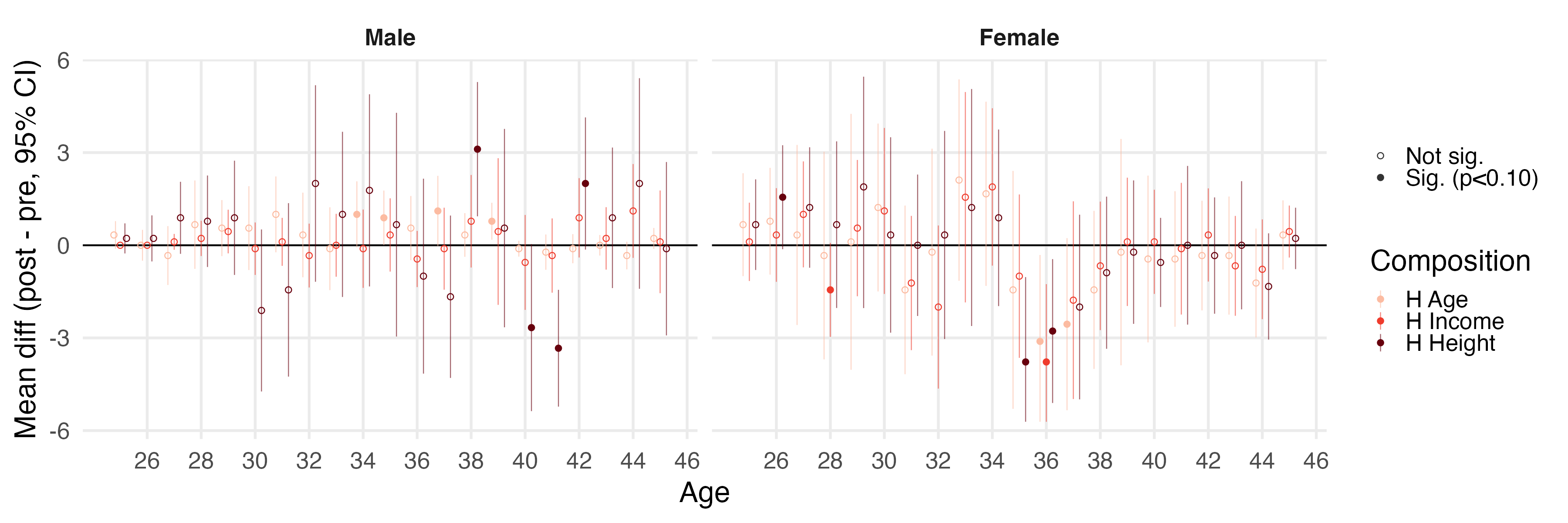}}\\[0.2in]
  \subfloat[Lower-type proposers (younger / lower-income / shorter)]{\includegraphics[width = 0.90\textwidth, height = 0.25\textheight, keepaspectratio]{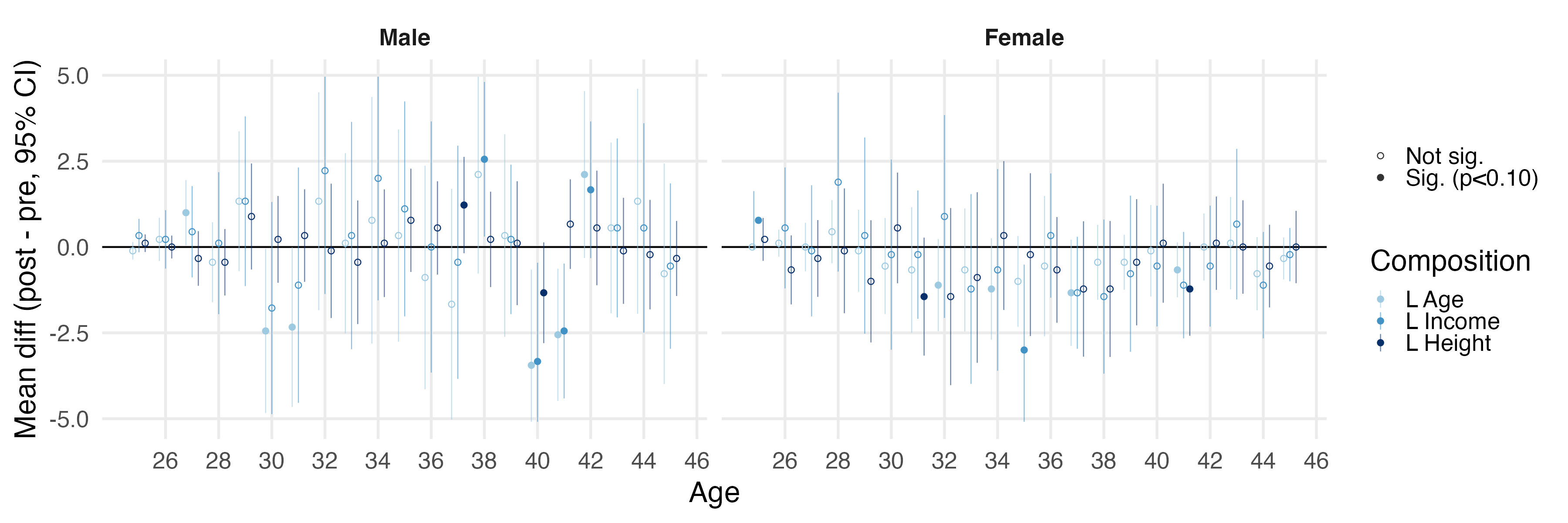}}\\[0.2in]
  \subfloat[By proposer stated child preference (Want / No Preference / Do Not Want)]{\includegraphics[width = 0.90\textwidth, height = 0.25\textheight, keepaspectratio]{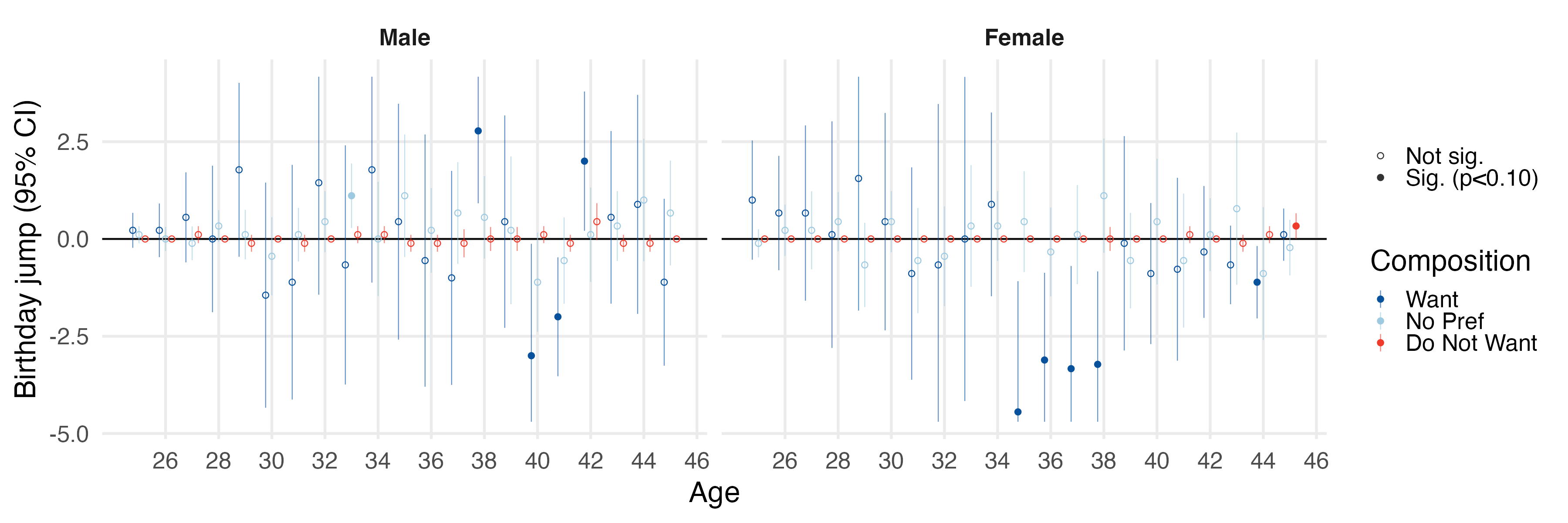}}
  \caption{Proposal-Stage Pre--Post Difference by Proposer Characteristics and Child Preference}
  \label{fg:app_characteristics_proposal}
  \end{center}
  \footnotesize
  Note: Segment-level Welch pre--post difference at the proposal stage by receiver age and gender. Panel (a) shows higher-type proposers; panel (b) shows lower-type proposers; panel (c) splits by the proposer's stated child preference.
\end{figure}

This subsection reports the later-stage counterparts to Figure~\ref{fg:characteristics_composition}. Figure~\ref{fg:app_characteristics_prerelation} uses the same within-member birthday-jump specification as the main pre-relationship estimates. Figures~\ref{fg:app_characteristics_serious} and~\ref{fg:app_characteristics_proposal} use the segment-level Welch pre--post differences used for the sparse serious-relationship and proposal outcomes. Across stages, the age split remains the most interpretable composition margin, while income, height, and child-preference cells are noisier and should be read as heterogeneity diagnostics.

\subsection{Proposer's Birthday Cutoff}\label{app:proposer_birthday_cutoff}

The two birthday designs are not symmetric and should not be read as the same identification strategy. The clean quasi-experimental design is the receiver birthday acting on opposite-side application behavior: proposers observe the receiver's displayed age but not the exact birthday, so the displayed-age update is a shock to the information faced by the other side. A proposer's own birthday is instead an own-age event that the proposer knows in advance and may respond to strategically through search effort, expectations, or psychology. The proposer-own-birthday analysis is therefore secondary, included as a behavioral and robustness exercise rather than as a co-equal causal estimate. It is also useful as indirect evidence on whether users generally increase their own platform actions at birthdays, since historical login and profile-update records are not available in the research extract.

\begin{figure}[!htbp]
  \begin{center}
  \includegraphics[width = 1.0\textwidth]{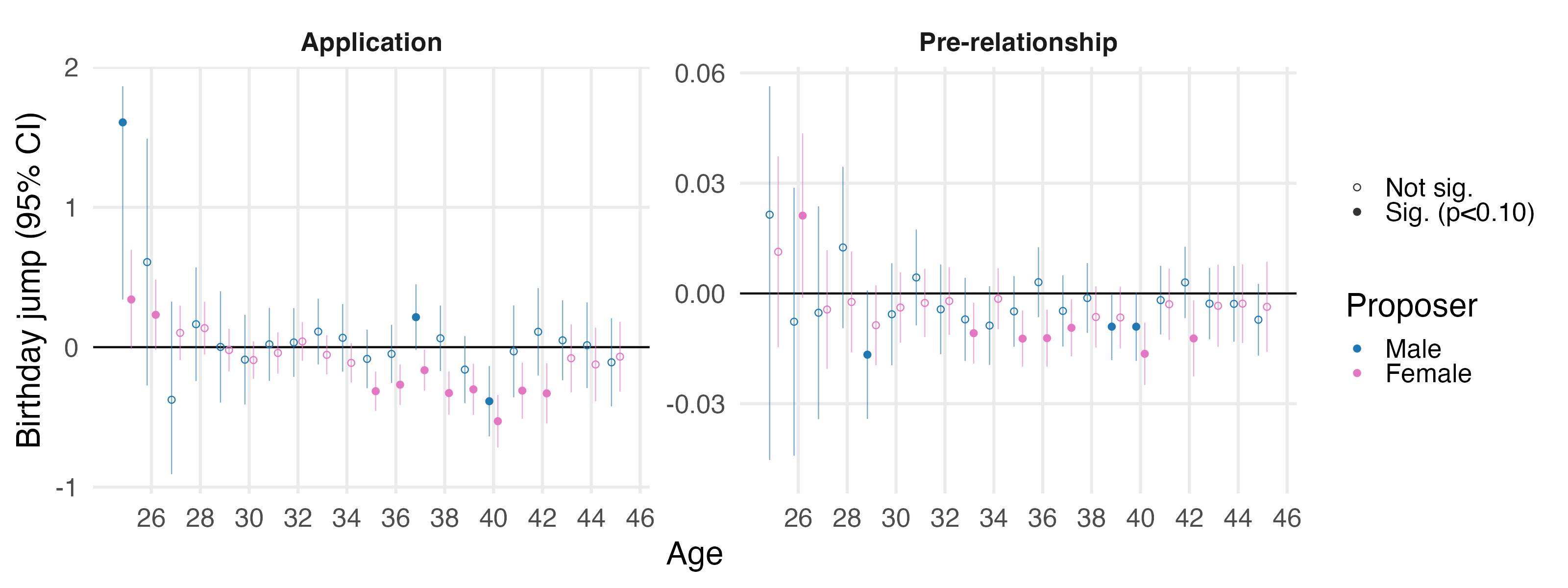}
  \caption{Within-Member Birthday Jump for Proposer Birthday at Application and Pre-relationship Stages}
  \label{fg:eventstudy_proposer_application_heatmap}
  \end{center}
  \footnotesize
  Note: Each point is the within-member birthday jump at the proposer's own birthday, by proposer age, with 95\% confidence intervals. The two panels are the application and pre-relationship stages (each on its own vertical scale). Filled markers are statistically significant at the 10\% level; hollow markers are not.
\end{figure}

Figure~\ref{fg:eventstudy_proposer_application_heatmap} repeats the within-member birthday-jump specification using the proposer's own birthday as the cutoff. Female proposers' application counts rise significantly at ages 25--26 and fall at ages 35--42; their pre-relationship counts rise at age 26 and fall at ages 33, 35--37, 40, and 42. Male-proposer effects are isolated and mixed: application jumps are positive at ages 25 and 37 and negative at 40, while pre-relationship jumps are negative at 29, 39, and 40. Thus, own-birthday behavior is not a general increase in proposer activity and differs from the receiver-side re-sorting response.

\section{Attribution and Pass-Through of the Birthday Response}\label{app:attribution_passthrough}

This appendix connects the re-sorting of inbound applications to the downstream female penalty in the pre-relationship transition. Section~\ref{app:who_ended} attributes the pre-relationship margin using both parties' recorded replies. Section~\ref{app:delta2_reweighting} decomposes the change in the application-tracked transition rate into proposer-composition and within-type terms. Section~\ref{app:realized_match} traces counterparty composition across funnel stages. Section~\ref{app:pooled_inference} reports pooled-age estimates.

\subsection{Who Ends the Pair: Side Attribution of the Pre-Relationship Margin}\label{app:who_ended}

Institutionally, after a formal meeting both parties submit a continue-or-decline reply through their agencies, and both replies are recorded regardless of the counterparty's answer; among held meetings in the analysis window, one-sided missing replies are negligible (on the order of one hundred pairs out of roughly 1.6 million). Each side's willingness to continue is therefore observable with negligible censoring, and the transition from application into \textcolor{black}{pre-relationship} can be attributed step by step: whether the receiver accepts the meeting request, whether a meeting is held, and, conditional on a meeting, which side declines to continue.

\begin{figure}[!htbp]
  \begin{center}
  \includegraphics[width = 0.98\textwidth]{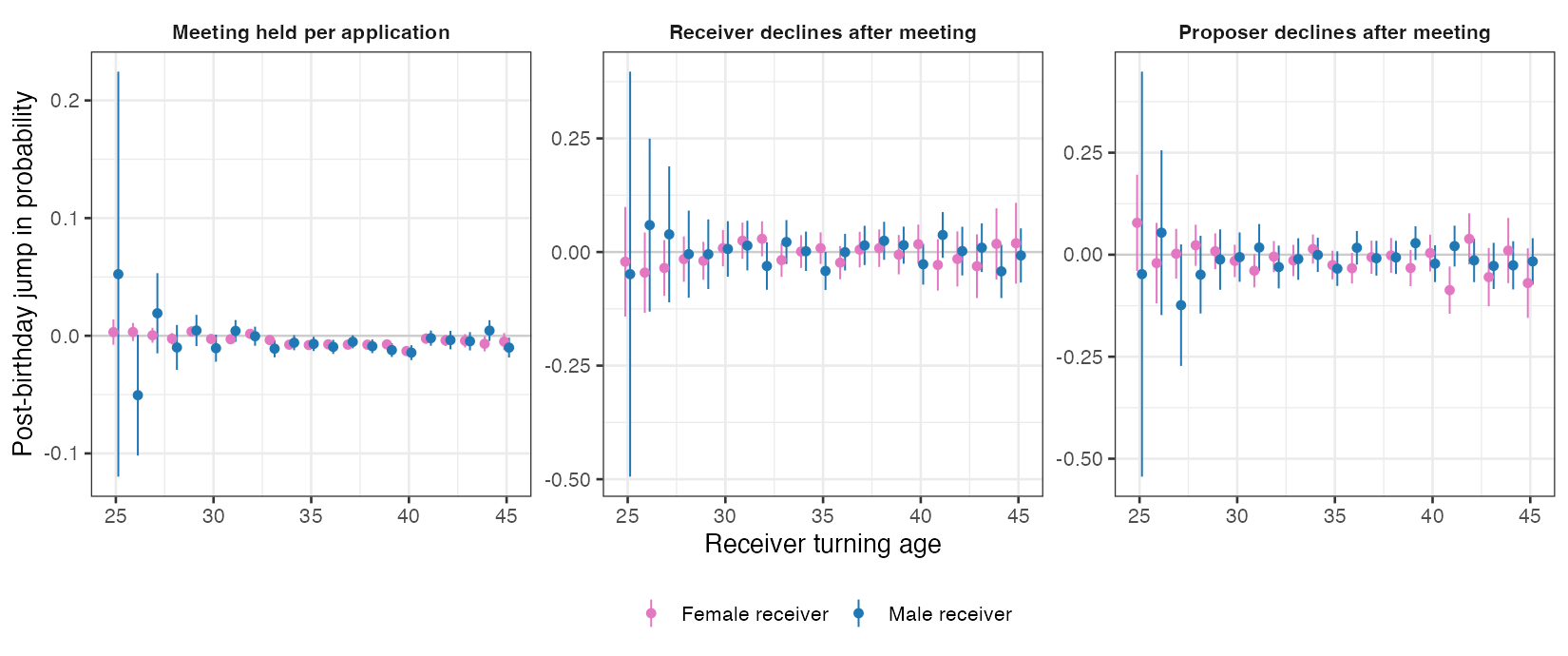}
  \caption{Side Attribution of the Application-to-Pre-Relationship Margin}
  \label{fg:meeting_attribution}
  \end{center}
  \footnotesize
  Note: Post-birthday jump in each probability by receiver turning age and gender, with 95\% confidence intervals. The left panel is the probability that a received application results in a held meeting; the middle and right panels condition on a held meeting and report each side's probability of declining to continue, from both parties' recorded replies. All estimates use the main specification with member, calendar-year, within-month-position, and tenure fixed effects, clustered by member.
\end{figure}

Figure~\ref{fg:meeting_attribution} plots the post-birthday change in three probabilities by receiver age, estimated with the main specification (member, calendar-year, within-month-position, and tenure fixed effects; member-clustered standard errors): the probability that an application results in a held meeting, and, conditional on a held meeting, the probability that the receiver declines and the probability that the proposer declines. Table~\ref{tab:who_ended_pooled} reports pooled estimates.

Three findings stand out for female receivers aged 31--40. First, held meetings per application fall by \(0.85\) percentage points from a pre-birthday mean of \(4.9\%\), a decline of about 17 percent. Second, receivers reject \(0.75\) percentage points fewer meeting requests after the birthday. Third, conditional on a held meeting, pre-relationship formation is nearly unchanged (\(+0.05\) percentage points): receiver declines rise by \(0.65\) percentage points and proposer declines fall by \(0.81\) percentage points. The aggregate loss therefore occurs before the meeting rather than in continuation after it. \textcolor{black}{The residual margin---requests neither rejected by the receiver nor followed by a held meeting---rises by \(1.60\) percentage points and is not further decomposed here (for example, into cancellations, scheduling failures, or expirations); the two post-meeting decline indicators can also overlap within a pair.}

\begin{table}[!htbp]
  \caption{Side Attribution of the Pre-Relationship Margin: Pooled Estimates}
  \label{tab:who_ended_pooled}
  \begin{center}
  \resizebox{\textwidth}{!}{\input{figuretable/birthday_project/who_ended_pair_attribution_pooled}}
  \end{center}
  \footnotesize
  Note: Post-birthday jumps in the indicated probabilities, pooled over receiver turning ages, with member-clustered standard errors in parentheses; SE in the column headings denotes standard error. The first two rows are estimated on all received applications; the remaining rows condition on a held meeting. Pre-birthday means are shown next to each estimate.
\end{table}

\subsection{Composition versus Within-Type Change in the Pre-Relationship Transition}\label{app:delta2_reweighting}

The re-sorting result and the downstream penalty are connected by an exact accounting identity. Let \(w_t\) denote the share of applications from proposers of relative-age type \(t\) (older, same-age, or younger than the receiver) and \(\pi_t\) the type-specific rate at which applications convert into pre-relationships within the application-tracked window, so that the aggregate rate is \(\pi = \sum_t w_t \pi_t\). The post-birthday change decomposes as
\[
\Delta\pi \;=\; \underbrace{\sum_t \left(w_t^{\mathrm{post}} - w_t^{\mathrm{pre}}\right)\pi_t^{\mathrm{pre}}}_{\text{composition}} \;+\; \underbrace{\sum_t w_t^{\mathrm{post}}\left(\pi_t^{\mathrm{post}} - \pi_t^{\mathrm{pre}}\right)}_{\text{within-type}}.
\]
The composition term is the accounting contribution of the shift in proposer-type shares; the within-type term captures changes in conversion rates within a fixed type.

\begin{figure}[!htbp]
  \begin{center}
  \includegraphics[width = 0.98\textwidth]{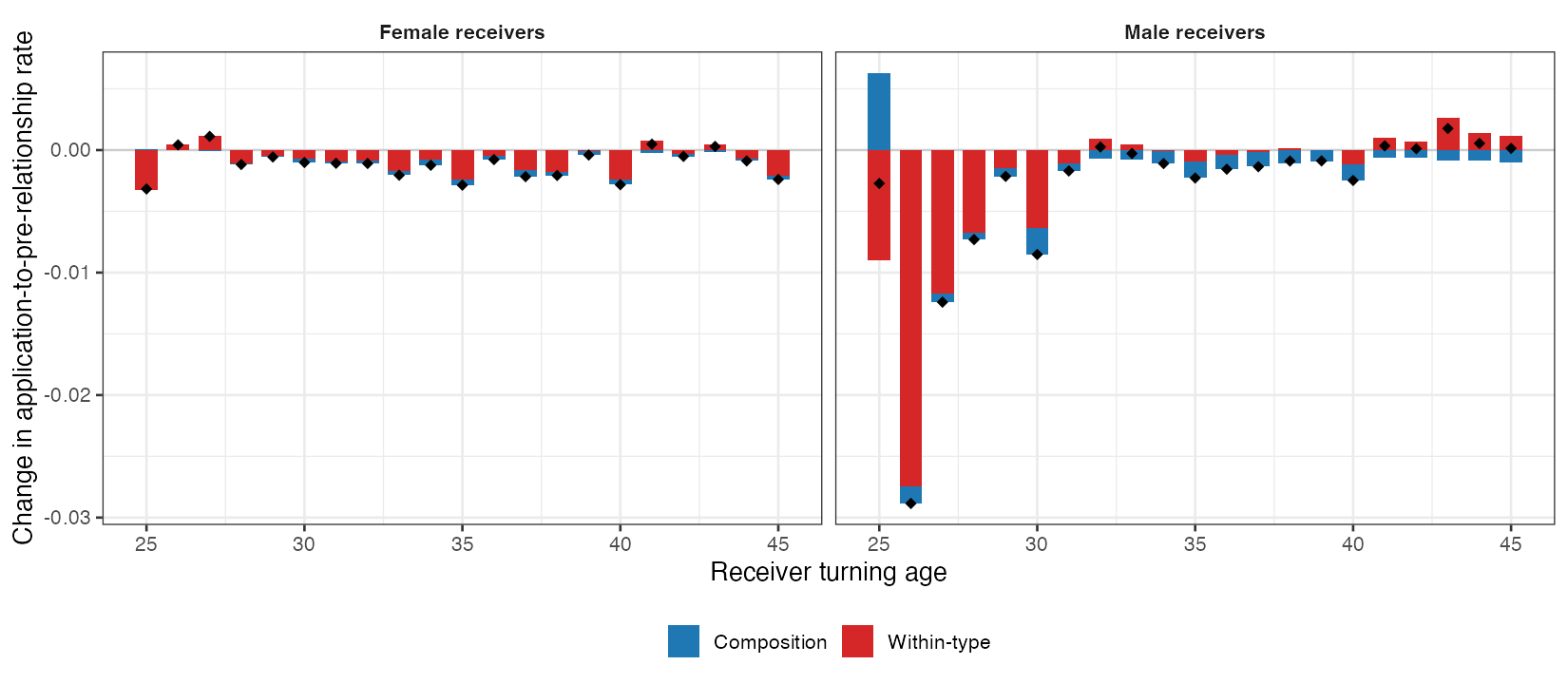}
  \caption{Shift-Share Decomposition of the Post-Birthday Change in the Pre-Relationship Transition Rate}
  \label{fg:delta2_reweighting}
  \end{center}
  \footnotesize
  Note: Stacked bars decompose the post-birthday change in the application-tracked application-to-pre-relationship rate into the proposer relative-age composition term and the within-type term, by receiver turning age and gender. Black diamonds mark the total observed change. The sample and follow-up windows are identical to the funnel decomposition in the main text.
\end{figure}

Figure~\ref{fg:delta2_reweighting} and Table~\ref{tab:delta2_reweighting_pooled} report the decomposition on the same application-tracked sample as the funnel decomposition. For female receivers aged 31--40, the transition rate falls from \(1.67\%\) to \(1.50\%\). The composition term accounts for \(22\%\) of this decline: the older-cohort application share rises from \(85.3\%\) to \(88.2\%\), and older-cohort applications reach pre-relationships at \(1.5\%\), compared with \(2.7\%\) for younger-cohort applications. The remaining 78\% is a within-type decline. For male receivers, the composition term more than accounts for the decline and is partly offset by a positive within-type term.

\begin{table}[!htbp]
  \caption{Shift-Share Decomposition: Pooled Cells}
  \label{tab:delta2_reweighting_pooled}
  \begin{center}
  \resizebox{\textwidth}{!}{\input{figuretable/birthday_project/delta2_reweighting_pooled}}
  \end{center}
  \footnotesize
  Note: Application-to-pre-relationship transition rates before and after the birthday, the total change, and its composition and within-type components, pooled over the indicated receiver ages. The composition share is the composition term divided by the total change; a value above one indicates that the composition term over-explains the total, with the within-type term positive.
\end{table}

\subsection{Counterparty Composition Across Funnel Stages}\label{app:realized_match}

\begin{figure}[!htbp]
  \begin{center}
  \includegraphics[width = 0.95\textwidth]{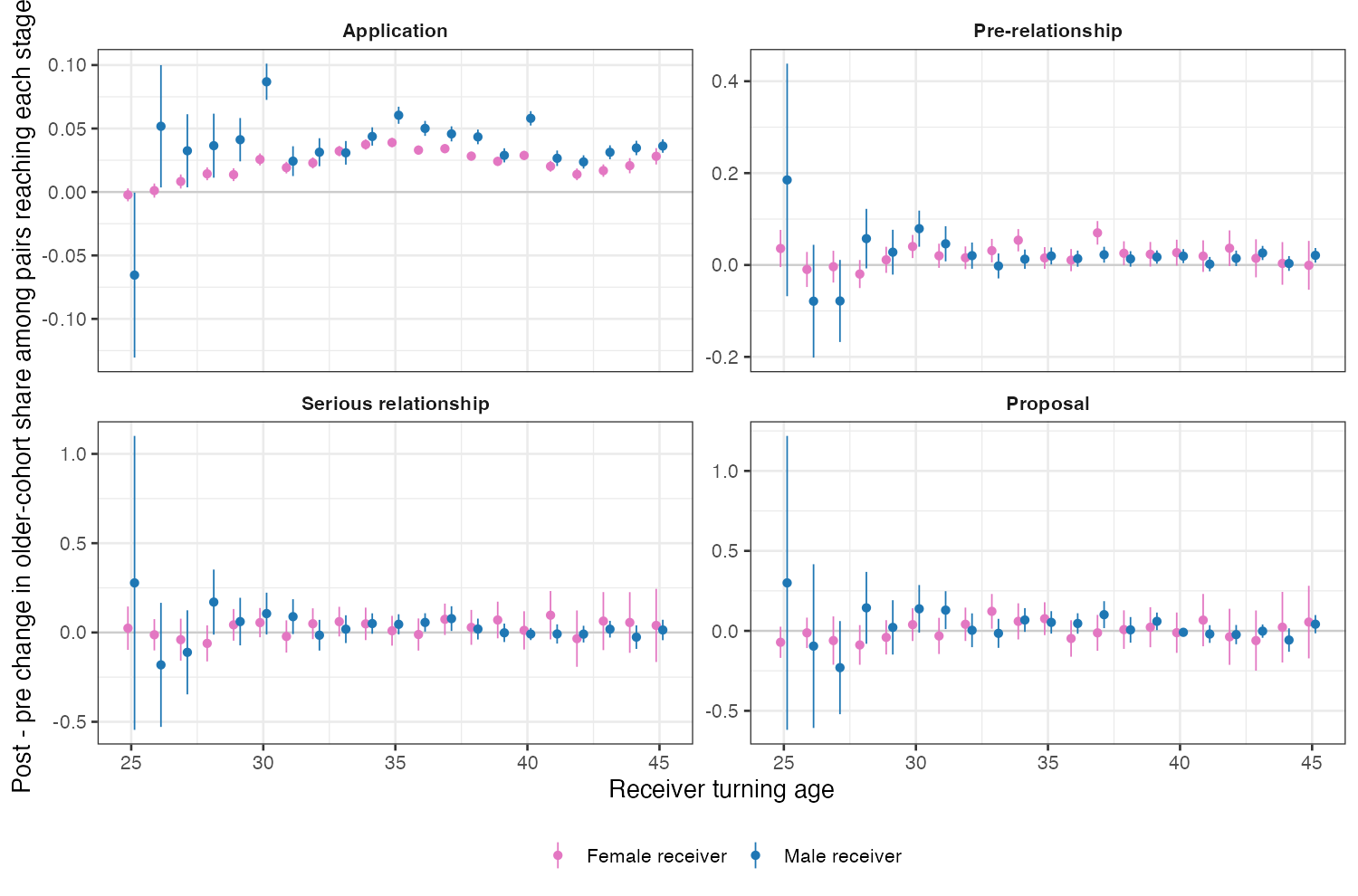}
  \caption{Post-Birthday Change in the Older-Cohort Counterparty Share Across Funnel Stages}
  \label{fg:realized_match_passthrough}
  \end{center}
  \footnotesize
  Note: Post-minus-pre difference in the older-cohort counterparty share among pairs reaching each stage, by receiver turning age and gender, \textcolor{black}{with 95\% confidence intervals clustered by receiver}.
\end{figure}

The main text documents re-sorting in inbound applications. This section traces the same counterparty attributes among pairs reaching each stage: the age gap, the older-cohort counterparty share, and the share of counterparties stating a preference for children. Figure~\ref{fg:realized_match_passthrough} reports the post-minus-pre difference in the older-cohort share by receiver age and stage; Table~\ref{tab:realized_match_pooled} reports pooled estimates for all three attributes.

For female receivers aged 31--40, the older-cohort share rises by \(3.00\) percentage points at applications, \(2.85\) at pre-relationships, \(3.12\) at serious relationships, and \(2.54\) among accepted proposals. The counterparty age gap rises by \(0.5815\), \(0.3683\), \(0.2393\), and \(0.1510\) years across the same stages, while the share stating a preference for children falls at every stage. The compositional shift therefore persists down the funnel but is not monotonic.

\begin{table}[!htbp]
  \caption{Counterparty Composition Across Funnel Stages: Pooled Estimates}
  \label{tab:realized_match_pooled}
  \begin{center}
  \resizebox{\textwidth}{!}{\input{figuretable/birthday_project/realized_match_composition_pooled}}
  \end{center}
  \footnotesize
  Note: Post-minus-pre differences in the indicated counterparty attributes among pairs reaching each stage, pooled over the indicated receiver ages, \textcolor{black}{with receiver-clustered standard errors in parentheses}.
\end{table}

\subsection{Pooled-Age Estimates}\label{app:pooled_inference}

Table~\ref{tab:pooled_jump_inference} reports pooled-age versions of the headline birthday jumps with member-clustered standard errors: the total inbound-count jump at the application and pre-relationship stages by gender, and the always-eligible re-sorting cells by the proposer's relative age. The final column reports a joint Wald test that the age-specific jumps are equal. Among female receivers aged 31--40, equality is rejected for applications (\(p=0.001\)) but not for pre-relationships (\(p=0.381\)); estimates over the full 25--45 range are more heterogeneous.

\begin{table}[!htbp]
  \caption{Pooled-Age Birthday Jumps with Clustered Inference}
  \label{tab:pooled_jump_inference}
  \begin{center}
  \resizebox{\textwidth}{!}{\input{figuretable/birthday_project/pooled_jump_inference}}
  \end{center}
  \footnotesize
  Note: Post-birthday jumps pooled over the indicated receiver turning ages, estimated with the main specification and clustered by member; SE in the column heading denotes standard error. The STAY-IN rows split applications from always-eligible proposers by the proposer's age relative to the receiver. Relative-age cells use the receiver's turning age: a same-age proposer is one year older than the receiver before the birthday and the same age afterward. The final column reports the \(p\)-value of a joint Wald test that the age-specific jumps are equal across ages within the cell.
\end{table}

\section{Robustness Check}\label{app:robustness}

This appendix collects the remaining robustness and diagnostic exercises. Appendix~\ref{app:tenure_diagnostic} compares the event-time path that omits tenure controls with the estimated tenure profile. Appendix~\ref{app:robust_shortrun} estimates the scalar jump close to the cutoff, and Appendix~\ref{app:robust_window} varies the birthday window and checks for a discontinuous exit response that could induce selection at the cutoff. Appendix~\ref{app:robust_anticipation} tests for birth-month anticipation, and Appendix~\ref{app:robust_placebo} uses artificial cutoffs.

\subsection{Near-Cutoff Birthday Jump}\label{app:robust_shortrun}

\begin{figure}[!htbp]
  \begin{center}
  \includegraphics[width = 0.95\textwidth, height = 0.40\textheight, keepaspectratio]{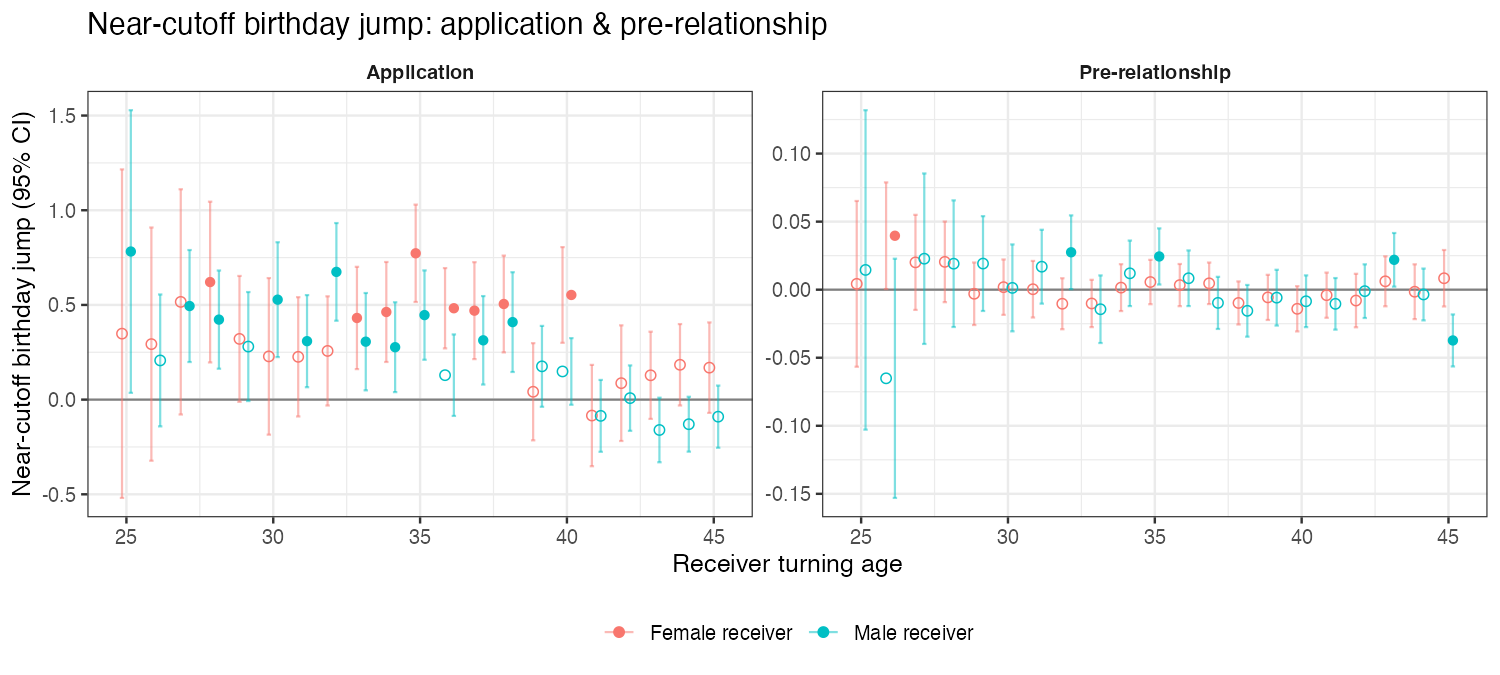}
  \caption{Near-Cutoff Birthday Jump at Application and Pre-relationship Stages}
  \label{fg:robust_shortrun_s1}
  \end{center}
  \footnotesize
  Note: Each point is the post-birthday coefficient from equation~\eqref{eq:rdd}, estimated on segments \(s\in\{-2,-1,1,2\}\). The specification includes member, calendar-year, within-month-position, and registration-tenure fixed effects and omits the birthday-containing segment. Bars are 95\% confidence intervals; solid markers denote estimates significant at the 5\% level.
\end{figure}

For applications and pre-relationships, I test whether the baseline \(\pm9\)-segment average masks a response concentrated near the cutoff by re-estimating equation~\eqref{eq:rdd} on \(s\in\{-2,-1,1,2\}\). Figure~\ref{fg:robust_shortrun_s1} reports the estimates by receiver age and gender.

\subsection{Birthday-Window Robustness}\label{app:robust_window}

\begin{figure}[!htbp]
  \begin{center}
  \includegraphics[width = 0.95\textwidth, height = 0.40\textheight, keepaspectratio]{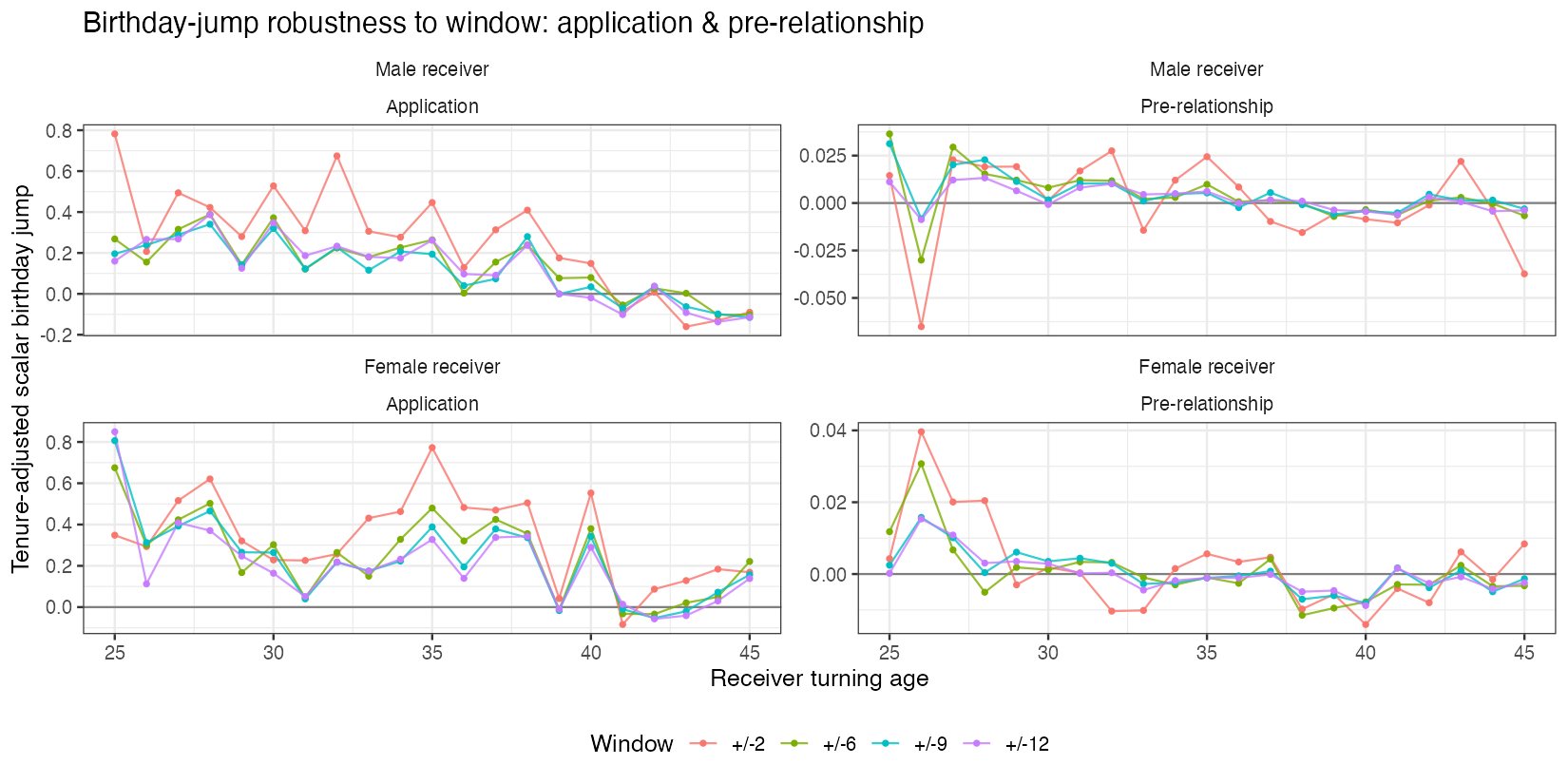}
  \caption{Birthday Jump Across Symmetric Event Windows, Application and Pre-relationship Stages}
  \label{fg:robust_window_s1}
  \end{center}
  \footnotesize
  Note: Scalar post-birthday coefficient by receiver turning age, estimated in symmetric \(\pm2\), \(\pm6\), \(\pm9\), and \(\pm12\) birthday-relative windows. All specifications include member, calendar-year, within-month-position, and registration-tenure fixed effects. The \(\pm9\) specification is the baseline.
\end{figure}

\begin{figure}[!htbp]
  \begin{center}
  \includegraphics[width = 0.95\textwidth, height = 0.40\textheight, keepaspectratio]{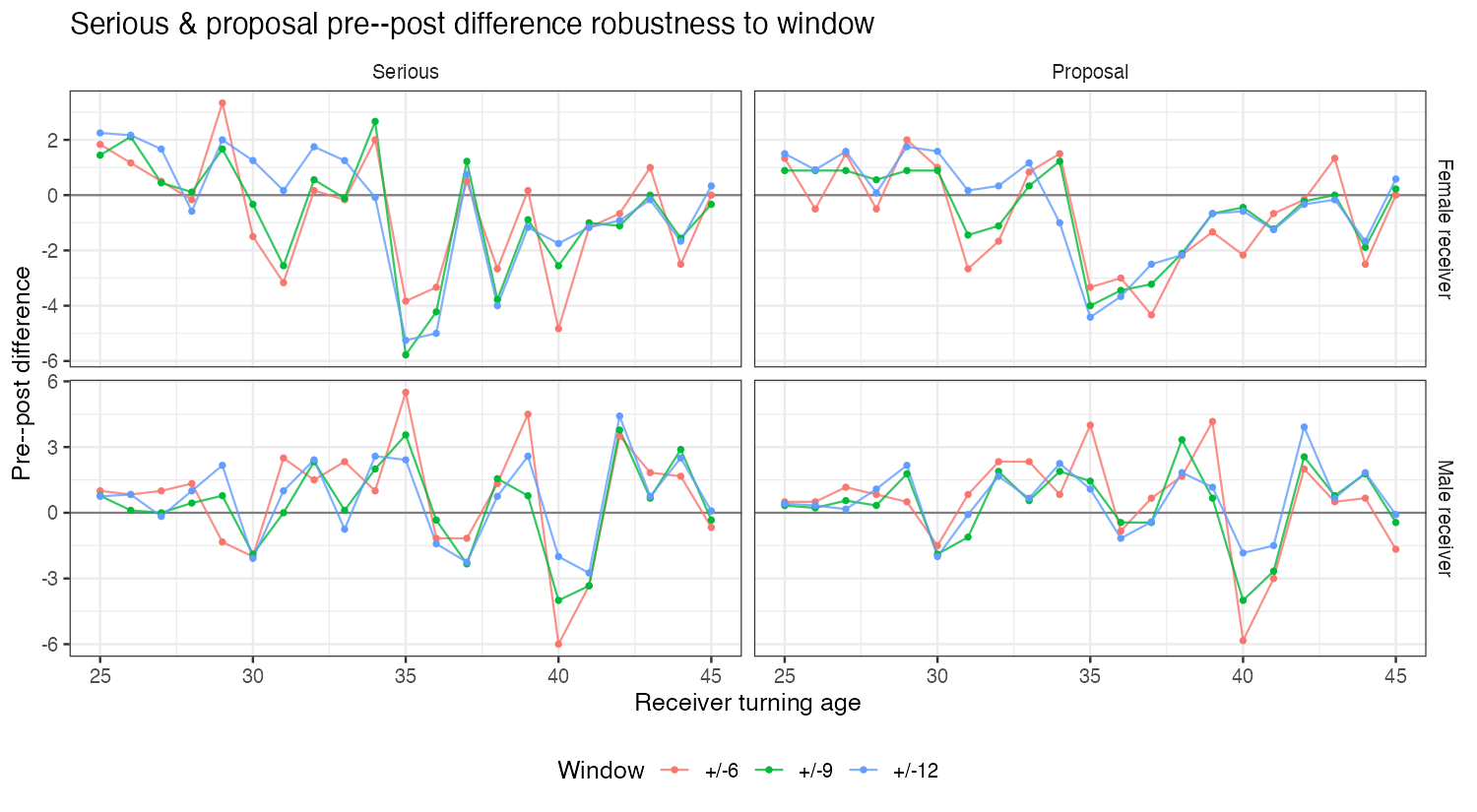}
  \caption{Pre--Post Difference Across Birthday Windows (\(\pm 6\), \(\pm 9\), \(\pm 12\)), Serious and Proposal Stages}
  \label{fg:robust_window_ttest}
  \end{center}
  \footnotesize
  Note: Welch pre--post difference by receiver turning age and gender at the serious-relationship and proposal stages, with the pre/post averaging window varied symmetrically. The \(\pm 9\) line is the baseline used in the main text.
\end{figure}

The main estimates use a \(\pm 9\)-segment window. Figure~\ref{fg:robust_window_s1} re-estimates the application and pre-relationship coefficients at \(\pm2\), \(\pm6\), \(\pm9\), and \(\pm12\). Figure~\ref{fg:robust_window_ttest} varies the symmetric averaging window for the serious-relationship and proposal contrasts.

Wider windows contain more active members and segments because activity is observed over longer support. Estimates from the \(\pm6\), \(\pm9\), and \(\pm12\) windows are similar across both exercises; the noisier \(\pm2\) application and pre-relationship estimates provide a separate near-cutoff check.

\begin{figure}[!htbp]
  \begin{center}
  \includegraphics[width = 0.85\textwidth, height = 0.42\textheight, keepaspectratio]{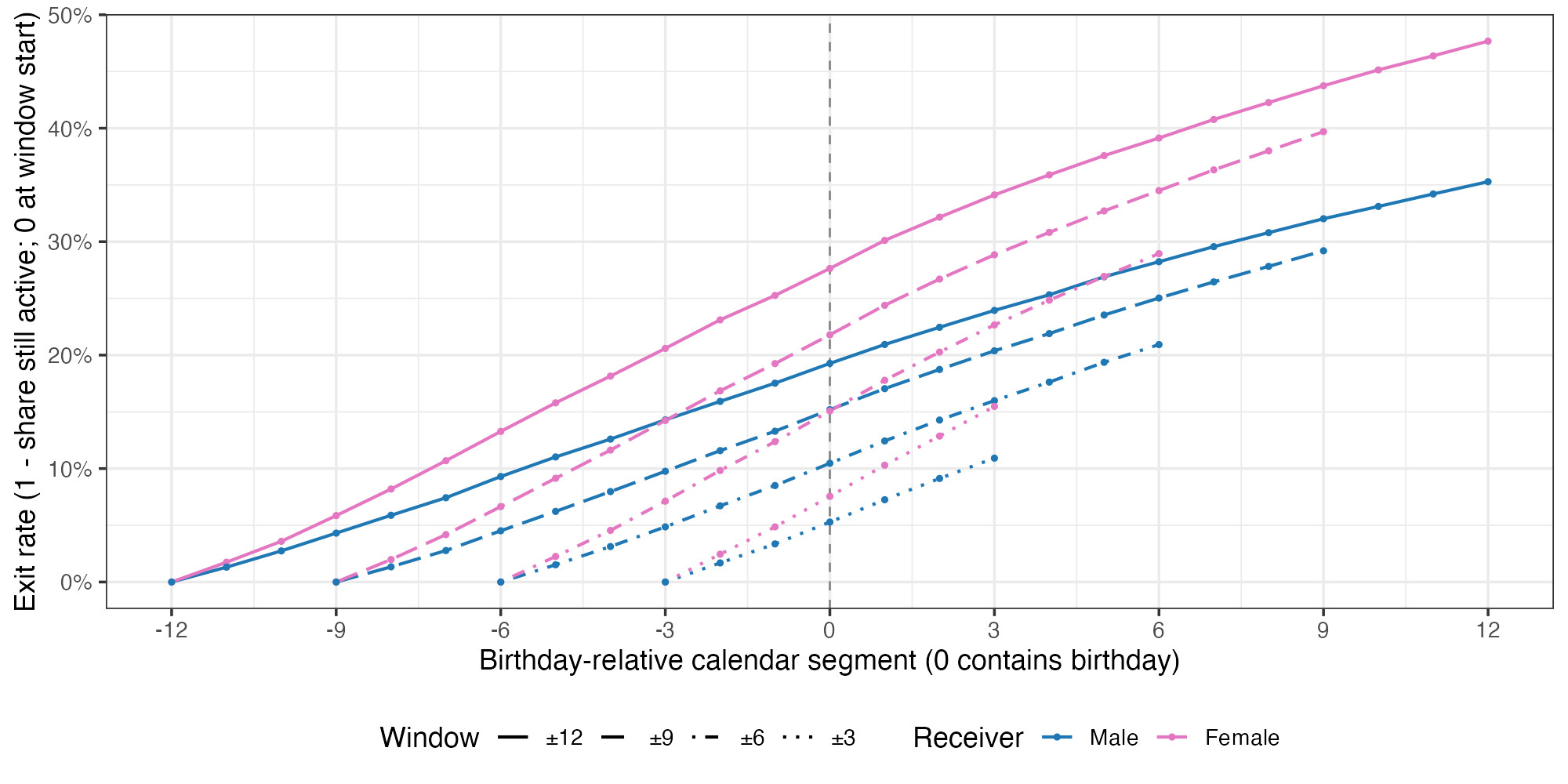}
  \caption{Platform Exit Rate Across Birthday Windows (\(\pm 3\), \(\pm 6\), \(\pm 9\), \(\pm 12\))}
  \label{fg:robust_window_exit}
  \end{center}
  \footnotesize
  Note: For each birthday-relative window with half-width \(W\), the sample is the members active at the window start (segment \(-W\)); the curve plots one minus the share of that cohort still active at each later segment, so every window begins at zero exit at its own start. The figure splits series by window and receiver gender. A birthday-driven exit would appear as a kink at segment zero common to all windows; none is present.
\end{figure}

A related concern is that conditioning on members active around the birthday could itself be post-treatment if the birthday drove platform exit. Figure~\ref{fg:robust_window_exit} addresses this directly. For each window it takes the members active at the window start and traces the share that has since left the platform, normalizing each cohort to zero exit at its own starting segment. The exit-rate curves rise smoothly and monotonically with no break at the birthday: each cohort loses a roughly constant share of members per segment, the narrower windows trace the early part of the same path that the wider windows extend, and women exit somewhat faster than men at every horizon. Because there is no discontinuous exit response just after the birthday segment, conditioning on the active sample does not appear to induce a post-treatment selection at the cutoff; the smoothly rising exit curves speak to the absence of a birthday jump in exit rather than to full independence of exit from the birthday.

\subsection{Calendar Controls: Month-of-Year and Year-Month Fixed Effects}\label{app:robust_fe_choice}

The main specification absorbs member, calendar-year, platform-tenure, and within-month-position fixed effects. The last control addresses the first/middle/final-segment pattern documented in Appendix~\ref{app:calendar_month_pattern}. In the six diagnostic application cells defined by gender and receiver ages 30, 35, and 40, adding this control changes \(\tau\) by at most \(0.007\), compared with standard errors of \(0.061\)--\(0.095\). Thus, the control absorbs an observed calendar pattern while leaving these estimates nearly unchanged.

Month-of-year effects address seasonal variation distinct from the within-month request cycle. Adding them to the main controls changes application-stage post-birthday coefficients by at most \(0.0123\); the resulting series correlate \(0.9997\) with the main estimates and have a mean absolute change of \(0.0035\). The reported specification therefore retains member, calendar-year, within-month-position, and tenure fixed effects.

\begin{figure}[!htbp]
  \begin{center}
  \includegraphics[width = 0.98\textwidth]{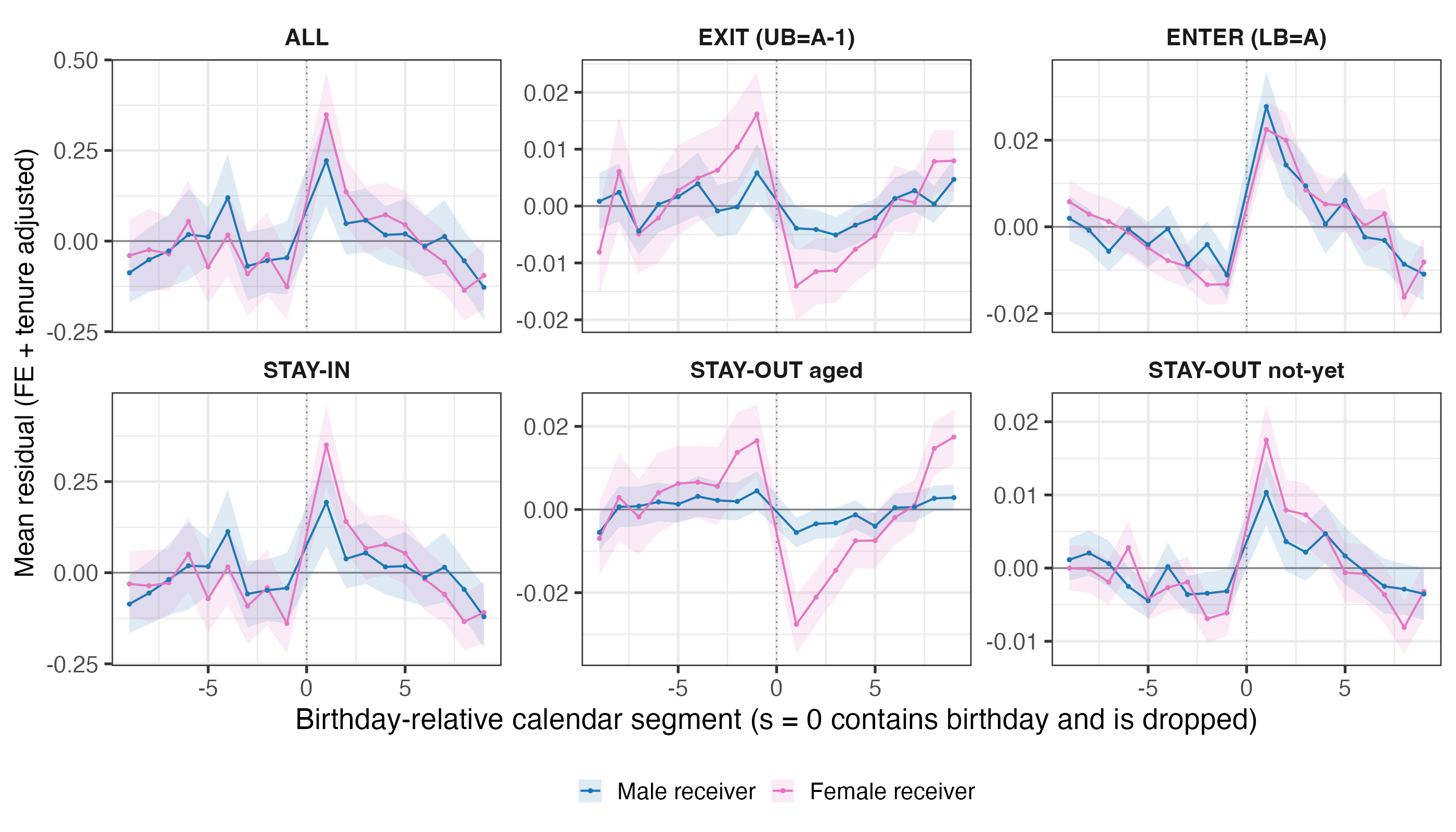}
  \caption{Birthday-Relative Application Patterns by Proposer Eligibility Type, with Month-of-Year Fixed Effects}
  \label{fg:event_study_age35_eligibility_cm}
  \end{center}
  \footnotesize
  Note: Month-of-year fixed-effect robustness counterpart of Figure~\ref{fg:event_study_age35_eligibility}, adding month-of-year fixed effects on top of the main controls (member, calendar-year, within-month-position, and tenure). Tenure-adjusted event study at receiver age 35 by proposer eligibility type.
\end{figure}

\begin{figure}[!htbp]
  \begin{center}
  \includegraphics[width = 1.0\textwidth]{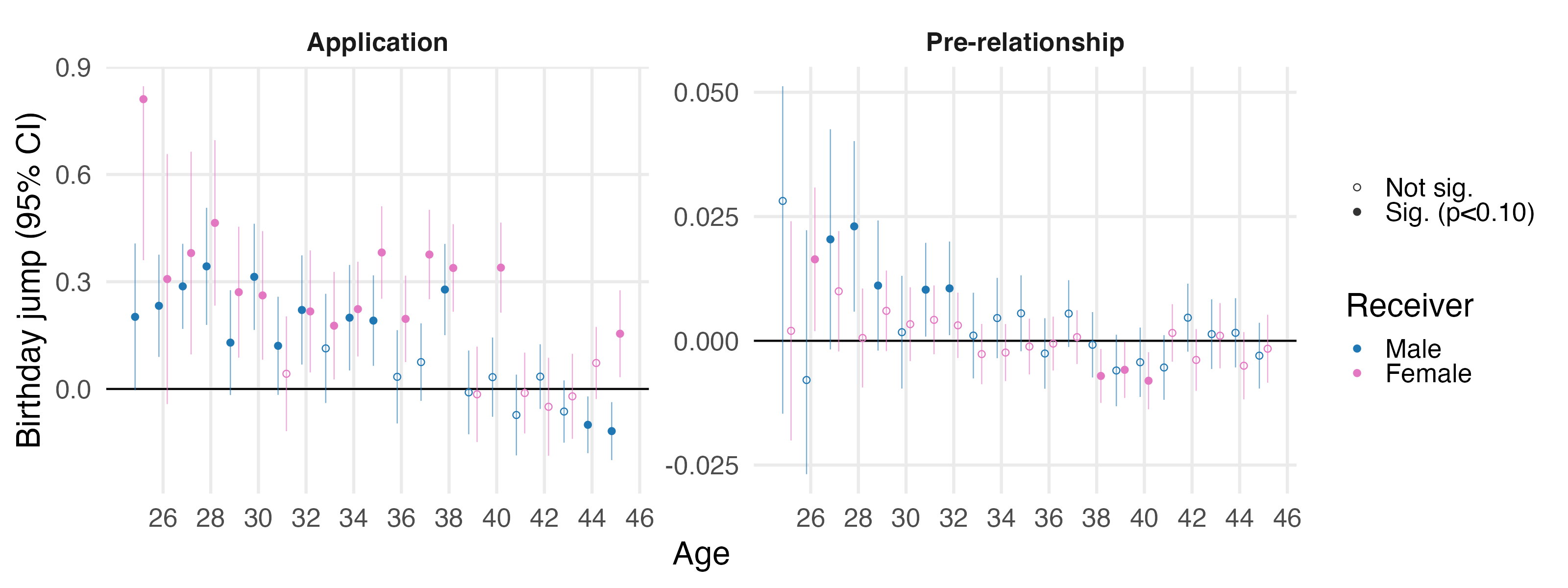}
  \caption{Birthday Jump at Application and Pre-relationship Stages, with Month-of-Year Fixed Effects}
  \label{fg:eventstudy_receiver_application_heatmap_cm}
  \end{center}
  \footnotesize
  Note: Month-of-year fixed-effect robustness counterpart of Figure~\ref{fg:eventstudy_receiver_application_heatmap}, adding month-of-year fixed effects on top of the main controls. Each point is the post-birthday coefficient by receiver age. Filled markers are significant at the 10\% level.
\end{figure}

\begin{figure}[!htbp]
  \begin{center}
  \includegraphics[width = 0.98\textwidth]{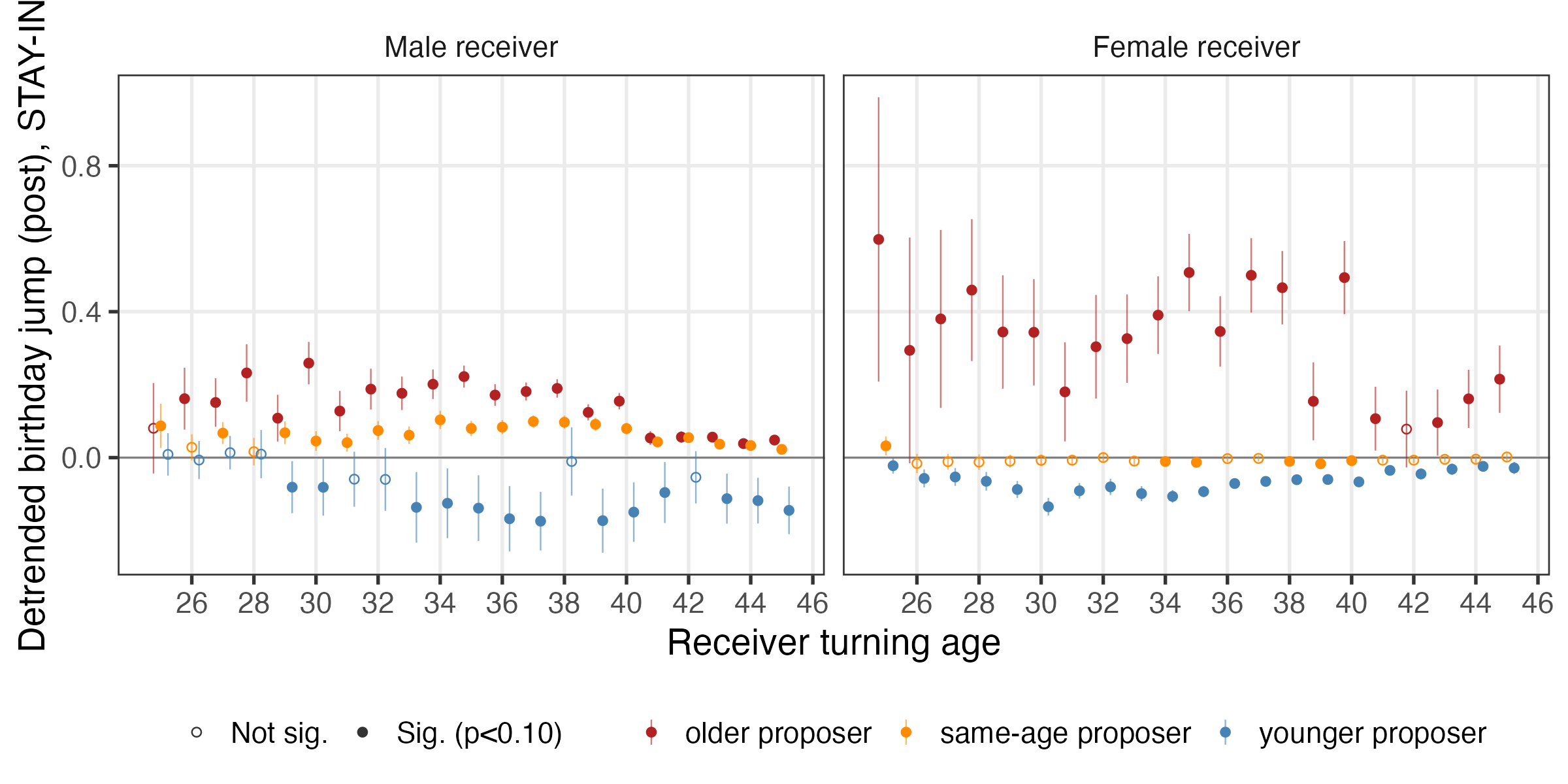}
  \caption{Proposer-Age Re-sorting at the Birthday within STAY-IN, with Month-of-Year Fixed Effects}
  \label{fg:resorting_E_cm}
  \end{center}
  \footnotesize
  Note: Month-of-year fixed-effect robustness counterpart of Figure~\ref{fg:resorting_E}, adding month-of-year fixed effects on top of the main controls. Within STAY-IN proposers, the tenure-adjusted birthday jump is split by the proposer's own age relative to the receiver; older-cohort applications rise and younger-cohort applications fall.
\end{figure}

Figures~\ref{fg:event_study_age35_eligibility_cm}--\ref{fg:resorting_E_cm} and Table~\ref{tab:eligibility_E_cm} reproduce the central receiver-side results with month-of-year fixed effects. The estimates remain very similar: EXIT falls, ENTER rises, STAY-IN carries most of the pooled jump, and applications re-sort toward older-cohort proposers.

\begin{table}[!htbp]
  \caption{Birthday Jump by Eligibility Type, Application Stage, Female Receivers, with Month-of-Year Fixed Effects}
  \label{tab:eligibility_E_cm}
  \begin{center}
  \resizebox{\textwidth}{!}{\input{figuretable/birthday_project/eventstudy_receiver_application_eligibility_E_female_calendarmonth.tex}}
  \par\vspace{0.4em}
  \begin{minipage}{0.95\textwidth}
  \footnotesize
  Note: Month-of-year fixed-effect robustness counterpart of Table~\ref{tab:eventstudy_receiver_application_eligibility_E}, adding month-of-year fixed effects on top of the main controls. Each cell reports the within-member application-stage birthday jump by receiver age and eligibility cell for female receivers, with standard errors in parentheses. STAYIN denotes eligibility on both sides of the cutoff; the two STAY-OUT columns are proposers ineligible on both sides (aged out or not yet reached). EXIT, ENTER, STAYIN, and the two STAY-OUT columns are mutually exclusive and sum to ALL, the all-proposer column, in each row.
  \end{minipage}
  \end{center}
\end{table}

Calendar-year and month-of-year fixed effects absorb a common annual level and a seasonal pattern shared across years, but not shocks specific to a calendar year-month. As a stronger check, I replace calendar-year effects with specific year-month effects \(\delta_{ym}\), leaving the remaining controls unchanged. Table~\ref{tab:eligibility_yearmonth} shows that the main and year-month profiles correlate above \(0.99\) for both genders. Mean absolute changes are about \(0.01\) for both genders, and maximum changes are \(0.03\) for female receivers and \(0.06\) for male receivers. The median jump is essentially unchanged, at \(0.22\) for women and \(0.12\) for men. Calendar year-month shocks therefore do not drive the application response.

\begin{table}[!htbp]
  \caption{Application-Stage Birthday Jump under Specific Year-Month Fixed Effects}
  \label{tab:eligibility_yearmonth}
  \begin{center}
  \small
  \input{figuretable/birthday_project/eventstudy_receiver_application_jump_yearmonth_compare}
  \par\vspace{0.4em}
  \begin{minipage}{0.9\textwidth}
  \footnotesize
  Note: Each cell is the within-member application-stage birthday jump for all proposers by receiver turning age, with standard errors in parentheses. ``Main'' is the reported specification, which absorbs member, calendar-year, within-month-position, and tenure fixed effects. The \(+\,\delta_{ym}\) columns replace the calendar-year fixed effect with specific year-month fixed effects (the absolute year-month index, which nests the calendar year), leaving the other controls unchanged.
  \end{minipage}
  \end{center}
\end{table}

\subsection{Birth-Month Anticipation}\label{app:robust_anticipation}

\begin{figure}[!htbp]
  \begin{center}
  \includegraphics[width = 0.95\textwidth, height = 0.34\textheight, keepaspectratio]{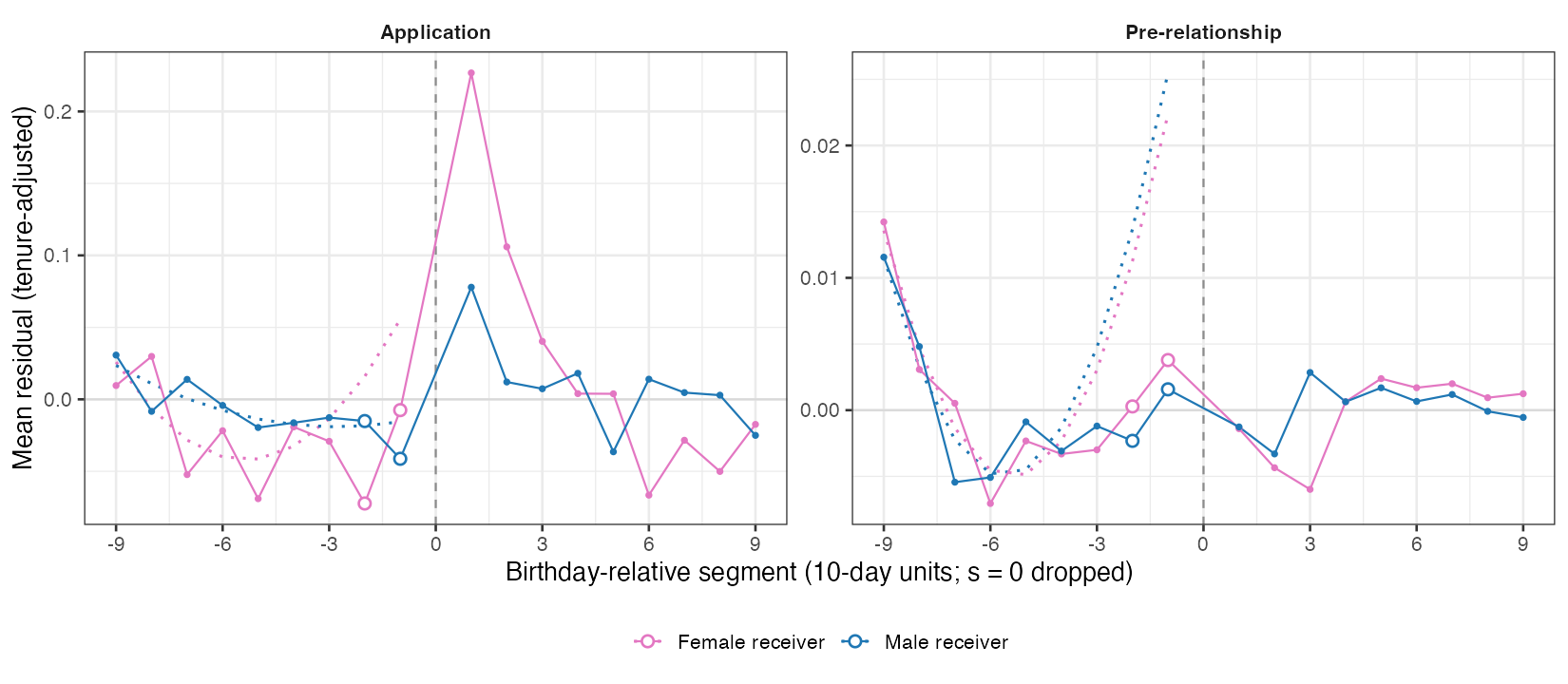}
  \caption{Birth-Month Anticipation Check: Tenure-Adjusted Inbound Response by Birthday-Relative Segment}
  \label{fg:robust_anticipation}
  \end{center}
  \footnotesize
  Note: Tenure-adjusted residual inbound count averaged by birthday-relative segment, by receiver gender and funnel stage. The dotted line extrapolates the far-pre-trend, fit on segments \(-9\) to \(-4\), into the near-pre period; open markers are the last two pre-birthday segments. The series account for member fixed effects, calendar-year fixed effects, the within-month calendar position, and the platform-tenure profile, and omit the segment containing the birthday.
\end{figure}

The exact birthday is masked, but birth month and year are visible after a click, so proposers could respond before the birthday within the birth month. Figure~\ref{fg:robust_anticipation} tests for such a run-up using the same tenure-adjusted series as the main figures. For applications, the first post-birthday coefficient is 3.07 times the largest absolute pre-birthday coefficient for female receivers and 3.23 times that for male receivers. Both near-pre coefficients are negative for each gender, providing no evidence of a positive run-up. A finer split by birthday position within the month is not separately identified because, conditional on that position, the within-month activity cycle is a deterministic function of event time.

\subsection{Pseudo-Birthday Placebo}\label{app:robust_placebo}

For each member, the placebo draws an artificial cutoff five to eight segments from the birthday within the observed \([-12,12]\) support and estimates a post-cutoff change in a symmetric \(\pm4\)-segment window that excludes the true birthday. Across 300 repetitions, the true application jump exceeds the placebo 95th percentile at 15 of 21 female ages and 10 of 20 male ages (the male age-25 cell lacks sufficient support for the placebo re-estimation); pooled centered placebo \(p\)-values are \(0.003\) for both genders.\footnote{\textcolor{black}{In each repetition, every eligible member draws uniformly from the feasible signed cutoffs \(\{-8,-7,-6,-5,5,6,7,8\}\). The pooled statistic re-estimates the gender-specific sample across all ages. The centered \(p\)-value is the two-sided randomization-style tail probability around the placebo mean, \([1+\sum_r\mathbbm{1}\{|\tau_r-\bar\tau_P|\geq|\widehat\tau-\bar\tau_P|\}]/(R+1)\); the age-specific 95th-percentile comparison is an upper-tail comparison.}} Pre-relationship estimates show no comparable systematic departure from the placebo distribution.

\begin{figure}[!htbp]
  \begin{center}
  \includegraphics[width = 0.85\textwidth, height = 0.52\textheight, keepaspectratio]{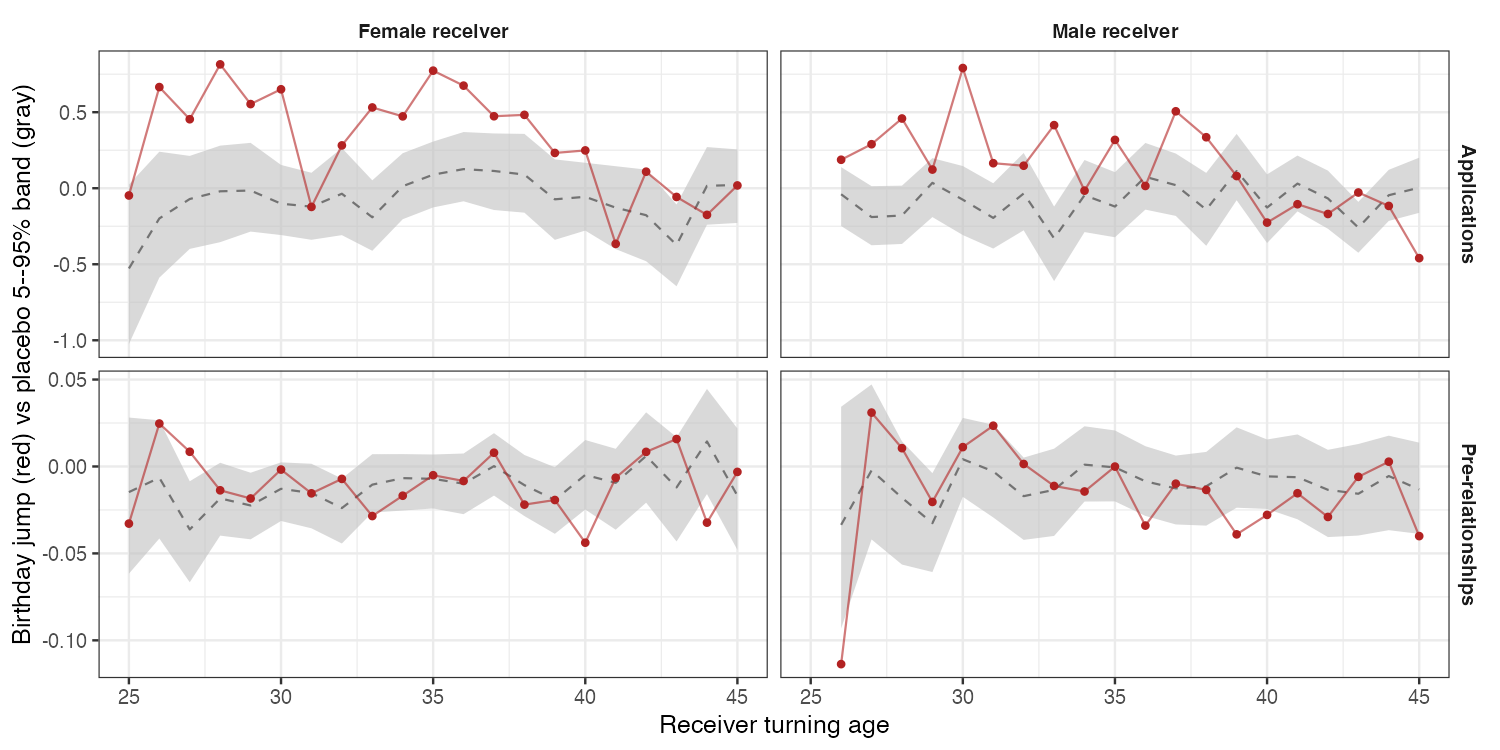}
  \caption{Pseudo-Birthday Placebo: Birthday Jump versus Placebo Distribution by Receiver Age}
  \label{fg:robust_placebo}
  \end{center}
  \footnotesize
  Note: Red points are true post-cutoff changes estimated in a symmetric \(\pm4\)-segment window; the gray band and dashed line are the placebo 5--95\% range and median from 300 draws. Applications and pre-relationships use the tenure-adjusted within-member estimator.
\end{figure}

\end{document}

%% file: figuretable/birthday_project/summary_statistics_male_female_receiver.tex
\begin{tabular}[t]{lrrrrrrrrrr}
\toprule
  & N & Mean & SD & Min & Max & N & Mean & SD & Min & Max\\
\midrule
 & Male &  &  &  &  & Female &  &  &  & \\
Age & 103784 & 39.40 & 8.62 & 20.00 & 93.00 & 138024 & 35.71 & 7.56 & 19.00 & 84.00\\
Income (upper limit) & 103410 & 753.95 & 351.43 & 300.00 & 2100.00 & 136751 & 490.90 & 214.06 & 300.00 & 2100.00\\
Height (cm) & 103784 & 171.44 & 5.76 & 142.00 & 198.00 & 138024 & 158.81 & 5.24 & 142.00 & 188.00\\
Upper age pref & 57945 & 38.92 & 7.02 & 20.00 & 90.00 & 81546 & 43.27 & 7.68 & 20.00 & 90.00\\
Lower age pref & 31350 & 29.47 & 6.52 & 20.00 & 69.00 & 54077 & 32.40 & 6.93 & 20.00 & 80.00\\
Application & 103784 & 70.05 & 158.04 & 1.00 & 6357.00 & 138024 & 122.21 & 160.73 & 1.00 & 3760.00\\
Pre-relation & 103784 & 2.06 & 3.14 & 0.00 & 105.00 & 138024 & 1.96 & 2.75 & 0.00 & 73.00\\
Serious & 103784 & 0.20 & 0.44 & 0.00 & 5.00 & 138024 & 0.17 & 0.40 & 0.00 & 5.00\\
Proposal & 103784 & 0.13 & 0.33 & 0.00 & 1.00 & 138024 & 0.11 & 0.31 & 0.00 & 1.00\\
Tenure (days) & 74477 & 417.38 & 348.63 & 1.00 & 1795.00 & 113465 & 357.74 & 317.25 & 1.00 & 1796.00\\
\bottomrule
\end{tabular}

%% file: figuretable/birthday_project/summary_statistics_male_female_proposer.tex
\begin{tabular}[t]{lrrrrrrrrrr}
\toprule
  & N & Mean & SD & Min & Max & N & Mean & SD & Min & Max\\
\midrule
 & Male &  &  &  &  & Female &  &  &  & \\
Age & 99974 & 38.88 & 7.74 & 23.00 & 60.00 & 116190 & 35.38 & 7.01 & 23.00 & 60.00\\
Income (upper limit) & 99659 & 734.01 & 342.84 & 300.00 & 2100.00 & 115177 & 499.28 & 216.12 & 300.00 & 2100.00\\
Height (cm) & 99974 & 171.32 & 5.82 & 142.00 & 198.00 & 116190 & 158.87 & 5.24 & 142.00 & 184.00\\
Upper age pref & 56135 & 38.72 & 6.48 & 20.00 & 90.00 & 69815 & 43.01 & 7.30 & 20.00 & 90.00\\
Lower age pref & 30391 & 29.04 & 5.79 & 20.00 & 61.00 & 45927 & 32.10 & 6.45 & 20.00 & 70.00\\
Application & 99974 & 169.85 & 439.12 & 1.00 & 43707.00 & 116190 & 63.16 & 125.15 & 1.00 & 5511.00\\
Pre-relation & 99974 & 2.72 & 4.62 & 0.00 & 112.00 & 116190 & 1.85 & 3.06 & 0.00 & 91.00\\
Serious & 99974 & 0.23 & 0.47 & 0.00 & 5.00 & 116190 & 0.18 & 0.41 & 0.00 & 5.00\\
Proposal & 99974 & 0.15 & 0.35 & 0.00 & 1.00 & 116190 & 0.11 & 0.32 & 0.00 & 1.00\\
\bottomrule
\end{tabular}

%% file: figuretable/birthday_project/stated_vs_actual_target_age_containment.tex
\begin{tabular}{lcc}
\toprule
  & Male Proposer & Female Proposer\\
\midrule
Stated $\supseteq$ Saved (incl.\ identical) & 71.2\% & 71.4\%\\
Saved $\supsetneq$ Stated & 12.5\% & 12.0\%\\
Partial overlap & 15.1\% & 16.1\%\\
Disjoint & 1.2\% & 0.5\%\\
$N$ (proposers) & 23,722 & 16,726\\
\bottomrule
\end{tabular}

%% file: figuretable/birthday_project/eventstudy_receiver_application_eligibility_E_female.tex
\begin{tabular}[t]{rllllll}
\toprule
age & ALL & EXIT & ENTER & STAYIN & STAYOUT (aged) & STAYOUT (not-yet)\\
\midrule
25 & 0.81 (0.23) & -0.00 (0.00) & 0.10 (0.02) & 0.61 (0.21) & -0.00 (0.00) & 0.10 (0.02)\\
26 & 0.31 (0.18) & 0.00 (0.00) & 0.02 (0.01) & 0.22 (0.17) & 0.00 (0.00) & 0.07 (0.02)\\
27 & 0.39 (0.14) & -0.00 (0.00) & 0.03 (0.01) & 0.33 (0.13) & 0.00 (0.00) & 0.04 (0.01)\\
28 & 0.47 (0.12) & -0.00 (0.00) & 0.05 (0.01) & 0.38 (0.11) & -0.00 (0.00) & 0.04 (0.01)\\
29 & 0.27 (0.09) & -0.02 (0.00) & 0.01 (0.00) & 0.24 (0.09) & -0.00 (0.00) & 0.03 (0.01)\\
30 & 0.26 (0.09) & -0.02 (0.00) & 0.07 (0.01) & 0.20 (0.09) & -0.01 (0.00) & 0.02 (0.00)\\
31 & 0.04 (0.08) & -0.03 (0.00) & 0.00 (0.00) & 0.08 (0.08) & -0.02 (0.00) & 0.01 (0.00)\\
32 & 0.22 (0.09) & -0.01 (0.00) & 0.01 (0.00) & 0.22 (0.08) & -0.02 (0.00) & 0.02 (0.00)\\
33 & 0.17 (0.08) & -0.04 (0.00) & 0.01 (0.00) & 0.22 (0.07) & -0.03 (0.00) & 0.02 (0.00)\\
34 & 0.22 (0.07) & -0.05 (0.00) & 0.01 (0.00) & 0.27 (0.06) & -0.03 (0.00) & 0.02 (0.00)\\
35 & 0.39 (0.07) & -0.03 (0.00) & 0.04 (0.00) & 0.41 (0.06) & -0.05 (0.00) & 0.02 (0.00)\\
36 & 0.19 (0.06) & -0.07 (0.01) & 0.01 (0.00) & 0.27 (0.05) & -0.04 (0.00) & 0.02 (0.00)\\
37 & 0.38 (0.06) & -0.02 (0.00) & 0.01 (0.00) & 0.43 (0.06) & -0.06 (0.01) & 0.02 (0.00)\\
38 & 0.34 (0.06) & -0.03 (0.00) & 0.02 (0.00) & 0.39 (0.06) & -0.06 (0.01) & 0.02 (0.00)\\
39 & -0.02 (0.07) & -0.03 (0.01) & 0.00 (0.00) & 0.08 (0.06) & -0.07 (0.01) & 0.01 (0.00)\\
40 & 0.34 (0.06) & -0.05 (0.01) & 0.03 (0.00) & 0.42 (0.06) & -0.07 (0.01) & 0.02 (0.00)\\
41 & -0.01 (0.06) & -0.04 (0.01) & 0.00 (0.00) & 0.07 (0.05) & -0.05 (0.01) & 0.01 (0.00)\\
42 & -0.05 (0.07) & -0.01 (0.00) & 0.00 (0.00) & 0.02 (0.06) & -0.07 (0.01) & 0.00 (0.00)\\
43 & -0.02 (0.06) & -0.03 (0.01) & 0.00 (0.00) & 0.06 (0.05) & -0.07 (0.01) & 0.01 (0.00)\\
44 & 0.07 (0.05) & -0.02 (0.01) & 0.00 (0.00) & 0.13 (0.04) & -0.06 (0.01) & 0.02 (0.00)\\
45 & 0.16 (0.06) & -0.01 (0.00) & 0.02 (0.00) & 0.19 (0.05) & -0.05 (0.01) & 0.01 (0.00)\\
\bottomrule
\end{tabular}

%% file: figuretable/birthday_project/counterfactual_table_male.tex
\begin{tabular}[t]{lrrr}
\toprule
Stage & Benchmark & CF1 (No stated age bounds) & CF2 (No birthday)\\
\midrule
Potential Proposer & 3972780 & 5262730 [+32.5\%] & 3969415 [-0.1\%]\\
Application & 822299 & 1055332 [+28.3\%] & 788220 [-4.1\%]\\
Pre-relationship & 21528 & 27854 [+29.4\%] & 20765 [-3.5\%]\\
Serious relationship & 1840 & 2372 [+28.9\%] & 1772 [-3.7\%]\\
Proposal & 1230 & 1586 [+28.9\%] & 1166 [-5.2\%]\\
\bottomrule
\end{tabular}

%% file: figuretable/birthday_project/counterfactual_table_female.tex
\begin{tabular}[t]{lrrr}
\toprule
Stage & Benchmark & CF1 (No stated age bounds) & CF2 (No birthday)\\
\midrule
Potential Proposer & 3838550 & 5139429 [+33.9\%] & 3888050 [+1.3\%]\\
Application & 1900131 & 2499021 [+31.5\%] & 1763720 [-7.2\%]\\
Pre-relationship & 29083 & 37829 [+30.1\%] & 29456 [+1.3\%]\\
Serious relationship & 2219 & 2870 [+29.3\%] & 2345 [+5.7\%]\\
Proposal & 1435 & 1863 [+29.9\%] & 1552 [+8.2\%]\\
\bottomrule
\end{tabular}

%% file: figuretable/birthday_project/counterfactual_headline_summary.tex
\begin{tabular}[t]{llccc}
\toprule
Quantity & Receiver & Benchmark & CF2 (No birthday) & Change\\
\midrule
Fitted proposals, ages 30--39 & Female & 944 & 1,066 & $+12.9\%$\\
 & Male & 792 & 739 & $-6.7\%$\\
Younger-suitor share, applications, ages 25--45 & Female & 6.2\% & 10.4\% & $+4.2$ pp\\
 & Male & 64.6\% & 75.2\% & $+10.6$ pp\\
Younger-suitor share, proposals, ages 25--45 & Female & 15.5\% & 19.7\% & $+4.2$ pp\\
 & Male & 77.5\% & 84.1\% & $+6.7$ pp\\
Younger-suitor share, proposals, ages 30--39 & Female & 17.1\% & 21.9\% & $+4.8$ pp\\
\bottomrule
\end{tabular}

%% file: figuretable/birthday_project/event_time_support_by_age_stage_gender.tex
\begin{tabular}[t]{rrrrrrrrr}
\toprule
\multicolumn{1}{c}{ } & \multicolumn{4}{c}{Male receiver} & \multicolumn{4}{c}{Female receiver} \\
\cmidrule(l{3pt}r{3pt}){2-5} \cmidrule(l{3pt}r{3pt}){6-9}
Age & App. & Pre-rel. & Serious & Proposal & App. & Pre-rel. & Serious & Proposal\\
\midrule
25 & 180 & 70 & 11 & 7 & 1,212 & 485 & 66 & 48\\
26 & 529 & 211 & 25 & 16 & 2,132 & 834 & 122 & 90\\
27 & 990 & 443 & 71 & 45 & 3,331 & 1,371 & 171 & 118\\
28 & 1,666 & 723 & 101 & 69 & 4,953 & 2,094 & 254 & 159\\
29 & 2,421 & 1,114 & 146 & 102 & 6,779 & 2,888 & 366 & 236\\
30 & 3,311 & 1,511 & 207 & 135 & 8,101 & 3,436 & 459 & 302\\
31 & 3,763 & 1,669 & 247 & 168 & 7,916 & 3,234 & 376 & 243\\
32 & 4,170 & 1,957 & 267 & 185 & 8,570 & 3,482 & 425 & 294\\
33 & 4,672 & 2,087 & 306 & 211 & 9,119 & 3,674 & 414 & 273\\
34 & 5,380 & 2,479 & 338 & 227 & 9,921 & 3,932 & 401 & 265\\
35 & 6,035 & 2,679 & 362 & 257 & 10,460 & 4,090 & 456 & 308\\
36 & 6,248 & 2,714 & 358 & 232 & 9,965 & 3,717 & 381 & 251\\
37 & 6,203 & 2,574 & 313 & 206 & 9,215 & 3,334 & 372 & 235\\
38 & 6,009 & 2,488 & 304 & 186 & 8,663 & 3,023 & 310 & 213\\
39 & 5,987 & 2,593 & 331 & 226 & 8,094 & 2,848 & 274 & 192\\
40 & 5,833 & 2,461 & 275 & 186 & 7,545 & 2,592 & 277 & 200\\
41 & 5,177 & 2,012 & 203 & 134 & 5,808 & 1,906 & 175 & 117\\
42 & 4,784 & 1,845 & 210 & 153 & 4,989 & 1,549 & 128 & 92\\
43 & 4,531 & 1,607 & 201 & 135 & 4,300 & 1,235 & 118 & 84\\
44 & 4,216 & 1,552 & 182 & 128 & 3,798 & 1,085 & 110 & 75\\
45 & 4,168 & 1,458 & 152 & 100 & 3,342 & 893 & 75 & 52\\
\midrule
25--45 & 86,273 & 36,247 & 4,610 & 3,108 & 138,213 & 51,702 & 5,730 & 3,847\\
\bottomrule
\end{tabular}

%% file: figuretable/birthday_project/eventstudy_receiver_application_eligibility_E_male.tex
\begin{tabular}[t]{rllllll}
\toprule
age & ALL & EXIT & ENTER & STAYIN & STAYOUT (aged) & STAYOUT (not-yet)\\
\midrule
25 & 0.20 (0.11) & NA (NA) & 0.01 (0.02) & 0.17 (0.09) & NA (NA) & 0.02 (0.02)\\
26 & 0.24 (0.07) & NA (NA) & 0.02 (0.01) & 0.19 (0.06) & 0.00 (0.00) & 0.03 (0.02)\\
27 & 0.29 (0.06) & NA (NA) & 0.02 (0.01) & 0.23 (0.05) & 0.00 (0.00) & 0.04 (0.01)\\
28 & 0.34 (0.08) & 0.00 (0.00) & 0.05 (0.01) & 0.25 (0.08) & 0.00 (0.00) & 0.03 (0.01)\\
29 & 0.14 (0.07) & -0.00 (0.00) & 0.02 (0.00) & 0.11 (0.07) & -0.00 (0.00) & 0.02 (0.01)\\
30 & 0.32 (0.07) & -0.00 (0.00) & 0.08 (0.01) & 0.23 (0.07) & -0.00 (0.00) & 0.02 (0.00)\\
31 & 0.12 (0.07) & -0.00 (0.00) & 0.01 (0.00) & 0.11 (0.07) & -0.00 (0.00) & 0.01 (0.00)\\
32 & 0.23 (0.08) & -0.00 (0.00) & 0.01 (0.00) & 0.21 (0.08) & -0.00 (0.00) & 0.01 (0.00)\\
33 & 0.12 (0.08) & -0.01 (0.00) & 0.01 (0.00) & 0.10 (0.08) & -0.00 (0.00) & 0.01 (0.00)\\
34 & 0.21 (0.08) & -0.01 (0.00) & 0.01 (0.00) & 0.19 (0.07) & -0.01 (0.00) & 0.02 (0.00)\\
35 & 0.19 (0.06) & -0.01 (0.00) & 0.04 (0.00) & 0.17 (0.06) & -0.01 (0.00) & 0.01 (0.00)\\
36 & 0.04 (0.07) & -0.05 (0.01) & 0.01 (0.00) & 0.09 (0.06) & -0.02 (0.00) & 0.01 (0.00)\\
37 & 0.07 (0.06) & -0.02 (0.00) & 0.01 (0.00) & 0.10 (0.05) & -0.03 (0.00) & 0.01 (0.00)\\
38 & 0.28 (0.07) & -0.02 (0.00) & 0.02 (0.00) & 0.28 (0.06) & -0.01 (0.00) & 0.01 (0.00)\\
39 & -0.00 (0.06) & -0.02 (0.00) & 0.00 (0.00) & 0.05 (0.06) & -0.03 (0.00) & 0.01 (0.00)\\
40 & 0.03 (0.06) & -0.04 (0.01) & 0.01 (0.00) & 0.09 (0.05) & -0.02 (0.00) & 0.00 (0.00)\\
41 & -0.07 (0.06) & -0.04 (0.01) & 0.00 (0.00) & 0.00 (0.05) & -0.04 (0.01) & 0.00 (0.00)\\
42 & 0.04 (0.05) & -0.01 (0.00) & 0.00 (0.00) & 0.06 (0.04) & -0.02 (0.01) & 0.00 (0.00)\\
43 & -0.06 (0.04) & -0.02 (0.00) & 0.00 (0.00) & -0.02 (0.04) & -0.03 (0.01) & 0.00 (0.00)\\
44 & -0.10 (0.04) & -0.02 (0.00) & 0.00 (0.00) & -0.04 (0.04) & -0.04 (0.01) & 0.00 (0.00)\\
45 & -0.12 (0.04) & -0.01 (0.00) & 0.01 (0.00) & -0.07 (0.04) & -0.04 (0.01) & 0.00 (0.00)\\
\bottomrule
\end{tabular}

%% file: figuretable/birthday_project/who_ended_pair_attribution_pooled.tex
\begin{tabular}[t]{lcccccc}
\toprule
\multicolumn{1}{c}{ } & \multicolumn{2}{c}{Female 25-45} & \multicolumn{2}{c}{Female 31-40} & \multicolumn{2}{c}{Male 25-45} \\
\cmidrule(l{3pt}r{3pt}){2-3} \cmidrule(l{3pt}r{3pt}){4-5} \cmidrule(l{3pt}r{3pt}){6-7}
Outcome & Pre mean & Jump (SE) & Pre mean & Jump (SE) & Pre mean & Jump (SE)\\
\midrule
Receiver rejects the meeting request & 0.637 & -0.0067 (0.0008) & 0.639 & -0.0075 (0.0010) & 0.674 & -0.0045 (0.0015)\\
Meeting held per application & 0.048 & -0.0075 (0.0003) & 0.049 & -0.0085 (0.0003) & 0.082 & -0.0082 (0.0005)\\
Receiver declines after the meeting & 0.579 & 0.0041 (0.0026) & 0.580 & 0.0065 (0.0031) & 0.487 & 0.0067 (0.0029)\\
Proposer declines after the meeting & 0.421 & -0.0085 (0.0026) & 0.426 & -0.0081 (0.0032) & 0.531 & -0.0015 (0.0029)\\
Pre-relationship forms per meeting & 0.337 & 0.0012 (0.0024) & 0.334 & 0.0005 (0.0029) & 0.336 & 0.0015 (0.0027)\\
\bottomrule
\end{tabular}

%% file: figuretable/birthday_project/delta2_reweighting_pooled.tex
\begin{tabular}[t]{lcccccc}
\toprule
Cell & Rate, pre & Rate, post & Change & Composition & Within-type & Composition share\\
\midrule
Female 31--40 & 0.0167 & 0.0150 & -0.0017 & -0.0004 & -0.0013 & 0.22\\
Female 25--45 & 0.0168 & 0.0156 & -0.0013 & -0.0003 & -0.0010 & 0.24\\
Male 25--45 & 0.0274 & 0.0264 & -0.0009 & -0.0010 & 0.0001 & 1.07\\
\bottomrule
\end{tabular}

%% file: figuretable/birthday_project/realized_match_composition_pooled.tex
\begin{tabular}[t]{llccc}
\toprule
Outcome & Stage & Female 25--45 & Female 31--40 & Male 25--45\\
\midrule
Counterparty age gap (years) & Application & 0.4801 (0.0147) & 0.5815 (0.0167) & 0.6053 (0.0154)\\
 & Pre-relationship & 0.3131 (0.0361) & 0.3683 (0.0446) & 0.4424 (0.0338)\\
 & Serious relationship & 0.2227 (0.1085) & 0.2393 (0.1377) & 0.3084 (0.0983)\\
 & Proposal & 0.0706 (0.1317) & 0.1510 (0.1678) & 0.4064 (0.1197)\\
Older-cohort counterparty share & Application & 0.0260 (0.0005) & 0.0300 (0.0006) & 0.0509 (0.0014)\\
 & Pre-relationship & 0.0255 (0.0033) & 0.0285 (0.0041) & 0.0257 (0.0029)\\
 & Serious relationship & 0.0316 (0.0116) & 0.0312 (0.0147) & 0.0320 (0.0090)\\
 & Proposal & 0.0155 (0.0143) & 0.0254 (0.0182) & 0.0370 (0.0110)\\
Counterparty wants children, share & Application & -0.0226 (0.0012) & -0.0333 (0.0011) & -0.0140 (0.0010)\\
 & Pre-relationship & -0.0197 (0.0034) & -0.0278 (0.0039) & -0.0126 (0.0034)\\
 & Serious relationship & -0.0144 (0.0109) & -0.0176 (0.0132) & -0.0245 (0.0107)\\
 & Proposal & -0.0173 (0.0134) & -0.0284 (0.0163) & -0.0218 (0.0132)\\
\bottomrule
\end{tabular}

%% file: figuretable/birthday_project/pooled_jump_inference.tex
\begin{tabular}[t]{llcccc}
\toprule
Cell & Receiver & Ages & Pre mean & Jump (SE) & Age-homogeneity $p$\\
\midrule
All proposers, application & Female & 25--45 & 2.984 & 0.21 (0.02) & $<0.001$\\
 & Female & 31--40 & 3.058 & 0.23 (0.02) & 0.001\\
 & Male & 25--45 & 1.848 & 0.08 (0.01) & $<0.001$\\
All proposers, pre-relationship & Female & 25--45 & 0.048 & -0.0007 (0.0008) & 0.014\\
 & Female & 31--40 & 0.049 & -0.0020 (0.0009) & 0.381\\
 & Male & 25--45 & 0.049 & 0.0019 (0.0009) & $<0.001$\\
STAY-IN, older proposers, application & Female & 25--45 & 2.298 & 0.33 (0.01) & \\
 & Female & 31--40 & 2.351 & 0.37 (0.02) & \\
 & Male & 25--45 & 0.288 & 0.14 (0.0038) & \\
STAY-IN, same-age proposers, application & Female & 25--45 & 0.132 & -0.01 (0.0014) & \\
 & Female & 31--40 & 0.133 & -0.01 (0.0017) & \\
 & Male & 25--45 & 0.166 & 0.07 (0.0023) & \\
STAY-IN, younger proposers, application & Female & 25--45 & 0.266 & -0.07 (0.0021) & \\
 & Female & 31--40 & 0.281 & -0.08 (0.0026) & \\
 & Male & 25--45 & 1.267 & -0.11 (0.01) & \\
\bottomrule
\end{tabular}

%% file: figuretable/birthday_project/eventstudy_receiver_application_eligibility_E_female_calendarmonth.tex
\begin{tabular}[t]{rllllll}
\toprule
age & ALL & EXIT & ENTER & STAYIN & STAYOUT (aged) & STAYOUT (not-yet)\\
\midrule
25 & 0.81 (0.23) & -0.00 (0.00) & 0.11 (0.02) & 0.61 (0.20) & -0.00 (0.00) & 0.10 (0.02)\\
26 & 0.31 (0.18) & 0.00 (0.00) & 0.02 (0.01) & 0.22 (0.17) & 0.00 (0.00) & 0.07 (0.02)\\
27 & 0.38 (0.14) & -0.00 (0.00) & 0.03 (0.01) & 0.32 (0.14) & 0.00 (0.00) & 0.04 (0.01)\\
28 & 0.46 (0.12) & -0.00 (0.00) & 0.05 (0.01) & 0.38 (0.11) & -0.00 (0.00) & 0.04 (0.01)\\
29 & 0.27 (0.09) & -0.02 (0.00) & 0.01 (0.00) & 0.25 (0.09) & -0.00 (0.00) & 0.03 (0.01)\\
30 & 0.26 (0.09) & -0.02 (0.00) & 0.07 (0.01) & 0.20 (0.09) & -0.01 (0.00) & 0.02 (0.00)\\
31 & 0.04 (0.08) & -0.03 (0.00) & 0.00 (0.00) & 0.08 (0.08) & -0.02 (0.00) & 0.01 (0.00)\\
32 & 0.22 (0.09) & -0.01 (0.00) & 0.01 (0.00) & 0.22 (0.08) & -0.02 (0.00) & 0.02 (0.00)\\
33 & 0.18 (0.08) & -0.04 (0.00) & 0.01 (0.00) & 0.22 (0.07) & -0.03 (0.00) & 0.02 (0.00)\\
34 & 0.22 (0.07) & -0.05 (0.00) & 0.01 (0.00) & 0.27 (0.06) & -0.03 (0.00) & 0.02 (0.00)\\
35 & 0.38 (0.07) & -0.03 (0.00) & 0.04 (0.00) & 0.40 (0.06) & -0.05 (0.00) & 0.02 (0.00)\\
36 & 0.20 (0.06) & -0.07 (0.01) & 0.01 (0.00) & 0.27 (0.05) & -0.04 (0.00) & 0.02 (0.00)\\
37 & 0.38 (0.06) & -0.02 (0.00) & 0.01 (0.00) & 0.43 (0.06) & -0.06 (0.01) & 0.02 (0.00)\\
38 & 0.34 (0.06) & -0.03 (0.00) & 0.02 (0.00) & 0.39 (0.06) & -0.06 (0.01) & 0.02 (0.00)\\
39 & -0.01 (0.07) & -0.03 (0.01) & 0.00 (0.00) & 0.08 (0.06) & -0.07 (0.01) & 0.01 (0.00)\\
40 & 0.34 (0.06) & -0.05 (0.01) & 0.03 (0.00) & 0.42 (0.06) & -0.07 (0.01) & 0.02 (0.00)\\
41 & -0.01 (0.06) & -0.04 (0.01) & 0.00 (0.00) & 0.06 (0.05) & -0.05 (0.01) & 0.01 (0.00)\\
42 & -0.05 (0.07) & -0.01 (0.00) & 0.00 (0.00) & 0.03 (0.06) & -0.07 (0.01) & 0.00 (0.00)\\
43 & -0.02 (0.06) & -0.03 (0.01) & 0.00 (0.00) & 0.06 (0.05) & -0.07 (0.01) & 0.01 (0.00)\\
44 & 0.07 (0.05) & -0.01 (0.01) & 0.00 (0.00) & 0.13 (0.04) & -0.06 (0.01) & 0.02 (0.00)\\
45 & 0.15 (0.06) & -0.01 (0.00) & 0.02 (0.00) & 0.19 (0.05) & -0.05 (0.01) & 0.01 (0.00)\\
\bottomrule
\end{tabular}

%% file: figuretable/birthday_project/eventstudy_receiver_application_jump_yearmonth_compare.tex
\begin{tabular}[t]{lcccc}
\toprule
\multicolumn{1}{c}{ } & \multicolumn{2}{c}{Male receiver} & \multicolumn{2}{c}{Female receiver} \\
\cmidrule(l{3pt}r{3pt}){2-3} \cmidrule(l{3pt}r{3pt}){4-5}
Age & Main & $+\,\delta_{ym}$ & Main & $+\,\delta_{ym}$\\
\midrule
25 & 0.196 (0.106) & 0.136 (0.081) & 0.806 (0.234) & 0.837 (0.237)\\
26 & 0.239 (0.073) & 0.225 (0.075) & 0.313 (0.178) & 0.279 (0.181)\\
27 & 0.289 (0.060) & 0.294 (0.062) & 0.392 (0.144) & 0.379 (0.144)\\
28 & 0.340 (0.084) & 0.344 (0.084) & 0.465 (0.119) & 0.442 (0.117)\\
29 & 0.141 (0.074) & 0.114 (0.075) & 0.266 (0.094) & 0.252 (0.094)\\
30 & 0.320 (0.074) & 0.317 (0.075) & 0.264 (0.092) & 0.241 (0.093)\\
31 & 0.123 (0.070) & 0.120 (0.069) & 0.040 (0.082) & 0.052 (0.083)\\
32 & 0.228 (0.078) & 0.222 (0.077) & 0.217 (0.087) & 0.214 (0.087)\\
33 & 0.116 (0.078) & 0.121 (0.078) & 0.175 (0.077) & 0.183 (0.077)\\
34 & 0.207 (0.076) & 0.195 (0.077) & 0.223 (0.068) & 0.222 (0.068)\\
35 & 0.194 (0.065) & 0.193 (0.065) & 0.388 (0.066) & 0.385 (0.066)\\
36 & 0.040 (0.067) & 0.037 (0.067) & 0.195 (0.062) & 0.196 (0.062)\\
37 & 0.073 (0.055) & 0.082 (0.055) & 0.378 (0.064) & 0.377 (0.064)\\
38 & 0.280 (0.065) & 0.287 (0.065) & 0.336 (0.062) & 0.340 (0.062)\\
39 & -0.000 (0.060) & 0.002 (0.060) & -0.017 (0.068) & -0.020 (0.068)\\
40 & 0.034 (0.056) & 0.037 (0.056) & 0.342 (0.064) & 0.348 (0.064)\\
41 & -0.070 (0.058) & -0.071 (0.058) & -0.010 (0.058) & -0.015 (0.058)\\
42 & 0.036 (0.046) & 0.034 (0.046) & -0.053 (0.070) & -0.046 (0.070)\\
43 & -0.063 (0.044) & -0.065 (0.045) & -0.020 (0.061) & -0.018 (0.061)\\
44 & -0.098 (0.041) & -0.104 (0.041) & 0.072 (0.051) & 0.072 (0.052)\\
45 & -0.115 (0.041) & -0.116 (0.041) & 0.157 (0.062) & 0.153 (0.062)\\
\bottomrule
\end{tabular}